---------------------------------------------------------------------------
############################## .TEX FILE ###################################
---------------------------------------------------------------------------

\documentstyle[aps]{revtex}

\newcommand{\bel}[1]{\begin{equation}\label{#1}}
\newcommand{\be}{\begin{equation}}
\newcommand{\ee}{\end{equation}}

\newcommand{\beal}[1]{\begin{eqnarray}\label{#1}}
\newcommand{\bea}{\begin{eqnarray}}
\newcommand{\eea}{\end{eqnarray}}

\newcommand{\bean}{\begin{eqnarray*}}
\newcommand{\eean}{\end{eqnarray*}}

\newcommand{\ba}{\begin{array}}
\newcommand{\ea}{\end{array}}

\newcommand{\bab}{\begin{abstract}}
\newcommand{\eab}{\end{abstract}}

\newcommand{\bml}{\begin{mathletters}}
\newcommand{\eml}{\end{mathletters}}

\newcommand{\q}{\quad}
\newcommand{\qq}{\quad\quad}

\newcommand{\bfm}[1]{\mbox{\boldmath $#1$}}

\newcommand{\dv}{\partial}

\newcommand{\bam}{\left( \begin{array}}
\newcommand{\eam}{\end{array} \right)}

\newcommand{\baq}[4]{\left( \begin{array}{c}{#1}\\{#2}\\{#3}\\
{#4}\end{array} \right)}

\newcommand{\bamd}[4]{\left( \begin{array}{cc}{#1}&{#2}\\
{#3}&{#4}\end{array} \right)}

\newcommand{\bamq}[4]{\left( \begin{array}{cccc}{#1}&{#2}&{#3}&{#4}\\}
\newcommand{\bamc}[5]{\left( \begin{array}{ccccc}{#1}&{#2}&{#3}&{#4}&{#5}\\}

\newcommand{\law}{\leftarrow}

\newcommand{\raw}{\rightarrow}
\newcommand{\lraw}{\longrightarrow}

\newcommand{\lrw}{\leftrightarrow}

\newcommand{\ag}{\alpha}
\newcommand{\bg}{\beta}
\newcommand{\cg}{\gamma}
\newcommand{\dg}{\delta}
\newcommand{\eg}{\epsilon}
\newcommand{\lgg}{\lambda}

\newcommand{\sg}{\sigma}
\newcommand{\pg}{\phi}
\newcommand{\vpg}{\varphi}

\def\xc{{\mathchoice {\setbox0=\hbox{$\displaystyle\rm C$}\hbox{\hbox
to0pt{\kern0.4\wd0\vrule height0.9\ht0\hss}\box0}}
{\setbox0=\hbox{$\textstyle\rm C$}\hbox{\hbox
to0pt{\kern0.4\wd0\vrule height0.9\ht0\hss}\box0}}
{\setbox0=\hbox{$\scriptstyle\rm C$}\hbox{\hbox
to0pt{\kern0.4\wd0\vrule height0.9\ht0\hss}\box0}}
{\setbox0=\hbox{$\scriptscriptstyle\rm C$}\hbox{\hbox
to0pt{\kern0.4\wd0\vrule height0.9\ht0\hss}\box0}}}}

\def\xg{{\mathchoice {\setbox0=\hbox{$\displaystyle\rm G$}\hbox{\hbox
to0pt{\kern0.4\wd0\vrule height0.9\ht0\hss}\box0}}
{\setbox0=\hbox{$\textstyle\rm G$}\hbox{\hbox
to0pt{\kern0.4\wd0\vrule height0.9\ht0\hss}\box0}}
{\setbox0=\hbox{$\scriptstyle\rm G$}\hbox{\hbox
to0pt{\kern0.4\wd0\vrule height0.9\ht0\hss}\box0}}
{\setbox0=\hbox{$\scriptscriptstyle\rm G$}\hbox{\hbox
to0pt{\kern0.4\wd0\vrule height0.9\ht0\hss}\box0}}}}

\def\xi{{\rm I\!I}}

\def\xo{{\mathchoice {\setbox0=\hbox{$\displaystyle\rm O$}\hbox{\hbox
to0pt{\kern0.4\wd0\vrule height0.9\ht0\hss}\box0}}
{\setbox0=\hbox{$\textstyle\rm O$}\hbox{\hbox
to0pt{\kern0.4\wd0\vrule height0.9\ht0\hss}\box0}}
{\setbox0=\hbox{$\scriptstyle\rm O$}\hbox{\hbox
to0pt{\kern0.4\wd0\vrule height0.9\ht0\hss}\box0}}
{\setbox0=\hbox{$\scriptscriptstyle\rm O$}\hbox{\hbox
to0pt{\kern0.4\wd0\vrule height0.9\ht0\hss}\box0}}}}

\def\xq{{\mathchoice {\setbox0=\hbox{$\displaystyle\rm
Q$}\hbox{\raise
0.15\ht0\hbox to0pt{\kern0.4\wd0\vrule height0.8\ht0\hss}\box0}}
{\setbox0=\hbox{$\textstyle\rm Q$}\hbox{\raise
0.15\ht0\hbox to0pt{\kern0.4\wd0\vrule height0.8\ht0\hss}\box0}}
{\setbox0=\hbox{$\scriptstyle\rm Q$}\hbox{\raise
0.15\ht0\hbox to0pt{\kern0.4\wd0\vrule height0.7\ht0\hss}\box0}}
{\setbox0=\hbox{$\scriptscriptstyle\rm Q$}\hbox{\raise
0.15\ht0\hbox to0pt{\kern0.4\wd0\vrule height0.7\ht0\hss}\box0}}}}

\def\xs{{\mathchoice
{\setbox0=\hbox{$\displaystyle     \rm S$}\hbox{\raise0.5\ht0\hbox
to0pt{\kern0.35\wd0\vrule height0.45\ht0\hss}\hbox
to0pt{\kern0.55\wd0\vrule height0.5\ht0\hss}\box0}}
{\setbox0=\hbox{$\textstyle        \rm S$}\hbox{\raise0.5\ht0\hbox
to0pt{\kern0.35\wd0\vrule height0.45\ht0\hss}\hbox
to0pt{\kern0.55\wd0\vrule height0.5\ht0\hss}\box0}}
{\setbox0=\hbox{$\scriptstyle      \rm S$}\hbox{\raise0.5\ht0\hbox
to0pt{\kern0.35\wd0\vrule height0.45\ht0\hss}\raise0.05\ht0\hbox
to0pt{\kern0.5\wd0\vrule height0.45\ht0\hss}\box0}}
{\setbox0=\hbox{$\scriptscriptstyle\rm S$}\hbox{\raise0.5\ht0\hbox
to0pt{\kern0.4\wd0\vrule height0.45\ht0\hss}\raise0.05\ht0\hbox
to0pt{\kern0.55\wd0\vrule height0.45\ht0\hss}\box0}}}}

\def\xt{{\mathchoice {\setbox0=\hbox{$\displaystyle\rm
T$}\hbox{\hbox to0pt{\kern0.3\wd0\vrule height0.9\ht0\hss}\box0}}
{\setbox0=\hbox{$\textstyle\rm T$}\hbox{\hbox
to0pt{\kern0.3\wd0\vrule height0.9\ht0\hss}\box0}}
{\setbox0=\hbox{$\scriptstyle\rm T$}\hbox{\hbox
to0pt{\kern0.3\wd0\vrule height0.9\ht0\hss}\box0}}
{\setbox0=\hbox{$\scriptscriptstyle\rm T$}\hbox{\hbox
to0pt{\kern0.3\wd0\vrule height0.9\ht0\hss}\box0}}}}

\def\xz{{\mathchoice {\hbox{$\sf\textstyle Z\kern-0.4em Z$}}
{\hbox{$\sf\textstyle Z\kern-0.4em Z$}}
{\hbox{$\sf\scriptstyle Z\kern-0.3em Z$}}
{\hbox{$\sf\scriptscriptstyle Z\kern-0.2em Z$}}}}

\newcommand{\fs}{\footnotesize}

\newcommand{\bii}{\begin{itemize}}
\newcommand{\eii}{\end{itemize}}

\newcommand{\ben}{\begin{enumerate}}
\newcommand{\een}{\end{enumerate}}

\newcommand{\bq}{\begin{quote}}
\newcommand{\eq}{\end{quote}}

\newcommand{\bc}{\begin{center}}
\newcommand{\ec}{\end{center}}

\newcommand{\btb}{\begin{table}}
\newcommand{\etb}{\end{table}}

\newcommand{\bt}{\begin{tabular}}
\newcommand{\et}{\end{tabular}}

\newcommand{\br}{\begin{flushright}}
\newcommand{\er}{\end{flushright}}

\newcommand{\bl}{\begin{flushleft}}
\newcommand{\el}{\end{flushleft}}

\newcommand{\vs}[1]{\vspace*{#1}}

\newcommand{\new}{\pagebreak}

\newcommand{\bref}{\begin{references}}
\newcommand{\eref}{\end{references}}

\newcommand{\bb}{\begin{thebibliography}{99}}
\newcommand{\eb}{\end{thebibliography}}
\newcommand{\bi}{\bibitem}

\newcommand{\btp}{\begin{titlepage}}
\newcommand{\etp}{\end{titlepage}}

\newcommand{\go}{\section{Introduction}}
\newcommand{\con}{\section{Conclusions}}

\newcommand{\axp}[3]{Ann.~Phys.~(NY)                    {\bf #1},  #2  (19#3)}

\newcommand{\ixa}[3]{Int.~J.~Mod.~Phys.~A               {\bf #1},  #2  (19#3)}

\newcommand{\jxe}[3]{J.~Math.~Phys.                     {\bf #1},  #2  (19#3)} 

\newcommand{\jxg}[3]{J.~Phys.~A                         {\bf #1},  #2  (19#3)}

\newcommand{\mxb}[3]{Mod.~Phys.~Lett.~A                 {\bf #1},  #2  (19#3)}

\newcommand{\nxb}[3]{Nucl.~Phys.                        {\bf #1},  #2  (19#3)}

\newcommand{\nxd}[3]{Nuovo Cimento                      {\bf #1},  #2  (19#3)}

\newcommand{\pxa}[3]{Phys.~Essay.                       {\bf #1},  #2  (19#3)}

\newcommand{\pxf}[3]{Phys.~Rev.~D                       {\bf #1},  #2  (19#3)}

\newcommand{\pxh}[3]{Phys.~Rev.~Lett.                   {\bf #1},  #2  (19#3)}
\newcommand{\pxi}[3]{Phys.~Lett.                        {\bf #1},  #2  (19#3)}

\newcommand{\pxxa}[3]{Prog.~Theor.~Phys.                {\bf #1},  #2  (19#3)}

\newcommand{\xxx}[3]{{\bf #1},  #2  (19#3)}

\newcommand{\co}{\mbox{\boldmath $\cal C$}}
\newcommand{\qu}{\mbox{\boldmath $\cal H$}}
\newcommand{\oct}{\mbox{\boldmath $\cal O$}}
\newcommand{\rea}{\mbox{\boldmath $\cal R$}}
\newcommand{\glr}{GL(8, \; \rea )}
\newcommand{\glc}{GL(4,  \; \co )}

\title{TOWARDS AN OCTONIONIC WORLD}

\author{Stefano 
De Leo\thanks{{\sl deleos@le.infn.it , deleo@ime.unicamp.br}}$^{(a,b)}$ 
and Khaled Abdel-Khalek\thanks{{\sl khaled@le.infn.it}}$^{(a)}$} 
\address{$^{(a)}$~Dipartimento di Fisica - Istituto Nazionale di Fisica 
Nucleare\\
- Lecce, 73100, Italy -\\
and\\
$^{(b)}$~Instituto de Matem\'atica, Estatist\'{\i}ca e Computa\c{c}\~ao 
Cient\'{\i}fica, 
IMECC-UNICAMP\\
- CP 6065, 13081-970, Campinas, S.P., Brasil -}

\date{October 1997}

\begin{document}

\maketitle

\bab
In order to obtain a consistent formulation of octonionic quantum mechanics 
(OQM), we introduce left-right barred operators. Such operators enable us to 
find the translation rules between octonionic numbers and $8\times 8$ real 
matrices (a translation is also given for $4\times 4$ complex matrices). 
The use of a complex geometry allows us to overcome the hermiticity problem 
and define an appropriate momentum operator within OQM. As an application of 
our results, we develop an octonionic relativistic free wave 
equation, linear in the derivatives. Even if the wave functions are only 
one-component we show that four independent solutions, corresponding to 
those of the Dirac equation, exist.         
\eab

\renewcommand{\thefootnote}{\sharp\arabic{footnote}}

\go

In the early thirties, in order to explain the novel 
phenomena of that time, namely $\beta$--decay
 and the strong interactions, Jordan~\cite{jor} introduced a 
nonassociative but commutative algebra as a basic block for a new quantum
theory.  With the discovery that 
$3 \times 3$ hermitian octonionic matrices realize the Jordan 
postulate~\cite{wig,alb1}, octonions appeared, in quantum mechanics, 
for the first time. 
The hope of applying nonassociative algebras to physics was soon dashed 
with the Fermi 
theory of the $\beta$--decay and with the Yukawa model of nuclear forces. 
Octonions disappears from physics as soon after being introduced. Banished 
from physics, octonions continued their career in 
mathematics~\cite{om1,om2,om3,om4}. 
Semi-simple Lie groups, classified in four categories: orthogonal 
groups, unitary groups, symplectic groups and exceptional groups, were 
respectively associated with real, complex, quaternionic and octonionic 
algebras. Thus, such algebras became the core of the classification of 
possible symmetries in physics.

From the sixties onwards, there has been renewed and intense interest in 
the use of 
octonions in physics~\cite{gur1}. The octonionic algebra has been in fact 
linked with a number of interesting subjects: structure 
of interactions~\cite{pais}, $SU(3)$ color symmetry and quark 
confinement~\cite{gur2,mor}, standard model gauge group~\cite{dix}, 
exceptional GUT groups~\cite{gur3}, Dirac-Clifford algebra~\cite{edm}, 
nonassociative Yang-Mills theories~\cite{jos1,jos2}, space-time symmetries in ten 
dimensions~\cite{dav}, supersymmetry and supergravity 
theories~\cite{sup1,sup2}. 
Moreover, the recent successful application of 
quaternionic numbers in quantum mechanics~\cite{adl,adl1,qua1,qua2,qua3}, 
in particular 
in formulating a quaternionic Dirac equation~\cite{dir1,dir2,dir3,dir4}, 
suggests 
going one step further and using octonions as underlying numerical field. 
Nonassociative numbers are difficult to manipulate and so the use of the 
octonionic field within OQM and in particular in formulating the Dirac 
equation~\cite{pen} is non-trivial. Obviously, if we are not able to 
construct a suitable OQM, octonions will remain beautiful ghosts in search of 
a physical incarnation.

In this work, we overcome the problems due to the nonassociativity of the 
octonionic algebra by introducing left-right barred operators 
(which will be sometimes called generalized octonions). Such operators 
complete the mathematical material introduced in the recent papers 
(on octonionic representations and nonassociative gauge theories) of Joshi 
{\it et al.}~\cite{jos1,jos2}. Then, we 
investigate their relations  to $\glr$ and $\glc$. Establishing this 
relation we find 
interesting translation rules, which gives us the opportunity to formulate 
a consistent OQM. Both the quantum mechanics postulates and the octonions 
nonassociativity property will be respected. 

The philosophy behind the translation can be concisely expressed by the 
following sentence: ``There exists at least one version of octonionic
quantum mechanics where the standard quantum mechanics is reproduced''. 
The use of a complex scalar product~\cite{hor} (or complex geometry as called 
by Rembieli\'nski~\cite{rem}) will be the main tool to obtain such an OQM. 

Is there any other acceptable octonionic quantum theory? Do octonionic quantum 
theories necessitate complex geometry? At this stage these questions lack 
answers and the aim of our work is to clarify these points. 

We wish to stress that translation rules don't imply that our octonionic 
quantum world (with complex geometry) is equivalent to the standard quantum 
world. When translation 
fails the two worlds are not equivalent. An interesting case is supersymmetry. 
Since the number of spinor components will be reduced from 4 to 1, the 
number of degrees of freedom between bosonic and fermionic fields matches. 
So we need just one fermion and one boson without any auxiliary field.

Similar translation rules, between quaternionic quantum mechanics (QQM)
with complex geometry and standard quantum mechanics, have been recently 
found~\cite{qua2}. As an application, such rules can be exploited in 
reformulating in a natural way the electroweak sector of the standard 
model~\cite{qua3}.

This article is organized as follows: In section II, we give a brief 
introduction to the octonionic division algebra. In section III, we  
discuss generalized numbers and introduce barred operators. 
Working with nonassociative numbers we need to distinguish between 
left-barred and right-barred operators. In section IV, we investigate the 
relation between generalized octonions and $8\times 8$ real matrices. 
In this section, we also give the translation rules 
between octonionic barred operators and $\glc$, which will be very useful in 
formulating our OQM. After these mathematical sections, in section V, we 
show how the complex geometry allows us to overcome the hermiticity 
problem. In this section we also introduce the appropriate definition for the 
momentum operator (which satisfy the required commutation rules with our 
octonionic hamiltonian) and the new completeness relations.  
As application of our results, in section VI, we explicitly develop an 
octonionic Dirac equation and 
suggest possible difference between complex 
and octonionic quantum theories. Our conclusion are drawn in the final 
section.

\section{Octonionic algebra}

A remarkable theorem of Albert~\cite{alb2} shows that the only algebras, 
$\cal A$, over the reals, with unit element and admitting a real modulus 
function 
$N(a)$ ($a \in {\cal A}$) with the following properties
\bml
\bea
N(0) & = & 0 \q ,\\
N(a) & >& 0 \qq \mbox{if} \q a\neq 0 \q ,\\
N(ra) & = & \vert r \vert ~N(a) \qq (~r \in \rea~) \q ,\\
N(a_{1}a_{2}) & \leq & N(a_{1})+N(a_{2}) \q ,
\eea
\eml
are the reals, $\rea$, the complex, $\co$, the quaternions,   
$\qu$ ($\qu$ in honour of Hamilton~\cite{ham}), and the octonions, $\oct$   
(or Graves-Cayley numbers~\cite{gra,cay}). Albert's theorem generalizes 
famous nineteenth-century results of Frobenius~\cite{fro} and 
Hurwitz~\cite{hur}, who first reached the same conclusion but with the 
additional assumption that $N(a)^{2}$ is a quadratic form.

In addition to Albert's theorem on algebras admitting a modulus function 
$N(a)$, we can characterize the algebras $\rea$, $\co$, $\qu$ and 
$\oct$ by the concept of {\tt division algebra} (in which one has no nonzero 
divisors of zero). A classical theorem~\cite{bot,ker} states that the only 
division algebra over the reals are algebras of dimensions 1, 2, 4 and 
8, the only associative division algebras over the reals are $\rea$, $\co$ 
and $\qu$, whereas the  {\tt nonassociative} algebras include the octonions 
$\oct$ (an interesting discussion concerning nonassociative algebras is 
presented 
in~\cite{oku}). For a very nice review of aspects of the quaternionic and 
octonionic algebras see ref.~\cite{gur1} and the recent book of 
Adler~\cite{adl}.
In this paper we will deal with octonions and their generalizations.

We now summarize our notation for the octonionic algebra and 
introduce useful elementary properties to manipulate the nonassociative 
numbers. There is a number of equivalent ways to represent the octonions 
multiplication table. Fortunately, it is always possible to choose an 
orthonormal basis $(e_0 , \ldots ,e_7 )$ such that
\be
o=r_{0}+\sum_{m=1}^{7} r_{m}e_{m} \qq (~r_{0,...,7}~~ \mbox{reals}~) \q , 
\ee 
where $e_{m}$ are elements obeying the noncommutative and nonassociative 
algebra
\be
e_{m}e_{n}=-\dg_{mn}+ \eg_{mnp}e_{p} \qq 
(~\mbox{{\fs $m, \; n, \; p =1,..., 7$}}~) 
\q ,
\ee
with $\eg_{mnp}$ totally antisymmetric and equal to unity for the seven 
combinations 
\[
123, \; 145, \; 176, \; 246, \; 257, \; 347 \; \mbox{and} \; 365  
\]
(each cycle represents a 
quaternionic subalgebra). The norm, $N(o)$, for the octonions is defined by
\be
N(o)=(o^{\dag}o)^{\frac{1}{2}}=(oo^{\dag})^{\frac{1}{2}}=
(r_{0}^{2}+  ... + r_{7}^{2})^{\frac{1}{2}} \q ,
\ee
with the octonionic conjugate $o^{\dag}$ given by
\be
o^{\dag}=r_{0}-\sum_{m=1}^{7} r_{m}e_{m} \q . 
\ee 
The inverse is then 
\be
o^{-1}=o^{\dag}/N(o) \qq (~o\neq 0~) \q .
\ee
We can define an {\tt associator} (analogous to the usual algebraic 
commutator) as follows
\bel{ass}
\{x, \; y , \; z\}\equiv (xy)z-x(yz) \q ,
\ee
where, in each term on the right-hand, we must, first of all, perform 
the multiplication in brackets. 
Note that for real, complex and quaternionic numbers the associator is 
trivially null. For octonionic imaginary units we have
\bel{eqass}
\{e_{m}, \; e_{n}, \; e_{p} \}\equiv(e_{m}e_{n})e_{p}-e_{m}(e_{n}e_{p})=
2 \eg_{mnps} e_{s} \q ,
\ee
with $\eg_{mnps}$ totally antisymmetric and equal to unity for the seven 
combinations 
\[ 
1247, \; 1265, \; 2345, \; 2376, \; 3146, \; 3157 \; \mbox{and} \; 4567 \q .
\]
Working with octonionic numbers the associator~(\ref{ass}) is in general 
non-vanishing, however, the ``alternative condition'' is fulfilled
\bel{rul}
\{ x, \; y, \; z\}+\{ z, \; y, \; x\}=0 \q .
\ee

\section{Left/Right-barred operators}

In 1989, writing a quaternionic Dirac equation~\cite{dir2}, Rotelli introduced
a {\tt barred}  momentum operator
\be
-\bfm{\dv}\mid i \qq [~(-\bfm{\dv}\mid i)\psi\equiv -\bfm{\dv}\psi i~] \q .
\ee
In a recent paper~\cite{qua2}, based upon the Rotelli operators, 
{\tt partially generalized quaternions} 
\be
q+p\mid i \qq [~q, \; p \in \qu~] \q ,
\ee
have been used to formulate a quaternionic quantum mechanics. From 
the viewpoint of group structure, these barred numbers are very similar to 
complexified quaternions~\cite{mor2}
\be
q+{\cal I}p
\ee
(the imaginary unit ${\cal I}$ commutes with the quaternionic 
imaginary units $i, \; j, \; k$), but in physical problems, like eigenvalue 
calculations, tensor 
products, relativistic equations solutions, they give different results. 

A complete generalization for quaternionic numbers is represented by 
the following barred operators
\be
q_{1} + q_{2}\mid i + q_{3}\mid j + q_{4}\mid k \qq
[~q_{1,...,4} \in \qu~] \q ,
\ee
which we call {\tt fully generalized quaternions}, or simply generalized 
quaternions. Fully generalized quaternions, with their 16 linearly 
independent elements, form
a basis of $GL(4, \; \rea )$. They are successfully used 
to reformulate Lorentz space-time transformations~\cite{rel} and write down a 
one-component Dirac equation~\cite{dir4}.

Thus, it seems to us natural to investigate the existence of
{\tt generalized octonions}
\be
o_{0}+ \sum_{m=1}^{7} o_{m}\mid e_{m} \q .
\ee
Nevertheless, we must observe that an octonionic {\tt barred} operator, 
\bfm{a\mid b}, which acts on octonionic wave functions, $\psi$, 
\[ [~a\mid b~]~\psi \equiv a\psi b \q , \]
is not a well defined object. For $a\neq b$ the triple product $a\psi b$ 
could be either $(a\psi)b$ or $a(\psi b)$. So, in order to avoid the 
ambiguity due to the nonassociativity of 
the octonionic numbers, we need to define left/right-barred 
operators. We will indicate {\tt left-barred} operators by 
\bfm{a~)~b}, with $a$ and $b$ which represent octonionic numbers. They 
act on octonionic functions $\psi$ as follows
\bml
\be
[~a~)~b~]~\psi = (a\psi)b \q .
\ee
In similar way we can introduce {\tt right-barred} operators, defined by 
\bfm{a~(~b} ,
\be
[~a~(~b~]~\psi = a(\psi b) \q .
\ee
\eml
Obviously, there are barred-operators in which the nonassociativity is not 
of relevance, like 
\[ 1~)~a = 1~(~a \equiv 1\mid a \q . \]
Furthermore, from eq.~(\ref{rul}), we have
\[ \{ x, \; y, \; x\}=0 \q ,\]
so  
\[ a~)~a = a~(~a  \equiv a\mid a \q .\]
At first glance it seems that we must consider the following 
106 barred-operators:
\bc
\bt{lr}
$1, \; e_{m}, 1\mid e_{m}$ & {\fs ~~~~~(15 elements)} ,\\
$e_{m}\mid e_{m}$ &{\fs (7)} ,\\
$e_{m}~)~e_{n}$~~~~{\fs $(m\neq n)$} & {\fs (42)} ,\\
$e_{m}~(~e_{n}$~~~~{\fs $(m\neq n)$} & {\fs (42)} ,\\
{\fs $(m, \; n =1, ..., 7) \q .$} & 
\et
\ec
Nevertheless, it is possible to prove that each right-barred operator can be 
expressed by a suitable combination of left-barred operators. For example, 
from eq.~(\ref{rul}), by posing $x=e_{m}$ and $z=e_{n}$, we quickly 
obtain 
\bel{+}
e_{m}~(~e_{n} + e_{n}~(~e_{m} ~~\equiv ~~e_{m}~)~e_{n} + e_{n}~)~e_{m} \q .
\ee
So we can represent the most general octonionic operator by only left-barred 
objects 
\bel{go}
o_{0}+\sum_{m=1}^{7} o_{m}~)~e_{m} \qq 
[~o_{0, ...,7}~~ \mbox{octonions}~] \q ,
\ee
reducing to 64 the previous 106 elements. This suggests a 
correspondence between our generalized  octonions~(\ref{go}) and 
$GL(8, \; \rea)$ (a complete discussion about the 
above-mentioned relationship is given in the following section).

\section{Translation Rules}

The nonassociativity of octonions represents a challenge.
We overcome the problems due to the octonions nonassociativity by introducing 
left/right-barred operators. We 
discuss in the next subsection their relation to $GL(8, \; \rea)$. In that  
subsection, we present our translation idea and give some explicit examples 
which allow us to establish the isomorphism between our octonionic left/right 
barred operators and $GL(8, \; \rea)$. In subsection IV-b, we focus our 
attention on the group $GL(4, \; \co) \subset GL(8, \; \rea)$. In doing so, 
we find that only particular combinations of octonionic barred operators 
give us suitable candidates for the $GL(4, \; \co)$-translation. Finally, 
in subsection IV-c, we explicitly give three octonionic representations for the 
gamma-matrices.

\subsection*{IV-a. Relation between barred operators and $8\times 8$ real 
matrices}

In order to explain the idea of translation, let us look 
explicitly at the action of the operators $1\mid e_1$ and $e_2$,  on a generic 
octonionic  function $\vpg$
\be
\vpg = \vpg_0
 + e_1 \vpg_1 + e_2 \vpg_2 + e_3 \vpg_3
+ e_4 \vpg_4 + e_5 \vpg_5 + e_6 \vpg_6 + e_7 \vpg_7
\q [~\vpg_{0,\dots ,7} \in \rea~] \q .
\ee
We have
\bml
\beal{opa}
[~1\mid e_{1}~]~\vpg ~ \equiv ~\vpg e_1 & ~=~ & 
e_1 \vpg_0 - \vpg_1 - e_3 \vpg_2 + e_2 \vpg_3
- e_5 \vpg_4 + e_4 \vpg_5 + e_7 \vpg_6 - e_6 \vpg_7 \q , \\
 e_{2}\vpg & ~=~ & e_2 \vpg_0 - e_3 \vpg_1 - \vpg_2 + e_1 \vpg_3
+ e_6 \vpg_4 + e_7 \vpg_5 - e_4 \vpg_6 - e_5 \vpg_7
\q .
\eea
\eml
If we represent our octonionic function $\vpg$ by the following real column 
vector
\be
\vpg ~ \lrw ~ \left( \begin{array}{c}
\vpg_0\\
\vpg_1\\
\vpg_2\\
\vpg_3\\
\vpg_4\\
\vpg_5\\
\vpg_6\\
\vpg_7
\end{array}
\right) \q ,
\ee
we can rewrite the eqs.~(\ref{opa}-b) in matrix form,
\bml
\bea
\left(
\begin{array}{cccccccc}
0 & $-1$ & 0 & 0 & 0 & 0 & 0 &0\\
1 & 0 & 0 & 0 & 0 & 0 & 0 &0\\
0 & 0 & 0 & 1 & 0 & 0 & 0 &0\\
0 & 0 & $-1$& 0 & 0 & 0 & 0 &0\\
0 & 0 & 0 & 0 & 0 & 1 & 0 &0\\
0 & 0 & 0 & 0 &$ -1$& 0 & 0 &0\\
0 & 0 & 0 & 0 & 0 & 0 & 0 &$-1$\\
0 & 0 & 0 & 0 & 0 & 0 & 1 &0
\end{array}
\right)
\left( \begin{array}{c}
\vpg_0\\
\vpg_1\\
\vpg_2\\
\vpg_3\\
\vpg_4\\
\vpg_5\\
\vpg_6\\
\vpg_7
\end{array}
\right) & = &
\left( \begin{array}{c}
$-$\vpg_1\\
\vpg_0\\
\vpg_3\\
$-$\vpg_2\\
\vpg_5\\
$-$\vpg_4\\
$-$\vpg_7\\
\vpg_6
\end{array} \right) \q , \\
\left(
\begin{array}{cccccccc}
0 & 0 &$-1$ & 0 & 0 & 0 & 0 &0\\
0 & 0 & 0 & 1 & 0 & 0 & 0 &0\\
1 & 0 & 0 & 0 & 0 & 0 & 0 &0\\
0 & $-1$& 0 & 0 & 0 & 0 & 0 &0\\
0 & 0 & 0 & 0 & 0 & 0 & $-1$&0\\
0 & 0 & 0 & 0 & 0 & 0 & 0 &$-1$\\
0 & 0 & 0 & 0 & 1 & 0 & 0 &0\\
0 & 0 & 0 & 0 & 0 & 1 & 0 &0
\end{array}
\right)
\left( \begin{array}{c}
\vpg_0\\
\vpg_1\\
\vpg_2\\
\vpg_3\\
\vpg_4\\
\vpg_5\\
\vpg_6\\
\vpg_7
\end{array}
\right) & = &
\left( \begin{array}{c}
$-$\vpg_2\\
\vpg_3\\
\vpg_0\\
$-$\vpg_1\\
$-$\vpg_6 \\
$-$\vpg_7\\
\vpg_4\\
\vpg_5
\end{array} \right) \q . 
\eea
\eml
In this way we can immediately obtain a real matrix representation for the 
octonionic barred operators $1\mid e_{1}$ and $e_{2}$. Following this 
procedure we can construct the complete set of translation rules for the 
imaginary units $e_{m}$ and the barred operators $1\mid e_{m}$ 
(table A1a in appendix A1). In this paper we will 
use the Joshi notation~\cite{jos1}: $L_m$ and $R_m$ will represent the matrix 
counterpart of the octonionic operators $e_m$ and $1\mid e_m$,
\bel{j1}
L_{m}~\lrw ~e_{m} \q \mbox{and} \q R_{m} ~\lrw ~1\mid e_{m} \q .
\ee

At first glance it seems that our translation doesn't work. If we extract, 
from the table A1a, the matrices corresponding to $e_{1}$, $e_{2}$ and 
$e_{3}$, namely,
\[
L_{1}= \left(
\begin{array}{cccccccc}
0 & $-1$ & 0 & 0 & 0 & 0 & 0 &0\\
1 & 0 & 0 & 0 & 0 & 0 & 0 &0\\
0 & 0 & 0 & $-1$ & 0 & 0 & 0 &0\\
0 & 0 & 1 & 0 & 0 & 0 & 0 &0\\
0 & 0 & 0 & 0 & 0 & $-1$ &0 &0\\
0 & 0 & 0 & 0 & 1 & 0 & 0 &0\\
0 & 0 & 0 & 0 & 0 & 0 & 0 &1\\
0 & 0 & 0 & 0 & 0 & 0 & $-1$ &0
\end{array}
\right) \q , \q 
L_{2} = \left(
\begin{array}{cccccccc}
0 & 0 & $-1$ & 0 & 0 & 0 & 0 &0\\
0 & 0 & 0 & 1 & 0 & 0 & 0 &0\\
1 & 0 & 0 & 0 & 0 & 0 & 0 &0\\
0 & $-1$ & 0 & 0 & 0 & 0 & 0 &0\\
0 & 0 & 0 & 0 & 0 & 0 &$-1$ &0\\
0 & 0 & 0 & 0 & 0 & 0 & 0 & $-1$\\
0 & 0 & 0 & 0 & 1 & 0 & 0 & 0\\
0 & 0 & 0 & 0 & 0 & 1 & 0 &0
\end{array}
\right) \q , \q 
L_{3} =  \left(
\begin{array}{cccccccc}
0 & 0 & 0 & $-1$ & 0 & 0 & 0 &0\\
0 & 0 & $-1$ & 0 & 0 & 0 & 0 &0\\
0 & 1 & 0 & 0 & 0 & 0 & 0 &0\\
1 & 0 & 0 & 0 & 0 & 0 & 0 &0\\
0 & 0 & 0 & 0 & 0 & 0 &0 & $-1$\\
0 & 0 & 0 & 0 & 0 & 0 & 1 &0\\
0 & 0 & 0 & 0 & 0 & $-1$ & 0 &0\\
0 & 0 & 0 & 0 & 1 & 0 & 0 &0
\end{array}
\right) \q ,
\]
we find 
\be
L_{1}L_{2} \neq L_{3} \q .
\ee
In obvious contrast with the octonionic relation
\be
e_1 e_2 = e_3 \q .
\ee
This bluff is soon explained. In deducing our translation rules, we 
understand octonions as operators, and so they must be applied to a 
certain octonionic function, $\vpg$. If we have the following octonionic 
relation
\bml
\be
(e_1 e_2)\vpg = e_3 \vpg 
\ee
the matrix counterpart will be  
\be
L_3 \vpg \q ,
\ee
\eml
since the matrix counterparts are defined by their action upon the ``wave 
function'' and {\tt not} upon another ``operator''. Whereas, 
\bml
\be
e_1 (e_2 \vpg) \neq e_3 \vpg 
\ee
will be translated by
\be
L_1 L_2 \vpg \neq L_3 \vpg \q .
\ee
\eml
We have to differentiate between two kinds of multiplication, 
``~$\cdot$~''
and ``~$\times $~''. At the level of octonions, one has 
\be
e_1 \cdot e_2 = e_3 \q ,
\ee
but at level of octonionic operators
\be
e_1 \times e_2 \neq e_3 
\ee
\[
~[~e_{1}\times e_{2} \equiv e_{3} + e_{1}~)~e_2 - e_1~(~e_2 \q 
~~\mbox{$\raw$ ~see below}~] \q .
\]
After completing our translation rules we will return to this point and  
discuss the multiplication rules for octonionic barred operators.

Working with left/right barred operators we show how the nonassociativity is 
inherent in our representation. Such operators enable us to reproduce the 
octonions nonassociativity by the matrix algebra. Consider for example
\be
[~e_{3}~)~e_{1}~]~ \vpg ~ \equiv ~ (e_3 \vpg)e_1 
~=~  e_2 \vpg_0 - e_3 \vpg_1 + \vpg_2 - e_1 
\vpg_3 - e_6 \vpg_4
- e_7 \vpg_5 + e_4 \vpg_6 +e_5 \vpg_7 \q .
\ee
This equation will be translated into
\be
\left(
\begin{array}{cccccccc}
0 & 0 & 1 & 0 & 0 & 0 & 0 &0\\
0 & 0 & 0 &$-1$ & 0 & 0 & 0 &0\\
1 & 0 & 0 & 0 & 0 & 0 & 0 &0\\
0 &$-1$ & 0 & 0 & 0 & 0 & 0 &0\\
0 & 0 & 0 & 0 & 0 & 0 &1 &0\\
0 & 0 & 0 & 0 & 0 & 0 & 0 &1\\
0 & 0 & 0 & 0 & $-1$ & 0 & 0 &0\\
0 & 0 & 0 & 0 & 0 & $-1$ & 0 &0
\end{array}
\right)
\left( \begin{array}{c}
\vpg_0\\
\vpg_1\\
\vpg_2\\
\vpg_3\\
\vpg_4\\
\vpg_5\\
\vpg_6\\
\vpg_7
\end{array}
\right) = 
\left( \begin{array}{c}
\vpg_2\\
$-$\vpg_3 \\
\vpg_0\\
$-$\vpg_1\\
\vpg_6 \\
\vpg_7\\
$-$\vpg_4\\
$-$\vpg_5
\end{array} \right) 
\q .
\ee
Whereas,
\be
[~e_{3}~(~e_{1}~]~ \vpg ~ \equiv ~ e_3 (\vpg e_1)
~=~  e_2 \vpg_0 - e_3 \vpg_1 + \vpg_2  - e_1 
\vpg_3   + e_6 \vpg_4 + e_7 \vpg_5 - e_4 \vpg_6 - e_5 \vpg_7 \q ,     
\ee
will become 
\be
\left(
\begin{array}{cccccccc}
0 & 0 & 1 & 0 & 0 & 0 & 0 &0\\
0 & 0 & 0 &$-1$ & 0 & 0 & 0 &0\\
1 & 0 & 0 & 0 & 0 & 0 & 0 &0\\
0 &$-1$ & 0 & 0 & 0 & 0 & 0 &0\\
0 & 0 & 0 & 0 & 0 & 0 & $-1$ &0\\
0 & 0 & 0 & 0 & 0 & 0 & 0 & $-1$\\
0 & 0 & 0 & 0 & 1 & 0 & 0 &0\\
0 & 0 & 0 & 0 & 0 & 1 & 0 &0
\end{array}
\right)
\left( \begin{array}{c}
\vpg_0\\
\vpg_1\\
\vpg_2\\
\vpg_3\\
\vpg_4\\
\vpg_5\\
\vpg_6\\
\vpg_7
\end{array}
\right) = 
\left( \begin{array}{c}
\vpg_2\\
$-$\vpg_3 \\
\vpg_0\\
$-$\vpg_1\\
$-$\vpg_6 \\
$-$\vpg_7\\
\vpg_4\\
\vpg_5
\end{array} \right)  
\q . 
\ee
The nonassociativity is then reproduced since left and right barred 
operators, like
\[ e_{3}~)~e_{1} \q \mbox{and} \q  e_{3}~(~e_{1} \]
are represented by different matrices. 
The complete set of translation rules for 
left/right-barred operators is given in the tables A1-L/R. Using 
Mathematica~\cite{math},  
we have proved the linear independence of
the 64 elements which appear in the tables ``A1a-b \& A1-L '' and ``A1 \& A1-R ''. 
So, our barred operators  form a complete basis for any
$8 \times 8$ real matrix and this establishes the isomorphism between
$\glr$ and generalized octonions. We provide the necessary tables for 
translating any generic $8\times 8$ real matrix into left/right-barred 
operators within appendix A2.

The matrix representation for left/right barred operators can be quickly 
obtained by suitable multiplications of the matrices $L_{m}$ and $R_{m}$. 
Let us clear up our assertion. From the tables A1-L/R we can extract the 
matrices which correspond to the operators
\[ e_m~)~e_n \qq \mbox{and} \qq e_{m}~(~e_n \qq ,\]
which we call, respectively,
\[ M^{L}_{mn} \qq \mbox{and} \qq M^{R}_{mn}  \q .\]
Our left/right barred operators can be represented by an ordered action of 
the operators $e_{m}$ and $1\mid e_{m}$, and so we can related the matrices 
$M^{L}_{mn}$ and  $M^{R}_{mn}$, quoted in tables A1-L/R, to the matrices 
$L_m$ and $R_m$, given in table A1a. Explicitly,
\bml
\bea
M^{L}_{mn} & ~\equiv ~& R_n L_m \q ,\\
M^{R}_{mn} & ~\equiv ~& L_m R_n \q .
\eea
\eml
The previous discussions concerning the octonions nonassociativity and the 
isomorphism between $\glr$ and generalized octonions, can be 
now, elegantly, presented as follows.\\
{\tt 1 - Matrix representation for octonions nonassociativity.} 
\bea
M^{L}_{mn} ~\neq ~ M^{R}_{mn} \qq 
[ ~R_n L_m  \neq L_m R_n ~~ \mbox{{\fs for $m\neq n$}} ] \q .
\eea
{\tt 2 - Isomorphism between} \bfm{\glr} {\tt and generalized octonions.}\\
If we rewrite our 106 barred operators by real matrices:
\bc
\bt{lr}
$1, \; L_{m}, \; R_{m}$ & {\fs ~~~~~(15 matrices)} ,\\
$M\equiv L_m R_m =R_m L_m $ &{\fs (7)} ,\\
$M^{L}_{mn}\equiv R_n L_{m}$ ~~~~{\fs $(m\neq n)$} & {\fs (42)} ,\\
$M^{R}_{mn}\equiv L_n R_{m}$ ~~~~{\fs $(m\neq n)$} & {\fs (42)} ,\\
{\fs $(m, \; n =1, ..., 7) \q ;$} & 
\et
\ec
we have two different basis for $\glr$:
\bc
\bt{cl}
{\fs (1)} & ~~~~~$1 \; , ~ L_{m} \; , ~ R_{m} \; , ~ R_n L_{m} \q ,$\\ 
 & \\
{\fs (2)} & ~~~~~$1 \; , ~ L_{m} \; , ~ R_{m} \; , ~ L_{m} R_n \q .$ 
\et
\ec

We now remark some difficulties deriving from the octonions nonassociativity. 
When we translate from generalized octonions
to $8 \times 8$ real matrices there is no problem. For example, in the 
octonionic equation
\bel{ww}
e_{4} \{[(e_{6}\vpg ) e_{1}]e_{5}\} \q , 
\ee
we quickly recognize the following left-barred operators,
\[
e_{4}~(~e_{5} \q \mbox{and} \q e_{6}~)~e_{1}  \q .
\]
Using our tables we can translate eq.~(\ref{ww}) into
\be
M^{L}_{45}~M^{L}_{61}~\vpg \q .
\ee

Nevertheless, in going from $8\times 8$ real matrices to octonions we should 
be careful in ordering. For example,
\bel{mm}
A B ~\vpg ~~~
\ee
can be understood as
\bml
\bel{mm1}
(AB) \vpg \q ,
\ee
or
\bel{mm2}
A (B \vpg) \q .
\ee
\eml
Which is the right equation? To find the solution let us, explicitly, 
use particular matrices. Defining
\bml
\bea
A ~\raw ~ L_6 =
\left( \begin{array}{cccccccc}
0 & 0 & 0 & 0 & 0 & 0 & $-1$& 0\\
0 & 0 & 0 & 0 & 0 & 0 & 0 & $-1$\\
0 & 0 & 0 & 0 & $-1$& 0 & 0 & 0\\
0 & 0 & 0 & 0 & 0 & 1 & 0 & 0\\
0 & 0 & 1 & 0 & 0 & 0 & 0 & 0\\
0 & 0 & 0 &$ -1$& 0 & 0 & 0 & 0\\
1 & 0 & 0 & 0 & 0 & 0 & 0 & 0\\
0 & 1 & 0 & 0 & 0 & 0 & 0 & 0
\end{array} \right) & ~\lrw ~&  e_{6} \q ,\\
B ~ \raw ~ M^{L}_{31} =
\left( \begin{array}{cccccccc}
0 & 0 & 1 & 0 & 0 & 0 & 0 & 0\\
0 & 0 & 0 & $-1$ & 0 & 0 & 0 & 0\\
1 & 0 & 0 &  0 & 0 & 0 & 0 & 0\\
0 & $-1$ & 0 & 0 & 0 & 0 & 0 & 0\\
0 & 0 & 0 & 0 & 0 & 0 & 1 & 0\\
0 & 0 & 0 & 0 & 0 & 0 & 0 & 1\\
0 & 0 & 0 & 0 & $-1$ & 0 & 0 & 0\\
0 & 0 & 0 & 0 & 0 & $-1$ & 0 & 0
\end{array} \right) &~ \lrw ~& e_{3}~)~e_{1} \q ,
\eea
\eml
the previous matrix eqs.~(\ref{mm1}-b), respectively, become
\bml
\be
e_{6} ~ \times ~[~e_{3}~)~e_{1}~]~ \vpg \q ,
\ee
and
\be
e_{6} [(e_{3}\vpg ) e_{1} ] \q .
\ee
\eml
We know that ``$~\times ~$'' multiplication is different from the standard 
octonionic multiplication, so
\[
e_{6}~  \times ~[~e_{3}~)~e_{1}~] ~\neq~ -~e_{5}~)~e_{1} \q .
\]
Using appendix A2 and translating the matrix $AB$, we can obtain the 
octonionic operators which corresponds to
\[
e_{6}~  \times ~[~e_{3}~)~e_{1}~] \q ,
\]
explicitly, we have
\[
~\{~ e_4 - 1\mid e_4 -2e_1~)~e_5 - e_5~)~e_1
-e_6~)~e_2 + e_2~)~e_6 + 2e_7~)~e_3 - e_3~)~e_7 ~\}/3 \q .
\]
Its complicated form suggests us to choose eq.~(\ref{mm2}) for translating 
eq.~(\ref{mm}). In general 
\be
ABC \ldots Z \vpg \equiv A(B(C \ldots (Z\vpg) \ldots )) \q .
\ee

Only for $e_m$ and $1\mid e_m$, we have  simple ``$~\times ~$''-multiplication 
rules. In fact, utilizing the associator properties we find
\bml
\beal{pp}
e_m(e_n\vpg) & ~=~ & (e_m e_n)\vpg + (e_m\vpg)e_n - e_m (\vpg e_n) \q ,\\ 
(\vpg e_m) e_n & ~=~ & \vpg (e_m e_n) - (e_m\vpg)e_n + e_m (\vpg e_n) \q .
\eea
\eml
Thus, 
\bml
\bea
e_m ~\times ~ e_n & ~\equiv~ & 
-\dg_{mn} + \eg_{mnp} e_p + e_m~)~e_n - e_m ~(~e_n \q ,\\ 
~[~1\mid e_n~] ~\times ~ [~1\mid e_m~] & ~\equiv~ & 
-\dg_{mn} + \eg_{mnp} e_p - e_m~)~e_n + e_m ~(~e_n \q .
\eea
\eml
At the beginning of this subsection, we noted that the correspondence 
between the matrices, $L_m$, and the octonionic imaginary units $e_m$ is in 
contrast with the standard octonionic relations
\bel{orel}
e_m e_n = -\dg_{mn} + \eg_{mnp} e_p \q ,
\ee
for example, look at
\[
L_1 L_2 \neq L_3 \q .
\]
Introducing a new matrix multiplication, ``~$\circ$~'', we can quickly 
reproduce the nonassociative octonionic algebra. From eq.~(\ref{pp}), we 
find
\be
L_m L_n ~ \vpg = L_m \circ L_n ~ \vpg + [R_n , \; L_m ] ~ \vpg \q ,
\ee
so we can relate the new matrix multiplication, ``~$\circ$~'', to the 
standard matrix multiplication (row by column) as follows
\bel{new}
L_m \circ L_n  \equiv L_m L_n + [R_n , \; L_m ] \q .
\ee
Eq.~(\ref{orel}) is then translated by
\be
L_m \circ L_n = -\dg_{mn} + \eg_{mnp} L_p \q .
\ee

\subsection*{IV-b. Relation between barred operators and $4\times 4$ complex 
matrices}

Some complex groups play a critical role
in physics. No one can deny the importance
of $U(1, \; \co)$ or $SU(2, \; \co)$. In relativistic
quantum mechanics, $\glc$ is essential in writing the 
Dirac equation. Having $\glr$, we should be able
to extract its subgroup $\glc$. So, we can 
translate the famous Dirac-gamma matrices and 
write down a new octonionic Dirac equation.

Let us show how we can isolate our 32 basis of $\glc$:\\
If we analyse the action of 
left-barred operators on our octonionic wave functions
\be
\psi = \psi_{1} + e_{2} \psi_{2} + e_{4} \psi_{3} + e_{6} \psi_{4} \qq
[~\psi_{1,  ..., 4} \in \bfm{\cal C}(1, \; e_{1})~] \q ,
\ee
we find, for example,
\bean
~[~1\mid e_{1}~]~\psi ~\equiv ~ & \psi e_{1} & ~=~ \psi_{1} + e_{2} (e_{1}\psi_{2}) + 
e_{4} (e_{1}\psi_{3}) + e_{6} (e_{1}\psi_{4}) \q ,\\
 & e_{2}\psi & ~=~  -\psi_{2} + e_{2} \psi_{1} -
e_{4} \psi_{4}^{*} + e_{6} \psi_{3}^{*} \q ,\\
~[~e_{3}~)~e_{1}~]~\psi ~\equiv ~ & (e_{3}\psi) e_{1} & ~=~ \psi_{2} + e_{2} \psi_{1} +
e_{4} \psi_{4}^{*} - e_{6} \psi_{3}^{*} \q ,
\eean
the action of our barred operators  is quoted in the tables B1a-b and B1-L/R, 
given in appendix B1.

Following the same methodology of the previous section,
we can immediately note a correspondence between 
the complex matrix $i {\bf 1}_{4\times 4}$ and the octonionic   
barred operator $1\mid e_{1}$
\be
\bamq{i}{0}{0}{0} 0 & i & 0 & 0\\ 0 & 0 & i & 0\\ 0 & 0 & 0 & i \eam
~ \lrw ~1\mid e_{1} \q .
\ee
Observe that we are working with the symplectic decomposition of octonions
\be
\baq{\psi_{1}}{\psi_{2}}{\psi_{3}}{\psi_{4}} ~ \lrw ~
\psi_{1} + e_{2} \psi_{2} + e_{4} \psi_{3} + e_{6} \psi_{4} \q .
\ee
Such an identification will be much clearer when we introduce a {\tt complex 
geometry}. In fact, choosing a complex projection for our scalar products, 
\[ \psi_{1}, \; e_{2} \psi_{2}, \;  e_{4} \psi_{3}, \; e_{6} \psi_{4} \]
will represent complex-orthogonal states.

The translation doesn't work for all barred operators. Let us show it, 
explicitly. For example, we cannot find a $4\times 4$ complex matrix which, 
acting on
\[
\baq{\psi_{1}}{\psi_{2}}{\psi_{3}}{\psi_{4}} \q , 
\]
gives the column vector
\[
\left( \begin{array}{c}
$-$\psi_{2} \\ \psi_{1} \\ $-$\psi_{4}^{*} \\ \psi_{3}^{*}
\end{array} \right) \qq \mbox{or} \qq 
\left( \begin{array}{c} 
\psi_{2} \\ \psi_{1} \\ \psi_{4}^{*} \\ $-$\psi_{3}^{*}
\end{array} \right) \q ,
\]
and so we have not the possibility to relate 
\[ e_{2} \qq \mbox{or} \qq  e_{3}~)~e_{1} \]
with a complex matrix. Nevertheless, a combined action of such operators
gives us
\[
e_{2}\psi + (e_{3}\psi)e_{1} = 2 e_{2}\psi_{1} \q ,
\]
and it allows us to represent the octonionic barred operator 
\bml
\be 
e_{2} \; + \; e_{3}~)~e_{1} \q , 
\ee
by the $4\times 4$ complex matrix
\be
\bamq{0}{0}{0}{0} 2 & 0 & 0 & 0\\ 0 & 0 & 0 & 0\\ 0 & 0 & 0 & 0 \eam \q .
\ee
\eml
Following this procedure we can represent a generic $4\times 4$ complex 
matrix by octonionic barred operators. The explicit correspondence-tables are 
given in appendix B2. 

We conclude this subsection discussing a point which 
will be relevant to an appropriate definition for the octonionic momentum 
operator (subsection V-b): The operator $1\mid e_{1}$ (represented by the 
matrix 
$i {\bf 1}_{4\times 4}$) commutes with all operators which can be translated 
by  $4\times 4$ complex matrices (see appendix B2). This is not generally true 
for a generic octonionic operator. For example, we can show that the 
operator $1\mid e_{1}$ doesn't commute with $e_{2}$, explicitly
\bml
\bea
e_2 ~ \{ ~[~1\mid e_1 ~] ~\psi ~\} &~\equiv  e_{2}(\psi e_{1}) & ~=~ -e_1 \psi_2 - e_3 \psi_1 - 
e_5 \psi_4^* - e_7 \psi_3^* \q ,\\
~[~1\mid e_1 ~] ~ \{ e_2 ~\psi ~\} & ~\equiv  (e_{2} \psi) e_{1} & ~=~ -e_1 \psi_2 - e_3 \psi_1 + 
e_5 \psi_4^* + e_7 \psi_3^*
  \q .
\eea
\eml
The interpretation is simple: $e_{2}$ cannot be represented by a $4\times 4$ 
complex matrix.

\subsection*{IV-c. Octonionic representations of the gamma-matrices.}
We conclude this section by showing explicitly three octonionic representation 
for the Dirac gamma-matrices~\cite{itz}:\\
\bc
1-{\tt Dirac representation,}
\ec
\bml
\beal{odgm1}
\cg^{0} & = & \frac{1}{3} -\frac{2}{3} \sum_{m=1}^{3} e_{m}\mid e_{m} +
\frac{1}{3} \sum_{n=4}^{7} e_{n} \mid e_{n} \q ,\\
\cg^{1} & = & -\frac{2}{3} e_{6} -\frac{1}{3}\mid e_{6} + e_{5}~)~e_{3} - 
e_{3}~)~e_{5}  - \frac{1}{3} \sum_{p, \; s =1}^{7} \eg_{ps6} e_{p}~)~e_{s} 
\q ,\\
\cg^{2} & = & -\frac{2}{3} e_{7} -\frac{1}{3}\mid e_{7} + e_{3}~)~e_{4} - 
e_{4}~)~e_{3}  - \frac{1}{3} \sum_{p, \; s =1}^{7} \eg_{ps7} e_{p}~)~e_{s} 
\q ,\\
\cg^{3} & = & -\frac{2}{3} e_{4} -\frac{1}{3}\mid e_{4} + e_{7}~)~e_{3} - 
e_{3}~)~e_{7}  - \frac{1}{3} \sum_{p, \; s =1}^{7} \eg_{ps4} e_{p}~)~e_{s} 
\q ;
\eea
\eml
\bc
2-{\tt Majorana representation,}
\ec
\bml
\beal{odgm2}
\cg^{0} & = & 
\frac{1}{3}  e_7 - 
\frac{1}{3} \mid e_7 
+ e_3~)~e_4 - e_5~)~e_2 + e_6~)~e_1
- \frac{1}{3} \sum_{p, \; s=1}^{7} \eg_{ps7} e_p~)~e_s \q ,\\
\cg^{1} & = & \frac{2}{3} e_{1}  + \frac{1}{3}\mid e_{1} 
+ e_{5}~)~e_{4} 
 - e_{4}~)~e_{5}  + \frac{1}{3} \sum_{p, \; s =1}^{7} 
\eg_{ps1} e_{p}~)~e_{s} 
\q ,\\
\cg^{2} & = & \frac{2}{3} e_{7} +\frac{1}{3}\mid e_{7} 
+ e_{4}~)~e_{3} -  
e_{3}~)~e_{4}  + \frac{1}{3} \sum_{p, \; s =1}^{7} \eg_{ps7} e_{p}~)~e_{s} 
\q ,\\
\cg^{3} & = & \frac{2}{3} e_{3} +\frac{1}{3}\mid e_{3} + 
e_{7}~)~e_{4} - 
e_{4}~)~e_{7}  + \frac{1}{3} \sum_{p, \; s =1}^{7} \eg_{ps3} e_{p}~)~e_{s} 
\q ;
\eea
\eml
\bc
3-{\tt Chiral representation,}
\ec
\bml
\beal{odgm3}
\cg^{0} & = & 
\frac{1}{3} e_4 -\frac{1}{3}\mid e_4 
+  e_7~)~e_3 - e_2~)~e_6 + e_5~)~e_1 
- \frac{1}{3}
\sum_{p, \; s =1}^{7} \eg_{ps4} e_{p}~)~e_{s} 
\q ,\\
\cg^{1} & = & -\frac{2}{3} e_{6} -\frac{1}{3}\mid e_{6} + e_{5}~)~e_{3} - 
e_{3}~)~e_{5}  - \frac{1}{3} \sum_{p, \; s =1}^{7} \eg_{ps6} e_{p}~)~e_{s} 
\q ,\\
\cg^{2} & = & -\frac{2}{3} e_{7} -\frac{1}{3}\mid e_{7} + e_{3}~)~e_{4} - 
e_{4}~)~e_{3}  - \frac{1}{3} \sum_{p, \; s =1}^{7} \eg_{ps7} e_{p}~)~e_{s} 
\q ,\\
\cg^{3} & = & -\frac{2}{3} e_{4} -\frac{1}{3}\mid e_{4} + e_{7}~)~e_{3} - 
e_{3}~)~e_{7}  - \frac{1}{3} \sum_{p, \; s =1}^{7} \eg_{ps4} e_{p}~)~e_{s} 
\q .
\eea
\eml

\section{Octonionic physical world}

We organize this section in three subsections. In subsection V-a, we discuss 
the Dirac algebra and its problems related to the 
nonassociativity of the octonionic numbers. In subsection V-b, we introduce 
the concept of complex geometry and define an appropriate momentum operator. 
In the final subsection we present the octonionic completeness relations.

\subsection*{V-a. Dirac algebra}

In the previous section we have given the gamma-matrices in three different 
octonionic representations. Obviously, we can investigate the 
possibility of having a more simpler representation for our octonionic 
$\cg^{\mu}$-matrices, without translation. 

Why not
\[ e_{1} \; , \q e_{2} \; , \q e_{3} \q \mbox{and} \q e_{4}\mid e_{4} ~~~~\]
or
\[ e_{1} \; , \q e_{2} \; , \q e_{3} \q \mbox{and} \q e_{4}~)~e_{1} \q ?\]
Apparently, they represent suitable choices. Nevertheless, the octonionic 
world is full of hidden traps and so we must proceed with prudence. 
Let us start from the standard Dirac equation
\be 
\cg^\nu p_{\nu} \psi=m\psi \q ,
\ee
(we will discuss the momentum operator in the following subsection, 
for the moment, $p_{\nu}$ represents the ``real'' eigenvalue of the 
momentum operator) and apply 
$\cg^{\mu} p_{\mu}$ to our equation 
\be 
\cg^{\mu} p_{\mu}(\cg^{\nu} p_{\nu} \psi)=m \cg^{\mu} p_{\mu} \psi \q .
\ee
The previous equation can be concisely rewritten as
\be 
p^{\mu} p_{\nu} \cg^{\mu} (\cg^{\nu} \psi)=m^{2} \psi \q .
\ee
Requiring that each component of $\psi$ satisfy the standard Klein-Gordon 
equation  we find the Dirac condition, which becomes in the octonionic world
\bel{odc}
\cg^{\mu}(\cg^{\nu}\psi)+\cg^{\nu}(\cg^{\mu}\psi)=2g^{\mu \nu} \psi \q ,
\ee
(where the parenthesis are relevant because of the octonions nonassociative 
nature). Using octonionic numbers and no barred operators we can obtain, 
from~(\ref{odc}), the standard Dirac condition
\bel{sdc}
\{ \cg^{\mu}, \; \cg^{\nu} \} = 2 g^{\mu \nu} \q .
\ee
In fact, recalling the associator property 
[which follows from eq.~(\ref{eqass})] 
\[ \{a, \; b, \; \psi \} = - \{b, \; a, \; \psi \} \qq 
[~a, \; b \q \mbox{octonionic numbers}~] \q , \]
we quickly find the following correspondence relation 
\[ (ab+ba)\psi=a(b\psi)+b(a\psi) \q . \]
We have no problem to write down three suitable gamma-matrices which 
satisfy the Dirac condition~(\ref{sdc}),
\be
(\cg^{1}, \; \cg^{2}, \; \cg^{3}) \equiv (e_{1}, \; e_{2}, \; e_{3}) \q ,
\ee
but, barred operators like 
\[ e_{4}\mid e_{4} \q \mbox{or} \q e_{4}~)~e_{1} \]
cannot represent the matrix $\cg^{0}$. From the tables B1 and B2 (appendix B),
after straightforward algebraic manipulations, one can prove that 
the barred operator, $e_{4}\mid e_{4}$,  doesn't anticommute 
with $e_{1}$, 
\bea 
e_{1}(e_{4}\psi e_{4})+e_{4}(e_{1}\psi)e_{4} & = & 
-2 (e_{3}\psi_2 + e_7 \psi_4 ) \neq  0 \q ,
\eea
whereas $e_{4}~)~e_{1}$ anticommutes with $e_1$
\bml
\bea 
e_{1}[(e_{4}\psi) e_{1}]+[e_{4}(e_{1}\psi)]e_{1} & = &  0 \q ,
\eea
but
\bea
\{ e_{4}[(e_{4}\psi) e_{1} ] \} e_1  & = & \psi_1 -e_2 \psi_2 +e_4 \psi_3
 -e_6 \psi_4 \neq \psi \q .
\eea
\eml

Thus, we must be satisfied with the octonionic 
representations given in the previous section. In the following subsection, 
we discuss two interesting questions: Do the octonionic imaginary units 
$e_{1}$, $e_{2}$, $e_{3}$ satisfy all the gamma-matrices properties? What 
about their hermiticity?

\subsection*{V-b. Complex geometry and octonionic momentum operator}

We begin this subsection by presenting an apparently hopeless problem related 
to the nonassociativity of the octonionic field. Working in quantum 
mechanics we require that an antihermitian operator satisfies the following 
relation
\be
\int d{\bf  x}~ \psi^{\dag} (A\pg)= -\int d{\bf  x}~ (A\psi)^{\dag} \pg \q .
\ee
In octonionic quantum mechanics (OQM) we can immediately verify that 
\bfm{\dv} represents an antihermitian operator with all the properties of a 
translation operator. Nevertheless, while in complex (CQM) and quaternionic 
(QQM) quantum mechanics we can define a corresponding hermitian operator 
multiplying by an imaginary unit the operator \bfm{\dv}, one encounters in 
OQM the following problem:
\bc
{\tt no imaginary unit,} \bfm{e_m} {\tt , represents an antihermitian operator \q .}
\ec
In fact, the nonassociativity of the octonionic algebra implies, in general 
(for arbitrary $\psi$ and $\pg$)
\be
\int d{\bf  x}~ \psi^{\dag} (e_{m}\pg) \neq 
-\int d{\bf  x}~ (e_{m}\psi)^{\dag} \pg =
\int d{\bf  x}~ (\psi^{\dag} e_{m}) \pg \qq \mbox{{\fs $(~m=1, ...,7~)$}} \q .
\ee
This contrasts with the situation within complex and quaternionic 
quantum mechanics. Such a difficulty is overcome by using a 
complex projection of the scalar product (complex geometry), with respect to 
one of our imaginary 
units. We break the symmetry between the seven imaginary units 
$e_{1}$, ... , $e_{7}$ and choose as projection plane that one characterized by 
$(1, \; e_{1})$. The new scalar product is quickly obtained performing, in 
the standard definition, the following substitution
\[ 
\int d{\bf x} \lraw  \int_{c} d{\bf x} \equiv 
\frac{1-e_{1}\mid e_{1}}{2}~ \int d{\bf x} \q .
\]
Working in OQM with {\tt complex geometry}, 
$e_{1}$ represents an antihermitian operator. In order to simplify the 
proof we write the octonionic functions $\psi$ and $\pg$ as follows:
\bean
\psi & ~=~ & \psi_{1} + e_{2} \psi_{2} + e_{4} \psi_{3} + e_{6} \psi_{4} \q ,\\
\pg & ~=~ & \pg_{1} + e_{2} \pg_{2} + e_{4} \pg_{3} + e_{6} \pg_{4} \q ,\\
    & & ~~~~~~~~~~~[~\psi_{1, ..., 4}~\mbox{and}~\pg_{1, ..., 4} 
          \in \bfm{\cal C}(1, \; e_{1})~] \q .
\eean
The antihermiticity of $e_{1}$ is shown if
\be
\int_{c} d{\bf  x}~ \psi^{\dag} (e_{1}\pg) = 
-\int_{c} d{\bf  x}~ (e_{1}\psi)^{\dag} \pg \q .
\ee
In the previous equation the only nonvanishing terms are represented by 
{\tt diagonal} terms ($\sim \psi_{1}^{\dag}\pg_{1}, 
\; \psi_{2}^{\dag}\pg_{2}, \; \psi_{3}^{\dag}\pg_{3}, 
\; \psi_{4}^{\dag}\pg_{4}$). In fact, {\tt off-diagonal} terms, like 
$\psi_{2}^{\dag}\pg_{3}, \; \psi_{3}^{\dag}\pg_{4}$, are killed by the 
complex projection,
\bean
~(\psi_{2}^{\dag} e_{2})[e_{1}(e_{4}\pg_{3})] & ~\sim ~ & 
(\ag_{2}e_{2}+\ag_{3}e_{3})(\ag_{4}e_{4}+\ag_{5}e_{5}) \sim  
\ag_{6}e_{6} + \ag_{7} e_{7} \q , \\ 
~[(\psi_{3}^{\dag} e_{4}) e_{1}](e_{6}\pg_{4}) & ~\sim ~& 
~(\bg_{4}e_{4}+\bg_{5}e_{5})(\bg_{6}e_{6}+\bg_{7}e_{7}) \sim  
\bg_{2}e_{2} + \bg_{3} e_{3} \q , \\
 & & ~~~~~~~~~~~[~\ag_{2, ..., 7}~\mbox{and}~\bg_{2, ..., 7} \in  
     \bfm{\cal R}~] \q .
\eean
The diagonal terms give
\bml
\beal{dia}
\int_{c} d{\bf  x}~ \psi^{\dag} (e_{1}\pg) & = &
\psi_{1}^{\dag}(e_{1}\pg_{1})
-(\psi_{2}^{\dag}e_{2})[e_{1}(e_{2}\pg_{2})]
-(\psi_{3}^{\dag}e_{4})[e_{1}(e_{4}\pg_{3})]
-(\psi_{4}^{\dag}e_{6})[e_{1}(e_{6}\pg_{4})] \q ,\\
-\int_{c} d{\bf  x}~ (e_{1}\psi)^{\dag} \pg  & = &
(\psi_{1}^{\dag}e_{1})\pg_{1}
-[(\psi_{2}^{\dag}e_{2})e_{1}](e_{2}\pg_{2})
-[(\psi_{3}^{\dag}e_{4})e_{1}](e_{4}\pg_{3})
-[(\psi_{4}^{\dag}e_{6})e_{1}](e_{6}\pg_{4}) \q .
\eea
\eml
The parenthesis in~(\ref{dia}-b) are not of relevance since
\bc
\bt{lll}
$\psi_{1}^{\dag}e_{1}\pg_{1}$ & {\fs ~~~$(1, \; e_{1})$} & ~~~is a complex 
number , \\ 
$\psi_{2}^{\dag} e_{2} e_{1} e_{2}\pg_{2}$ & {\fs ~~~(subalgebra 123)} , & \\ 
$\psi_{3}^{\dag} e_{4} e_{1} e_{4}\pg_{3}$ & {\fs ~~~(subalgebra 145)} , & \\
$\psi_{4}^{\dag} e_{6} e_{1} e_{6}\pg_{4}$ & {\fs ~~~(subalgebra 176)}   & 
are quaternionic numbers .
\et
\ec
The above-mentioned demonstration does not work for the imaginary units 
$e_{2}$, ... , $e_{7}$ (breaking the symmetry between the seven 
octonionic imaginary units). 

Now, we can define an hermitian operator multiplying by $e_{1}$ 
the operator \bfm{\dv}. 
However, such an operator is not expected to commute with the Hamiltonian, 
which will be, in general, an octonionic quantity. The final step towards an 
appropriate definition of the momentum operator is represented by choosing 
as imaginary unit the barred operator $1\mid e_{1}$ (the 
antihermiticity proof is very similar to the previous one). In OQM with 
complex geometry the appropriate momentum operator is then given by
\be 
{\bf p} \equiv - \bfm{\dv} \mid e_{1} \q .
\ee
Obviously, in order to write equations relativistically covariant, we must 
treat the space components and time in the same way, hence we are obliged 
to modify the standard QM operator, $i\dv_{t}$, by the following substitution
\[ i\dv_{t} \lraw \dv_{t} \mid e_{1} \q ,\]
and so the octonionic Dirac equation becomes
\be
\dv_{t} \psi e_{1} = \bfm{\ag} \cdot ({\bf p}\psi)+ m \bg \psi 
\qq (~{\bf p} \equiv - \bfm{\dv} \mid e_{1}~) \q .
\ee
The possibility to write a consistent momentum operator represents for us 
an impressive argument in favor of the use of a complex geometry in 
formulation an OQM. Besides, such a complex geometry gives us a welcome 
{\tt quadrupling} of solutions. In fact, 
\[ \psi, \; e_{2}\psi, \; e_{4}\psi, \; e_{6}\psi \qq
\psi \in \bfm{\cal C}(1, \; e_{1}) \]
represent now complex-orthogonal solutions. Therefore, we have the 
possibility to write a one-component octonionic Dirac equation in which all 
four standard Dirac free-particle solutions appear.

\subsection*{V-c. Octonionic completeness relations}

We observe that the dimensionality of a complete set of states for complex 
inner product $<\psi \mid \phi>_{c}$ is {\em four times} 
that for the octonionic inner product $<\psi \mid \phi>$. 
Specifically if $\mid \eta_{l}>$ are a complete set of intermediate states for 
the octonionic inner product, so that
\[ <\psi \mid \phi> \; = \sum_{l} <\psi \mid \eta_{l}><\eta_{l} \mid \phi> \; 
\; ,\]
$\mid \eta_{l}>$, $\mid \eta_{l} \; e_{2}>$, $\mid \eta_{l} \;  e_{4}>$, 
$\mid \eta_{l} \; e_{6}>$ form a complete set of states for the complex 
inner product,
\bean 
\mid \phi> & = & \sum_{l} \;  ( \; 
\mid \eta_{l}><\eta_{l} \mid \phi>_{c}+ 
\mid \eta_{l} \; e_{2} ><\eta_{l} \; e_{2} \mid \phi>_{c}+\\
           &   & \; \; \; \; + 
\mid \eta_{l} \; e_{4}><\eta_{l} \; e_{4} \mid \phi>_{c}+ 
\mid \eta_{l} \; e_{6}><\eta_{l} \; e_{6} \mid \phi>_{c} \; )\\
 & = & \sum_{m} \mid \chi_{m}><\chi_{m} \mid \phi>_{c} \; \; ,
\eean
where $\chi_{m}$ represent {\em complex} orthogonal states. Thus the  
completeness relation can be written as (for further details on the 
completeness relation, one can consult an interesting work of Horwitz and 
Biedenharn, see~\cite{hor} - pag.~455)
\bean  
\stackrel{\raw}{\bf 1} & = & \sum_{m} \mid \chi_{m}>~~ << \chi_{m} \mid \; \; ,\\
\stackrel{\law}{\bf 1} & = & \sum_{m} \mid \chi_{m} >> ~~<\chi_{m} \mid \; \; ,
\eean
so in our formalism we generalize the Dirac's notation by the 
definitions
\bean <<\chi_{m} \mid \phi> & = & 
<\chi_{m} \mid \phi>_{c} \; \; ,\\
<\phi \mid \chi_{m}>> & = & 
<\phi \mid \chi_{m} >_{c} \; \; .
\eean

\section{Octonionic Dirac equation}

As remarked in section V, the appropriate momentum operator in OQM is 
\[
{\bf p} \equiv - \bfm{\dv} \mid e_{1} \q .
\]
Thus, the octonionic Dirac equation, in covariant form, is given by
\bel{ode}
\cg^{\mu}(\dv_{\mu}\psi e_{1})=m\psi \q ,
\ee
where $\cg^{\mu}$ are represented by octonionic barred  
operators~(\ref{odgm1}-d). We can now proceed in the standard manner. 
Plane wave solutions exist [${\bf p}~(\equiv -\bfm{\dv} \mid e_{1}$)  
commutes with a generic octonionic Hamiltonian] and are of the form
\be
\psi({\bf x}, \; t) = [~u_{1}({\bf p})+e_{2}u_{2}({\bf p})+
e_{4}u_{3}({\bf p})+e_{6}u_{4}({\bf p})~]~e^{-pxe_{1}} \qq
[~u_{1, ... , 4} \in \bfm{\cal C}(1, \; e_{1})~] \q .
\ee
Let's start with 
\[{\bf p} \equiv (0, \; 0, \; p_{z}) \q , \]
from~(\ref{ode}), we have
\bel{ode2}
E(\cg^{0} \psi) - p_{z}(\cg^{3} \psi) =m\psi \q .
\ee
Using the explicit form of the octonionic operators $\cg^{0, \; 3}$ and 
extracting their action (see subsection VI-a) from the tables quoted in 
appendix B1, we find  
\bel{appc}
E(u_{1}+e_{2}u_{2}-e_{4}u_{3}-e_{6}u_{4})
-p_{z}(u_{3}-e_{2}u_{4}-e_{4}u_{1}+e_{6}u_{2})
=m(u_{1}+e_{2}u_{2}+e_{4}u_{3}+e_{6}u_{4}) 
\ee
From~(\ref{appc}), we derive four complex equations:
\bean
(E-m)u_{1} & = & +p_{z}u_{3} \q ,\\
(E-m)u_{2} & = & -p_{z}u_{4} \q ,\\
(E+m)u_{3} & = & +p_{z}u_{1} \q ,\\
(E+m)u_{4} & = & -p_{z}u_{2} \q .
\eean
After simple algebraic manipulations we find the following octonionic 
Dirac solutions:
\bean
E=+\vert E \vert & ~~~~~~~~u^{(1)}=N \left( 1 + e_{4} 
         \frac{p_{z}}{\vert E \vert +m} \right) \q , 
\q u^{(2)}= N \left( e_{2}-e_{6} \frac{p_{z}}{\vert E \vert +m} \right)
                                =u^{(1)}e_{2} \q ; \\
E=-\vert E \vert & ~~~~~~~~u^{(3)}=N \left(\frac{p_{z}}{\vert E \vert +m} 
- e_{4} \right) \q , 
\q  u^{(4)}=N \left(e_{2}\frac{p_{z}}{\vert E \vert +m} +e_{6} \right) 
                                =u^{(3)}e_{2} \q ,
\eean
with $N$ real normalization constant. Setting the norm to 
$2\vert E \vert$, we find
\[ N=(\vert E \vert + m)^{\frac{1}{2}} \q . \]

We now observe (as for the quaternionic Dirac equation) a difference with 
respect to the standard Dirac equation. 
Working in our representation~(\ref{odgm1}-d) and introducing the 
octonionic spinor
\[ \bar{u}\equiv(\cg_{0} u)^{+}= u_{1}^{*}-e_{2}u_{2}+
e_{4}u_{3}+e_{6}u_{4} \qq [~u= u_{1}+e_{2}u_{2}+
e_{4}u_{3}+e_{6}u_{4}~] \q ,\]
we have
\be
\bar{u}^{(1)}u^{(1)}=u^{(1)}\bar{u}^{(1)}=
\bar{u}^{(2)}u^{(2)}=u^{(2)}\bar{u}^{(2)}=2(m+e_{4}p_{z}) \q .
\ee
Thus we find 
\bml
\bel{os}
u^{(1)}\bar{u}^{(1)}+u^{(2)}\bar{u}^{(2)}=4(m+e_{4}p_{z}) \q ,
\ee
instead of the expected relation
\bel{cs}
u^{(1)}\bar{u}^{(1)}+u^{(2)}\bar{u}^{(2)}=\cg^{0} E - \cg^{3} p_{z} + m \q .
\ee
\eml
Furthermore, the previous difference is compensated if we compare the 
complex projection of~(\ref{os}) with the trace of~(\ref{cs})
\be
[~(u^{(1)}\bar{u}^{(1)}+u^{(2)}\bar{u}^{(2)})^{OQM}~]_{c}~\equiv~
Tr~[~(u^{(1)}\bar{u}^{(1)}+u^{(2)}\bar{u}^{(2)})^{CQM}~]~=~4m \q .
\ee
We know that spinor relations like~(\ref{os}-b) are relevant in 
perturbation calculus, so the previous results suggest to analyze quantum 
electrodynamics in order to investigate possible differences between 
complex and octonionic quantum mechanics. This could represents the aim of 
a future work.

\subsection*{VI-a. $\cg^{0, \; 3}$-action on octonionic spinors}

In the following tables,  we explicitly show the action on the octonionic 
spinor 
\[
u=u_{1}+e_{2}u_{2}+e_{4}u_{3}+e_{6}u_{4} \qq [~u_{1,...,4} \in 
\bfm{\cal C}(1, \; e_{1})~] \q , \]
of the barred operators which appear in $\cg^{0}$ and $\cg^{3}$. Using such 
tables, after straightforwards algebraic manipulations we find
\bean
\cg^{0}u & ~=~ & u_{1}+e_{2}u_{2}-e_{4}u_{3}-e_{6}u_{4} \q ,\\
\cg^{3}u & ~=~ & u_{3}-e_{2}u_{4}-e_{4}u_{1}+e_{6}u_{2} \q . 
\eean
\vs{.5cm}
\bc
\bt{l|rrrr}
 & & & & \\
$\cg^{0}$-action~~~ & 
$~~~~~~~u_{1}$ & $~~~~~~~e_{2}u_{2}$ & $~~~~~~~e_{4}u_{3}$ & 
$~~~~~~~e_{6}u_{4}$\\
 & & & & \\
\hline \hline
 & & & & \\
$e_{1}\mid e_{1}$ &
$-u_{1}$ & $e_{2}u_{2}$ & $e_{4}u_{3}$ & $e_{6}u_{4}$\\
$e_{2}\mid e_{2}$ &
$-u_{1}^{*}$ & $-e_{2}u_{2}^{*}$ & $e_{4}u_{3}$ & $e_{6}u_{4}$\\
$e_{3}\mid e_{3}$ &
$-u_{1}^{*}$ & $e_{2}u_{2}^{*}$ & $e_{4}u_{3}$ & $e_{6}u_{4}$\\
$e_{4}\mid e_{4}$ &
$-u_{1}^{*}$ & $e_{2}u_{2}^{*}$ & $-e_{4}u_{3}^{*}$ & $e_{6}u_{4}$\\
$e_{5}\mid e_{5}$ &
$-u_{1}^{*}$ & $e_{2}u_{2}$ & $e_{4}u_{3}^{*}$ & $e_{6}u_{4}$\\
$e_{6}\mid e_{6}$ &
$-u_{1}^{*}$ & $e_{2}u_{2}$ & $e_{4}u_{3}$ & $-e_{6}u_{4}^{*}$\\
$e_{7}\mid e_{7}$ &
$-u_{1}^{*}$ & $e_{2}u_{2}$ & $e_{4}u_{3}$ & $e_{6}u_{4}^{*}$
\et
\ec
\vs{.5cm}
\bc
\bt{l|rrrr}
 & & & & \\
$\cg^{3}$-action~~~ & 
$~~~~~~~u_{1}$ & $~~~~~~~e_{2}u_{2}$ & $~~~~~~~e_{4}u_{3}$ & 
$~~~~~~~e_{6}u_{4}$\\
 & & & & \\
\hline \hline 
 & & & & \\
$e_{4}$ &
$e_{4}u_{1}$ &  $-e_{6}u_{2}^{*}$ & $-u_{3}$ & $e_{2}u_{4}$\\
$1\mid e_{4}$ &
$e_{4}u_{1}^{*}$ &  $e_{6}u_{2}^{*}$ & $-u_{3}^{*}$ & $-e_{2}u_{4}^{*}$\\
$e_{7}~)~e_{3}$ &
$e_{4}u_{1}^{*}$ &  $e_{6}u_{2}$ & 
$u_{3}$ & $-e_{2}u_{4}^{*}$\\
$e_{3}~)~e_{7}$ &
$-e_{4}u_{1}^{*}$ &  $-e_{6}u_{2}^{*}$ & 
$-u_{3}$ & $e_{2}u_{4}$\\
$e_{6}~)~e_{2}$ &
$e_{4}u_{1}^{*}$ &  $-e_{6}u_{2}$ & 
$u_{3}$ & $-e_{2}u_{4}^{*}$\\
$e_{2}~)~e_{6}$ &
$-e_{4}u_{1}^{*}$ &  $-e_{6}u_{2}^{*}$ & 
$-u_{3}$ & $-e_{2}u_{4}$\\
$e_{5}~)~e_{1}$ &
$e_{4}u_{1}$ &  $e_{6}u_{2}^{*}$ & $u_{3}$ & 
$-e_{2}u_{4}^{*}$\\
$e_{1}~)~e_{5}$ &
$-e_{4}u_{1}^{*}$ &  $-e_{6}u_{2}^{*}$ & $-u_{3}^{*}$ & 
$e_{2}u_{4}^{*}$
\et
\ec

\con

This paper aimed to give a clear exposition of the potentiality of 
generalized numbers in quantum mechanics. We know that quantum mechanics is 
the basic tool for different physical applications. Many physicists believe 
that imaginary numbers are related to the deep secret of 
quantization. Penrose~\cite{penr} thinks that the quantization is completely 
based on complex numbers. Trying to overcome the problem of quantum gravity, 
he proposed to complexify the Minkowskian space-time. This represents the 
main assumption behind the twistor program. Adler~\cite{adl} believes that 
quantization processes should not be limited to complex numbers but should be 
extended to another member of the division algebras rank, the quaternionic 
field. He postulates that a successful unification of the fundamental 
forces will require a generalization beyond complex quantum mechanics. 
Adler envisages a two-level correspondence principle:
\bc
\bt{ccl}
 $\vert $       &                  & ~~~~~~classical physics and fields ,\\
 $\vert $       & ~distance scale  & ~~~~~~complex quantum mechanics and fields ,\\
 $\downarrow$  &                  & ~~~~~~quaternionic quantum field 
                                    dynamics (preonic level~\cite{adl2}) ,
\et
\ec
with quaternionic quantum dynamics interfacing with complex quantum theory, 
and then with complex quantum theory interfacing in the familiar manner 
with classical physics (~\cite{adl}, pag.~498).

Following this approach, we are tempted to postulate that octonionic quantum 
field theory may play an essential role in an even deeper fundamental level 
of physical structure.

Quaternionic quantum mechanics, using complex geometry~\cite{qua1,qua2,qua3} 
or quaternionic geometry~\cite{adl,adl1,adl2}, seems to be consistent from the 
mathematical point of view. Due to the octonions nonassociativity property, 
octonionic quantum mechanics seems to be a puzzle. In the physical 
literature, we find a method to partially overcome the issues 
relating to the octonions nonassociativity. Some people introduces a ``new'' 
imaginary units ``~$i=\sqrt{-1}$~'' which commutes with all others 
octonionic imaginary units, $e_{m}$. The new field is often called 
{\tt complexified octonionic field}. Different papers have been written in 
such a formalism: Quark Structure and Octonions~\cite{gur2}, 
Octonions, Quark and QCD~\cite{mor}, Octonions and 
Isospin~\cite{pen}, Dirac-Clifford algebra~\cite{edm}, and so on. 
Nevertheless, we don't like complexifying the octonionic field and so we 
have presented in this paper an alternative 
way to look at the octonionic world. 
No new imaginary unit is necessary to formulate in a consistent way an 
octonionic quantum mechanics.

A first motivation, in using octonions numbers in physics, can be concisely 
resumed as follows: We hope to get a better understanding of standard 
theories if we have more than one concrete realization. In this way we can 
recognize the fundamental postulates which hold for any generic numerical 
field.

Having a nonassociative algebra needs special care. In this work, we 
introduced a ``trick'' which allowed us to manipulate octonions without useless 
efforts. We summarize the more important results found in previous sections:
\bc
{\tt M - Mathematical Contents :}
\ec

{\tt M1} - The introduction of barred operators (natural objects if one works 
with noncommutative numbers) facilitate our job and enable us to formulate 
a ``friendly'' connection between $8 \times 8$ real matrices and octonions;

{\tt M2} - The nonassociativity is reproduced by left/right barred 
operators. We consider these operators the natural extension of generalized 
quaternions, recently introduced in literature~\cite{qua2};

{\tt M3} - We tried to investigate the properties of our generalized 
numbers and studied their special characteristics in order to use them in a 
proper way. After having established their isomorphism to $\glr$, life became 
easier;

{\tt M4} -  The connection between $\glr$ and generalized octonions gives us 
the possibility to extract the octonionic generators corresponding to the 
complex subgroup $\glc$. This step represents the main tool to manipulate 
octonions in quantum mechanics;

{\tt M5} - To the best of our knowledge, for the first time, an octonionic 
representation for the 4-dimensional Clifford algebra, appears in print.
\bc
{\tt P - Physical Contents :}
\ec

{\tt P1} - We emphasize that a characteristic of our formalism is the 
{\em absolute need of a complex scalar product} (in QQM the use of a 
complex geometry is not obligatory and thus a question of choice). 
Using a complex geometry we 
overcame the hermiticity problem and gave the appropriate and unique 
definition of momentum operator;

{\tt P2} - A positive feature of this octonionic version of quantum 
mechanics, is the appearance of all four standard Dirac free-particle 
solutions notwithstanding the one-component structure of the wave functions. 
We have the following situation for the division algebras:
\bc
\bt{lcccl}
{\sf field} :~~~ & ~~complex,~~ & ~~quaternions,~~ & ~~octonions,~~& \\
{\sf Dirac Equation} :~~~ & $4\times 4$, &  $2\times 2$, &  $1\times 1$ & 
~~~{\fs ( matrix dimension ) ;} 
\et
\ec

{\tt P3} - Many physical result can be reobtained by translation, so we 
have one version of octonionic quantum mechanics where a part of the standard 
quantum mechanics is reproduced. This represents for the authors a first 
fundamental step towards an octonionic world. We remark that 
our translation will not be possible in 
all situations, so it is only partial, consistent with the fact that the 
octonionic version could provide additional physical predictions. 
\bc
{\tt I - Further Investigations :}
\ec

We conclude with a listing of open questions for future investigations, 
whose study lead to further insights.

{\tt I1} - How may we complete the translation? Note that translation, as 
presented in this paper, works for $4n\times 4n$ matrices. What about 
odd-dimensional matrices?

{\tt I2} - From the translation tables we can extract the multiplication 
rules for the octonionic barred operators. This will allow us to work 
directly with octonions without translations. 

{\tt I3} - Inspired from eq.~(\ref{new}), we could look for a more 
convenient way to express the new nonassociative multiplication 
(for example we can try to modify the standard 
multiplication rule: row by column);

{\tt I4} - The reproduction in octonionic calculations of the standard QED 
results will be a nontrivial objective, due to the explicit differences in 
certain spinorial identities (see subsection V-c). We are going to study 
this problem in a forthcoming paper;

{\tt I5} - A very attractive point is to try to treat the strong field 
by octonions, and then to formulate in a suitable manner a standard 
model, based on our octonionic dynamical Dirac equation;

{\tt I6} - A last interesting research topic could be to generalize the group 
theoretical structure by our barred octonionic operators.

Many of the problems on this list deal with technical details although the 
answers to some will be important for further development of the subject.

We hope that the work presented in this paper, demonstrates that octonionic 
quantum mechanics may constitute a coherent and well-defined branch of 
theoretical physics. We are convinced that octonionic quantum mechanics 
represents largely uncharted and potentially very interesting, terrain in 
theoretical physics.

We conclude emphasizing that the core of our paper is surely represented by 
absolute need of adopting a complex geometry within a quantum octonionic 
world.

\new

\section*{Appendix A1}

In this appendix we give the translation rules between octonionic 
left-right barred operators and $8\times 8$ real matrices. In order to 
simplify our translation tables we introduce the following notation:
\bml
\bea
\{~a, \; b, \; c, \; d~\}_{(1)} ~ \equiv ~\bamq{a}{0}{0}{0} 0 & b & 0 & 0\\
0 & 0 & c & 0\\ 0 & 0 & 0 & d \eam   & \q , \q &
\{~a, \; b, \; c, \; d~\}_{(2)} ~ \equiv ~\bamq{0}{a}{0}{0} b & 0 & 0 & 0\\
0 & 0 & 0 & c\\ 0 & 0 & d & 0 \eam \q ,\\
\{~a, \; b, \; c, \; d~\}_{(3)} ~ \equiv ~\bamq{0}{0}{a}{0} 0 & 0 & 0 & b\\
c & 0 & 0 & 0\\ 0 & d & 0 & 0 \eam   & \q , \q &
\{~a, \; b, \; c, \; d~\}_{(4)} ~ \equiv ~\bamq{0}{0}{0}{a} 0 & 0 & b & 0\\
0 & c & 0 & 0\\ d & 0 & 0 & 0 \eam \q ,
\eea
\eml
where $a, \; b, \; c, \; d$ and $0$ represent $2\times 2$ real matrices.

In the following tables $\sg_{1}$, $\sg_{2}$, $\sg_{3}$ represent the 
standard Pauli matrices:
\be
\sg_{1} = \bamd{0}{1}{1}{0} \q , \q  
\sg_{2} = \bamd{0}{-i}{i}{0} \q , \q  
\sg_{3} = \bamd{1}{0}{0}{-1} \q .  
\ee

\vs{.5cm}
\bc
{\tt TABLE A1a}\\
{\fs Translation rules between $8\times 8$ real matrices and octonionic 
barred operators $e_{m}$, $1\mid e_{m}$}\\
\vs{.2cm}
\bt{|rcrrrrcrcrrrrc|}
\hline 
 & & & & & & & & & & & & & \\ 
~~~~~~~~\bfm{e_{1}} & $~\lrw~\{$ & $-i\sg_{2}$, & ~$-i\sg_{2}$, & ~$-i\sg_{2}$, & 
~$i\sg_{2} ~\}_{(1)}$
&~~~,~~~~& 
\bfm{1\mid e_{1}} & $~\lrw~\{$ & $-i\sg_{2}$, & ~$i\sg_{2}$, & ~$i\sg_{2}$, & 
~$-i\sg_{2} ~\}_{(1)}$
&~~~,~~~~\\
\bfm{e_{2}} & $~\lrw~\{$ & $ -\sg_{3}$, & ~$\sg_{3}$, & ~$-1$, & ~$1 ~\}_{(2)}$
&~~~,~~~~& 
\bfm{1\mid e_{2}} & $~\lrw~\{$ & $ -1$, & ~$1$, & ~$1$, & 
~$-1 ~\}_{(2)}$
&~~~,~~~~\\
\bfm{e_{3}} & $~\lrw~\{$ & $ -\sg_{1}$, & ~$\sg_{1}$, & ~$-i\sg_{2}$, & 
~$-i\sg_{2} ~\}_{(2)}$
&~~~,~~~~& 
\bfm{1\mid e_{3}} & $~\lrw~\{$ & $ -i\sg_{2}$, & ~$-i\sg_{2}$, & ~$i\sg_{2}$, & 
~$i\sg_{2} ~\}_{(2)}$
&~~~,~~~~\\
\bfm{e_{4}} & $~\lrw~\{$ & $ -\sg_{3}$, & ~$1$, & ~$\sg_{3}$, & ~$-1 ~\}_{(3)}$
&~~~,~~~~& 
\bfm{1\mid e_{4}} & $~\lrw~\{$ & $ -1$, & ~$-1$, & ~$1$, & 
~$1 ~\}_{(3)}$
&~~~,~~~~\\
\bfm{e_{5}} & $~\lrw~\{$ & $ -\sg_{1}$, & ~$i\sg_{2}$, & ~$\sg_{1}$, & 
~$i\sg_{2} ~\}_{(3)}$
&~~~,~~~~& 
\bfm{1\mid e_{5}} & $~\lrw~\{$ & $ -i\sg_{2}$, & ~$-i\sg_{2}$, & ~$-i\sg_{2}$, & 
~$-i\sg_{2} ~\}_{(3)}$
&~~~,~~~~\\
\bfm{e_{6}} & $~\lrw~\{$ & $ -1$, & ~$-\sg_{3}$, & ~$\sg_{3}$, & ~$1 ~\}_{(4)}$
&~~~,~~~~& 
\bfm{1\mid e_{6}} & $~\lrw~\{$ & $ -\sg_{3}$, & ~$\sg_{3}$, & ~$-\sg_{3}$, & 
~$\sg_{3} ~\}_{(4)}$
&~~~,~~~~\\
\bfm{e_{7}} & $~\lrw~\{$ & $ -i\sg_{2}$, & ~$-\sg_{1}$, & ~$\sg_{1}$, & 
~$-i\sg_{2} ~\}_{(4)}$
&~~~,~~~~& 
\bfm{1\mid e_{7}} & $~\lrw~\{$ & $ -\sg_{1}$, & ~$\sg_{1}$, & ~$-\sg_{1}$, & 
~$\sg_{1} ~\}_{(4)}$
&~~~,~~~~\\
 & & & & & & & & & & & & & \\
\hline
\et
\ec

\vs{1cm}
\bc
{\tt TABLE A1b}\\
{\fs Translation rules between $8\times 8$ real matrices and octonionic barred o
operators
$e_{m} \mid e_{m}$}\\
\vs{.2cm}
\bt{|rcrrrrcrcrrrrc|}
\hline 
 & & & & & & & & & & & & & \\ 
~~~~\bfm{e_{1}\mid e_{1}} & $~\lrw~\{$ & $ -1$, & ~$1$, & ~$1$, & ~$1 ~\}_{(1)}$
&~~~,~~~~& 
\bfm{e_{2}\mid e_{2}} & $~\lrw~\{$ & $ -\sg_{3}$, & ~$-\sg_{3}$, & 
~$1$, & ~$1 ~\}_{(1)}$
&~~~,~~~~\\
\bfm{e_{3}\mid e_{3}} & $~\lrw~\{$ & $ -\sg_{3}$, & ~$\sg_{3}$, & 
~$1$, & ~$1 ~\}_{(1)}$
&~~~,~~~~& 
\bfm{e_{4}\mid e_{4}} & $~\lrw~\{$ & $ -\sg_{3}$, & ~$1$, & ~$-\sg_{3}$, & 
~$1 ~\}_{(1)}$
&~~~,~~~~\\
\bfm{e_{5}\mid e_{5}} & $~\lrw~\{$ & $ -\sg_{3}$, & ~$1$, & 
~$\sg_{3}$, & ~$1 ~\}_{(1)}$
&~~~,~~~~& 
\bfm{e_{6}\mid e_{6}} & $~\lrw~\{$ & $ -\sg_{3}$, & ~$1$, & ~$1$, & 
~$-\sg_{3} ~\}_{(1)}$
&~~~,~~~~\\
\bfm{e_{7}\mid e_{7}} & $~\lrw~\{$ & $ -\sg_{3}$, & ~$1$, & ~$1$, & 
~$\sg_{3} ~\}_{(1)}$
&~~~.~~~~& & & & & & & \\
 & & & & & & & & & & & & & \\
\hline
\et
\ec
\new
%\vs{1.5cm}
\bc
{\tt TABLE A1-L}\\
{\fs Translation rules between $8\times 8$ real matrices and octonionic 
left-barred operators}\\
\vs{.2cm}
\bt{|rcrrrrcrcrrrrc|}
\hline 
 & & & & & & & & & & & & & \\
~~~~\bfm{e_{1}~)~e_{2}} & 
$~\lrw~\{$ & $ i\sg_{2}$, & ~$-i\sg_{2}$, & ~$i\sg_{2}$, & 
~$i\sg_{2} ~\}_{(2)}$
&~~~,~~~~& 
\bfm{e_{1}~)~e_{3}} & $~\lrw~\{$ & $ -1$, & ~$-1$, & ~$-1$, & ~$1 ~\}_{(2)}$
&~~~,~~~~\\
\bfm{e_{1}~)~e_{4}} & $~\lrw~\{$ & $ i\sg_{2}$, & ~$-i\sg_{2}$, & ~$-i\sg_{2}$, & 
~$-i\sg_{2} ~\}_{(3)}$
&~~~,~~~~& 
\bfm{e_{1}~)~e_{5}} & $~\lrw~\{$ & $ -1$, & ~$1$, & ~$-1$, & ~$-1 ~\}_{(3)}$
&~~~,~~~~\\
\bfm{e_{1}~)~e_{6}} & $~\lrw~\{$ & $ -\sg_{1}$, & ~$-\sg_{1}$, & ~$\sg_{1}$, & 
~$-\sg_{1} ~\}_{(4)}$
&~~~,~~~~& 
\bfm{e_{1}~)~e_{7}} & $~\lrw~\{$ & $ \sg_{3}$, & ~$\sg_{3}$, & ~$-\sg_{3}$, & 
~$\sg_{3} ~\}_{(4)}$
&~~~,~~~~\\
\bfm{e_{2}~)~e_{1}} & $~\lrw~\{$ & $ -\sg_{1}$, & ~$-\sg_{1}$, & ~$-i\sg_{2}$, & 
~$-i\sg_{2}~\}_{(2)}$
&~~~,~~~~& 
\bfm{e_{2}~)~e_{3}} & $~\lrw~\{$ & $ \sg_{1}$, & ~$-\sg_{1}$, & ~$i\sg_{2}$, & 
~$-i\sg_{2} ~\}_{(1)}$
&~~~,~~~~\\
\bfm{e_{2}~)~e_{4}} & $~\lrw~\{$ & $ 1$, & ~$-1$, & ~$-\sg_{3}$, & 
~$\sg_{3} ~\}_{(4)}$
&~~~,~~~~& 
\bfm{e_{2}~)~e_{5}} & $~\lrw~\{$ & $ i\sg_{2}$, & ~$-i\sg_{2}$, & ~$-\sg_{1}$, & 
~$\sg_{1} ~\}_{(4)}$
&~~~,~~~~\\
\bfm{e_{2}~)~e_{6}} & $~\lrw~\{$ & $ -\sg_{3}$, & ~$-\sg_{3}$, & ~$-1$, & 
~$-1 ~\}_{(3)}$
&~~~,~~~~& 
\bfm{e_{2}~)~e_{7}} & $~\lrw~\{$ & $ -\sg_{1}$, & ~$-\sg_{1}$, & ~$i\sg_{2}$, & 
~$i\sg_{2} ~\}_{(3)}$
&~~~,~~~~\\
\bfm{e_{3}~)~e_{1}} & $~\lrw~\{$ & $ \sg_{3}$, & ~$\sg_{3}$, & ~$1$, & 
~$-1 ~\}_{(2)}$
&~~~,~~~~& 
\bfm{e_{3}~)~e_{2}} & $~\lrw~\{$ & $ -\sg_{1}$, & ~$-\sg_{1}$, & ~$-i\sg_{2}$, & 
~$i\sg_{2} ~\}_{(1)}$
&~~~,~~~~\\
\bfm{e_{3}~)~e_{4}} & $~\lrw~\{$ & $ i\sg_{2}$, & ~$i\sg_{2}$, & ~$-\sg_{1}$, & 
~$\sg_{1} ~\}_{(4)}$
&~~~,~~~~& 
\bfm{e_{3}~)~e_{5}} & $~\lrw~\{$ & $ -1$, & ~$-1$, & ~$\sg_{3}$, & 
~$-\sg_{3} ~\}_{(4)}$
&~~~,~~~~\\
\bfm{e_{3}~)~e_{6}} & $~\lrw~\{$ & $ \sg_{1}$, & ~$-\sg_{1}$, & ~$-i\sg_{2}$, & 
~$-i\sg_{2} ~\}_{(3)}$
&~~~,~~~~& 
\bfm{e_{3}~)~e_{7}} & $~\lrw~\{$ & $ -\sg_{3}$, & ~$\sg_{3}$, & ~$-1$, & 
~$-1 ~\}_{(3)}$
&~~~,~~~~\\
\bfm{e_{4}~)~e_{1}} & $~\lrw~\{$ & $ -\sg_{1}$, & ~$i\sg_{2}$, & ~$-\sg_{1}$, & 
~$i\sg_{2} ~\}_{(3)}$
&~~~,~~~~& 
\bfm{e_{4}~)~e_{2}} & $~\lrw~\{$ & $ -1$, & ~$-\sg_{3}$, & ~$-1$, & ~$-1 ~\}_{(4)}$
&~~~,~~~~\\
\bfm{e_{4}~)~e_{3}} & $~\lrw~\{$ & $ -i\sg_{2}$, & ~$-\sg_{1}$, & ~$-i\sg_{2}$, & 
~$-\sg_{1} ~\}_{(4)}$
&~~~,~~~~& 
\bfm{e_{4}~)~e_{5}} & $~\lrw~\{$ & $ \sg_{1}$, & ~$i\sg_{2}$, & ~$-\sg_{1}$, & 
~$-i\sg_{2} ~\}_{(1)}$
&~~~,~~~~\\
\bfm{e_{4}~)~e_{6}} & $~\lrw~\{$ & $ \sg_{3}$, & ~$1$, & ~$-\sg_{3}$, & 
~$-1 ~\}_{(2)}$
&~~~,~~~~& 
\bfm{e_{4}~)~e_{7}} & $~\lrw~\{$ & $ \sg_{1}$, & ~$-i\sg_{2}$, & ~$-\sg_{1}$, & 
~$i\sg_{2} ~\}_{(2)}$
&~~~,~~~~\\
\bfm{e_{5}~)~e_{1}} & $~\lrw~\{$ & $ \sg_{3}$, & ~$-1$, & ~$\sg_{3}$, & 
~$1 ~\}_{(3)}$
&~~~,~~~~& 
\bfm{e_{5}~)~e_{2}} & $~\lrw~\{$ & $ -i\sg_{2}$, & ~$-\sg_{1}$, & ~$i\sg_{2}$, & 
~$-\sg_{1} ~\}_{(4)}$
&~~~,~~~~\\
\bfm{e_{5}~)~e_{3}} & $~\lrw~\{$ & $1$, & ~$\sg_{3}$, & ~$-1$, & 
~$\sg_{3} ~\}_{(4)}$
&~~~,~~~~& 
\bfm{e_{5}~)~e_{4}} & $~\lrw~\{$ & $ -\sg_{1}$, & ~$-i\sg_{2}$, & ~$-\sg_{1}$, & 
~$i\sg_{2} ~\}_{(1)}$
&~~~,~~~~\\
\bfm{e_{5}~)~e_{6}} & $~\lrw~\{$ & $ -\sg_{1}$, & ~$i\sg_{2}$, & ~$-\sg_{1}$, & 
~$-i\sg_{2} ~\}_{(2)}$
&~~~,~~~~& 
\bfm{e_{5}~)~e_{7}} & $~\lrw~\{$ & $ \sg_{3}$, & ~$1$, & ~$\sg_{3}$, & 
~$-1 ~\}_{(2)}$
&~~~,~~~~\\
\bfm{e_{6}~)~e_{1}} & $~\lrw~\{$ & $ i\sg_{2}$, & ~$\sg_{1}$, & ~$-\sg_{1}$, & 
~$-i\sg_{2} ~\}_{(4)}$
&~~~,~~~~& 
\bfm{e_{6}~)~e_{2}} & $~\lrw~\{$ & $ \sg_{3}$, & ~$-1$, & ~$1$, & 
~$-\sg_{3} ~\}_{(3)}$
&~~~,~~~~\\
\bfm{e_{6}~)~e_{3}} & $~\lrw~\{$ & $ -\sg_{1}$, & ~$i\sg_{2}$, & ~$i\sg_{2}$, & 
~$-\sg_{1} ~\}_{(3)}$
&~~~,~~~~& 
\bfm{e_{6}~)~e_{4}} & $~\lrw~\{$ & $ -\sg_{3}$, & ~$-1$, & ~$-1$, & 
~$-\sg_{3} ~\}_{(2)}$
&~~~,~~~~\\
\bfm{e_{6}~)~e_{5}} & $~\lrw~\{$ & $ \sg_{1}$, & ~$-i\sg_{2}$, & ~$i\sg_{2}$, & 
~$-\sg_{1} ~\}_{(2)}$
&~~~,~~~~& 
\bfm{e_{6}~)~e_{7}} & $~\lrw~\{$ & $ -\sg_{1}$, & ~$-i\sg_{2}$, & ~$-i\sg_{2}$, & 
~$-\sg_{1} ~\}_{(1)}$
&~~~,~~~~\\
\bfm{e_{7}~)~e_{1}} & $~\lrw~\{$ & $ -1$, & ~$-\sg_{3}$, & ~$\sg_{3}$, & 
~$-1 ~\}_{(4)}$
&~~~,~~~~& 
\bfm{e_{7}~)~e_{2}} & $~\lrw~\{$ & $ \sg_{1}$, & ~$-i\sg_{2}$, & ~$-i\sg_{2}$, & 
~$-\sg_{1} ~\}_{(3)}$
&~~~,~~~~\\
\bfm{e_{7}~)~e_{3}} & $~\lrw~\{$ & $ \sg_{3}$, & ~$-1$, & ~$1$, & 
~$\sg_{3}~\}_{(3)}$
&~~~,~~~~& 
\bfm{e_{7}~)~e_{4}} & $~\lrw~\{$ & $ -\sg_{1}$, & ~$\sg_{2}$, & ~$-i\sg_{2}$, & 
~$-\sg_{1} ~\}_{(2)}$
&~~~,~~~~\\
\bfm{e_{7}~)~e_{5}} & $~\lrw~\{$ & $ -\sg_{3}$, & ~$-1$, & ~$-1$, & 
~$\sg_{3} ~\}_{(2)}$
&~~~,~~~~& 
\bfm{e_{7}~)~e_{6}} & $~\lrw~\{$ & $ \sg_{1}$, & ~$i\sg_{2}$, & ~$i\sg_{2}$, & 
~$-\sg_{1} ~\}_{(1)}$
&~~~.~~~~\\
 & & & & & & & & & & & & & \\
\hline
\et
\ec
\vs{1.5cm}
\bc
{\tt TABLE A1-R}\\
{\fs Translation rules between $8\times 8$ real matrices and octonionic 
right-barred operators}\\
\vs{.2cm}
\bt{|rcrrrrcrcrrrrc|}
\hline
 & & & & & & & & & & & & & \\
~~~~\bfm{e_{1}~(~e_{2}} & 
$~\lrw~\{$ & $i\sg_{2}$, & ~$-i\sg_{2}$, & ~$-i\sg_{2}$, & 
~$-i\sg_{2} ~\}_{(2)}$
&~~~,~~~~& 
\bfm{e_{1}~(~e_{3}} & $~\lrw~\{$ & $ -1$, & ~$-1$, & ~$1$, & 
~$-1 ~\}_{(2)}$
&~~~,~~~~\\
\bfm{e_{1}~(~e_{4}} & $~\lrw~\{$ & $ i\sg_{2}$, & ~$i\sg_{2}$, & ~$-i\sg_{2}$, & 
~$i\sg_{2} ~\}_{(3)}$
&~~~,~~~~& 
\bfm{e_{1}~(~e_{5}} & $~\lrw~\{$ & $ -1$, & ~$-1$, & ~$-1$, & ~$1 ~\}_{(3)}$
&~~~,~~~~\\
\bfm{e_{1}~(~e_{6}} & $~\lrw~\{$ & $ -\sg_{1}$, & ~$\sg_{1}$, & ~$-\sg_{1}$, & 
~$-\sg_{1} ~\}_{(4)}$
&~~~,~~~~& 
\bfm{e_{1}~(~e_{7}} & $~\lrw~\{$ & $ \sg_{3}$, & ~$-\sg_{3}$, & ~$\sg_{3}$, & 
~$\sg_{3} ~\}_{(4)}$
&~~~,~~~~\\
\bfm{e_{2}~(~e_{1}} & $~\lrw~\{$ & $ -\sg_{1}$, & ~$-\sg_{1}$, & ~$i\sg_{2}$, & 
~$i\sg_{2} ~\}_{(2)}$
&~~~,~~~~& 
\bfm{e_{2}~(~e_{3}} & $~\lrw~\{$ & $ \sg_{1}$, & ~$-\sg_{1}$, & ~$-i\sg_{2}$, & 
~$i\sg_{2}~\}_{(1)}$
&~~~,~~~~\\
\bfm{e_{2}~(~e_{4}} & $~\lrw~\{$ & $ \sg_{3}$, & ~$-\sg_{3}$, & ~$-1$, & ~$1 ~\}_{(4)}$
&~~~,~~~~& 
\bfm{e_{2}~(~e_{5}} & $~\lrw~\{$ & $ \sg_{1}$, & ~$-\sg_{1}$, & ~$i\sg_{2}$, & 
~$-i\sg_{2} ~\}_{(4)}$
&~~~,~~~~\\
\bfm{e_{2}~(~e_{6}} & $~\lrw~\{$ & $ -1$, & ~$-\sg_{3}$, & ~$-\sg_{3}$, & 
~$-\sg_{3} ~\}_{(3)}$
&~~~,~~~~& 
\bfm{e_{2}~(~e_{7}} & $~\lrw~\{$ & $ -i\sg_{2}$, & ~$-i\sg_{2}$, & ~$-\sg_{1}$, & 
~$-\sg_{1} ~\}_{(3)}$
&~~~,~~~~\\
\bfm{e_{3}~(~e_{1}} & $~\lrw~\{$ & $ \sg_{3}$, & ~$\sg_{3}$, & ~$-1$, & ~$1 ~\}_{(2)}$
&~~~,~~~~& 
\bfm{e_{3}~(~e_{2}} & $~\lrw~\{$ & $ -\sg_{1}$, & ~$-\sg_{1}$, & ~$i\sg_{2}$, & 
~$-i\sg_{2} ~\}_{(1)}$
&~~~,~~~~\\
\bfm{e_{3}~(~e_{4}} & $~\lrw~\{$ & $ \sg_{1}$, & ~$-\sg_{1}$, & ~$-i\sg_{2}$, & 
~$-i\sg_{2} ~\}_{(4)}$
&~~~,~~~~& 
\bfm{e_{3}~(~e_{5}} & $~\lrw~\{$ & $ -\sg_{3}$, & ~$\sg_{3}$, & ~$-1$, & 
~$-1 ~\}_{(4)}$
&~~~,~~~~\\
\bfm{e_{3}~(~e_{6}} & $~\lrw~\{$ & $ i\sg_{2}$, & ~$i\sg_{2}$, & ~$\sg_{1}$, & 
~$-\sg_{1} ~\}_{(3)}$
&~~~,~~~~& 
\bfm{e_{3}~(~e_{7}} & $~\lrw~\{$ & $ -1$, & ~$-1$, & ~$-\sg_{3}$, & 
~$\sg_{3} ~\}_{(3)}$
&~~~,~~~~\\
\bfm{e_{4}~(~e_{1}} & $~\lrw~\{$ & $ -\sg_{1}$, & ~$-i\sg_{2}$, & ~$-\sg_{1}$, & 
~$-i\sg_{2} ~\}_{(3)}$
&~~~,~~~~& 
\bfm{e_{4}~(~e_{2}} & $~\lrw~\{$ & $ -\sg_{3}$, & ~$-1$, & ~$-\sg_{3}$, & 
~$-1 ~\}_{(4)}$
&~~~,~~~~\\
\bfm{e_{4}~(~e_{3}} & $~\lrw~\{$ & $ -\sg_{1}$, & ~$i\sg_{2}$, & ~$-\sg_{1}$, & 
~$i\sg_{2} ~\}_{(4)}$
&~~~,~~~~& 
\bfm{e_{4}~(~e_{5}} & $~\lrw~\{$ & $ \sg_{1}$, & ~$-i\sg_{2}$, & ~$-\sg_{1}$, & 
~$i\sg_{2} ~\}_{(1)}$
&~~~,~~~~\\
\bfm{e_{4}~(~e_{6}} & $~\lrw~\{$ & $ 1$, & ~$\sg_{3}$, & ~$-1$, & 
~$-\sg_{3} ~\}_{(2)}$
&~~~,~~~~& 
\bfm{e_{4}~(~e_{7}} & $~\lrw~\{$ & $ i\sg_{2}$, & ~$\sg_{1}$, & ~$-i\sg_{2}$, & 
~$-\sg_{1} ~\}_{(2)}$
&~~~,~~~~\\
\bfm{e_{5}~(~e_{1}} & $~\lrw~\{$ & $ \sg_{3}$, & ~$1$, & ~$\sg_{3}$, & 
~$-1 ~\}_{(3)}$
&~~~,~~~~& 
\bfm{e_{5}~(~e_{2}} & $~\lrw~\{$ & $ -\sg_{1}$, & ~$-i\sg_{2}$, & ~$-\sg_{1}$, & 
~$i\sg_{2} ~\}_{(4)}$
&~~~,~~~~\\
\bfm{e_{5}~(~e_{3}} & $~\lrw~\{$ & $ \sg_{3}$, & ~$-1$, & ~$\sg_{3}$, & ~$1 ~\}_{(4)}$
&~~~,~~~~& 
\bfm{e_{5}~(~e_{4}} & $~\lrw~\{$ & $ -\sg_{1}$, & ~$i\sg_{2}$, & ~$-\sg_{1}$, & 
~$-i\sg_{2} ~\}_{(1)}$
&~~~,~~~~\\
\bfm{e_{5}~(~e_{6}} & $~\lrw~\{$ & $ -i\sg_{2}$, & ~$-\sg_{1}$, & ~$i\sg_{2}$, & 
~$-\sg_{1} ~\}_{(2)}$
&~~~,~~~~& 
\bfm{e_{5}~(~e_{7}} & $~\lrw~\{$ & $ 1$, & ~$\sg_{3}$, & ~$-1$, & ~$\sg_{3} ~\}_{(2)}$
&~~~,~~~~\\
\bfm{e_{6}~(~e_{1}} & $~\lrw~\{$ & $i\sg_{2}$, & ~$-\sg_{1}$, & ~$\sg_{1}$, & 
~$-i\sg_{2} ~\}_{(4)}$
&~~~,~~~~& 
\bfm{e_{6}~(~e_{2}} & $~\lrw~\{$ & $ 1$, & ~$-\sg_{3}$, & ~$\sg_{3}$, & ~$-1 ~\}_{(3)}$
&~~~,~~~~\\
\bfm{e_{6}~(~e_{3}} & $~\lrw~\{$ & $ -i\sg_{2}$, & ~$-\sg_{1}$, & ~$-\sg_{1}$, & 
~$-i\sg_{2} ~\}_{(3)}$
&~~~,~~~~& 
\bfm{e_{6}~(~e_{4}} & $~\lrw~\{$ & $ -1$, & ~$-\sg_{3}$, & ~$-\sg_{3}$, & 
~$-1 ~\}_{(2)}$
&~~~,~~~~\\
\bfm{e_{6}~(~e_{5}} & $~\lrw~\{$ & $ i\sg_{2}$, & ~$\sg_{1}$, & ~$-\sg_{1}$, & 
~$-i\sg_{2} ~\}_{(2)}$
&~~~,~~~~& 
\bfm{e_{6}~(~e_{7}} & $~\lrw~\{$ & $ -\sg_{1}$, & ~$i\sg_{2}$, & ~$i\sg_{2}$, & 
~$-\sg_{1} ~\}_{(1)}$
&~~~,~~~~\\
\bfm{e_{7}~(~e_{1}} & $~\lrw~\{$ & $ -1$, & ~$\sg_{3}$, & ~$-\sg_{3}$, & 
~$-1 ~\}_{(4)}$
&~~~,~~~~& 
\bfm{e_{7}~(~e_{2}} & $~\lrw~\{$ & $ i\sg_{2}$, & ~$-\sg_{1}$, & ~$\sg_{1}$, & 
~$i\sg_{2} ~\}_{(3)}$
&~~~,~~~~\\
\bfm{e_{7}~(~e_{3}} & $~\lrw~\{$ & $ 1$, & ~$\sg_{3}$, & ~$\sg_{3}$, & 
~$-1 ~\}_{(3)}$
&~~~,~~~~& 
\bfm{e_{7}~(~e_{4}} & $~\lrw~\{$ & $ -i\sg_{2}$, & ~$-\sg_{1}$, & ~$-\sg_{1}$, & 
~$i\sg_{2} ~\}_{(2)}$
&~~~,~~~~\\
\bfm{e_{7}~(~e_{5}} & $~\lrw~\{$ & $ -1$, & ~$-\sg_{3}$, & ~$\sg_{3}$, & 
~$-1 ~\}_{(2)}$
&~~~,~~~~& 
\bfm{e_{7}~(~e_{6}} & $~\lrw~\{$ & $ \sg_{1}$, & ~$-i\sg_{2}$, & ~$-i\sg_{2}$, & 
~$-\sg_{1} ~\}_{(1)}$
&~~~.~~~~\\
 & & & & & & & & & & & & & \\
\hline
\et
\ec

\new

\section*{Appendix A2}

In this appendix we explicitly give the rules which enable us, 
given a generic $8\times 8$ real matrix, to quickly obtain its octonionic 
counterpart.
\bel{a2}
M_{8\times 8}= \left( \begin{array}{cccccccc} 
\ag_{1} & \bg_{1} & \cg_{1} & \dg_{1} & \eg_{1} & \vpg_{1} & \eta_{1} & 
\lgg_{1}\\
\ag_{2} & \bg_{2} & \cg_{2} & \dg_{2} & \eg_{2} & \vpg_{2} & \eta_{2} & 
\lgg_{2}\\
\ag_{3} & \bg_{3} & \cg_{3} & \dg_{3} & \eg_{3} & \vpg_{3} & \eta_{3} & 
\lgg_{3}\\
\ag_{4} & \bg_{4} & \cg_{4} & \dg_{4} & \eg_{4} & \vpg_{4} & \eta_{4} & 
\lgg_{4}\\
\ag_{5} & \bg_{5} & \cg_{5} & \dg_{5} & \eg_{5} & \vpg_{5} & \eta_{5} & 
\lgg_{5}\\
\ag_{6} & \bg_{6} & \cg_{6} & \dg_{6} & \eg_{6} & \vpg_{6} & \eta_{6} & 
\lgg_{6}\\
\ag_{7} & \bg_{7} & \cg_{7} & \dg_{7} & \eg_{7} & \vpg_{7} & \eta_{7} & 
\lgg_{7}\\
\ag_{8} & \bg_{8} & \cg_{8} & \dg_{8} & \eg_{8} & \vpg_{8} & \eta_{8} & 
\lgg_{8}
\end{array} \right) \q \lrw \q O = \sum_{m=1}^{64}x_{m}\rho_{m} \q ,
\ee
where $x_{m}$ are real numbers and
\bc
\bt{lclclclc}
$\rho_{1}=\bfm{1}$  & ~~,~~~ &$ \rho_{2}=\bfm{e_{1}} $  & ~~,~~ &
$\rho_{3}=\bfm{e_{2}}$  & ~~,~~~ & $\rho_{4}=\bfm{e_{3}}$   & ~~,~~\\
$\rho_{5}=\bfm{e_{4}}$  & ~~,~~~ & $\rho_{5}=\bfm{e_{5}}$   & ~~,~~&
$\rho_{7}=\bfm{e_{6}}$  & ~~,~~~ & $\rho_{8}=\bfm{e_{7}}$   & ~~,~~\\
$\rho_{9}=\bfm{1\mid e_{1}} $ & ~~,~~~ & $\rho_{10}=\bfm{1\mid e_{2}}$  & ~~,~~&
$\rho_{11}=\bfm{1\mid e_{3}}$ & ~~,~~~ & $\rho_{12}=\bfm{1\mid e_{4}}$  & ~~,~~\\
$\rho_{13}=\bfm{1\mid e_{5}}$ & ~~,~~~ & $\rho_{14}=\bfm{1\mid e_{6}}$  & ~~,~~&
$\rho_{15}=\bfm{1\mid e_{7}}$ & ~~,~~~ & $\rho_{16}=\bfm{e_{1}\mid e_{1}}$  
& ~~,~~\\
$\rho_{17}=\bfm{e_{2}\mid e_{2}}$ & ~~,~~~ & $\rho_{18}=\bfm{e_{3}\mid e_{3}}$  
& ~~,~~&
$\rho_{19}=\bfm{e_{4}\mid e_{4}}$ & ~~,~~~ & $\rho_{20}=\bfm{e_{5}\mid e_{5}}$  
& ~~,~~\\
$\rho_{21}=\bfm{e_{6}\mid e_{6}}$ & ~~,~~~ & $\rho_{22}=\bfm{e_{7}\mid e_{7}}$  
& ~~,~~&
$\rho_{23}=\bfm{e_{1}~)~e_{2}}$ & ~~,~~~ & $\rho_{24}=\bfm{e_{1}~)~e_{3}}$  
& ~~,~~\\
$\rho_{25}=\bfm{e_{1}~)~e_{4}}$ & ~~,~~~ & $\rho_{26}=\bfm{e_{1}~)~ e_{5}}$  
& ~~,~~&
$\rho_{27}=\bfm{e_{1}~)~e_{6}}$ & ~~,~~~ & $\rho_{28}=\bfm{e_{1}~)~e_{7}} $ 
& ~~,~~\\
$\rho_{29}=\bfm{e_{2}~)~e_{1}}$ & ~~,~~~ & $\rho_{30}=\bfm{e_{2}~)~ e_{3}}$  
& ~~,~~&
$\rho_{31}=\bfm{e_{2}~)~e_{4}}$ & ~~,~~~ & $\rho_{32}=\bfm{e_{2}~)~e_{5}} $ 
& ~~,~~\\
$\rho_{33}=\bfm{e_{2}~)~e_{6}}$ & ~~,~~~ & $\rho_{34}=\bfm{e_{2}~)~ e_{7}}$  
& ~~,~~&
$\rho_{35}=\bfm{e_{3}~)~e_{1}}$ & ~~,~~~ & $\rho_{36}=\bfm{e_{3}~)~e_{2}} $ 
& ~~,~~\\
$\rho_{37}=\bfm{e_{3}~)~e_{4}}$ & ~~,~~~ & $\rho_{38}=\bfm{e_{3}~)~ e_{5}}$  
& ~~,~~&
$\rho_{39}=\bfm{e_{3}~)~e_{6}}$ & ~~,~~~ & $\rho_{40}=\bfm{e_{3}~)~e_{7}} $ 
& ~~,~~\\
$\rho_{41}=\bfm{e_{4}~)~e_{1}}$ & ~~,~~~ & $\rho_{42}=\bfm{e_{4}~)~ e_{2}}$  
& ~~,~~&
$\rho_{43}=\bfm{e_{4}~)~e_{3}}$ & ~~,~~~ & $\rho_{44}=\bfm{e_{4}~)~e_{5}} $ 
& ~~,~~\\
$\rho_{45}=\bfm{e_{4}~)~e_{6}}$ & ~~,~~~ & $\rho_{46}=\bfm{e_{4}~)~ e_{7}}$  
& ~~,~~&
$\rho_{47}=\bfm{e_{5}~)~e_{1}}$ & ~~,~~~ & $\rho_{48}=\bfm{e_{5}~)~e_{2}} $ 
& ~~,~~\\
$\rho_{49}=\bfm{e_{5}~)~e_{3}}$ & ~~,~~~ & $\rho_{50}=\bfm{e_{5}~)~ e_{4}}$  
& ~~,~~&
$\rho_{51}=\bfm{e_{5}~)~e_{6}}$ & ~~,~~~ & $\rho_{52}=\bfm{e_{5}~)~e_{7}} $ 
& ~~,~~\\
$\rho_{53}=\bfm{e_{6}~)~e_{1}}$ & ~~,~~~ & $\rho_{54}=\bfm{e_{6}~)~ e_{2}}$  
& ~~,~~&
$\rho_{55}=\bfm{e_{6}~)~e_{3}}$ & ~~,~~~ & $\rho_{56}=\bfm{e_{6}~)~e_{4}} $ 
& ~~,~~\\
$\rho_{57}=\bfm{e_{6}~)~e_{5}}$ & ~~,~~~ & $\rho_{58}=\bfm{e_{6}~)~ e_{7}}$  
& ~~,~~&
$\rho_{59}=\bfm{e_{7}~)~e_{1}}$ & ~~,~~~ & $\rho_{60}=\bfm{e_{7}~)~e_{2}} $ 
& ~~,~~\\
$\rho_{61}=\bfm{e_{7}~)~e_{3}}$ & ~~,~~~ & $\rho_{62}=\bfm{e_{7}~)~ e_{4}}$  
& ~~,~~&
$\rho_{63}=\bfm{e_{7}~)~e_{5}}$ & ~~,~~~ & $\rho_{64}=\bfm{e_{7}~)~e_{6}} $ 
& ~~.~~
\et
\ec
\vs{1cm}
\bc
{\tt TABLE A2}\\
{\fs Real coefficients for the octonionic barred operators}\\
\vs{.2cm}
{\fs
\bt{|rclcrclc|}
\hline
 & & & & & & & \\ 
~~~$x_{1}$ & $=$ & 
$(5\ag_{1}+\bg_{2}+\cg_{3}+\dg_{4}+\eg_{5}+\vpg_{6}+\eta_{7}+\lgg_{8})/12$
&~~,~~~& 
$x_{2}$ & $=$ & 
$(4\ag_{1}-\bg_{1}-\cg_{4}-\eg_{6}+\vpg_{5}+\eta_{8}-\lgg_{7})/10$
&~~,~~~\\
$x_{3}$ & $=$ & 
$(5\ag_{3}+\bg_{4}-\cg_{1}-\dg_{2}-\eg_{7}-\vpg_{8}+\eta_{5}+\lgg_{6})/12$
&~~,~~~& 
$x_{4}$ & $=$ & 
$(5\ag_{4}-\bg_{3}+\cg_{2}-\dg_{1}-\eg_{8}+\vpg_{7}-\eta_{6}+\lgg_{5})/12$
&~~,~~~\\
$x_{5}$ & $=$ & 
$(5\ag_{5}+\bg_{6}+\cg_{7}+\dg_{8}-\eg_{1}-\vpg_{2}-\eta_{3}-\lgg_{4})/12$
&~~,~~~& 
$x_{6}$ & $=$ & 
$(5\ag_{6}-\bg_{5}+\cg_{8}-\dg_{7}+\eg_{2}-\vpg_{1}+\eta_{4}-\lgg_{3})/12$
&~~,~~~\\
$x_{7}$ & $=$ & 
$(5\ag_{7}-\bg_{8}-\cg_{5}+\dg_{6}+\eg_{3}-\vpg_{4}-\eta_{1}+\lgg_{2})/12$
&~~,~~~& 
$x_{8}$ & $=$ & 
$(3\ag_{8}-\bg_{7}-\cg_{6}-\dg_{5}+\eg_{4}+\vpg_{3}-\eta_{2}-\lgg_{1})/12$
&~~,~~~\\
$x_{9}$ & $=$ & 
$(\ag_{2}-4\bg_{1}+\cg_{4}+\eg_{6}-\vpg_{5}-\eta_{8}+\lgg_{7})/10$
&~~,~~~& 
$x_{10}$ & $=$ & 
$(\ag_{3}-\bg_{4}-5\cg_{1}+\dg_{2}+\eg_{7}+\vpg_{8}-\eta_{5}-\lgg_{6})/12$
&~~,~~~\\
~~~$x_{11}$ & $=$ & 
$(\ag_{4}+\bg_{3}-\cg_{2}-5\dg_{1}+\eg_{8}-\vpg_{7}+\eta_{6}-\lgg_{5})/12$
&~~,~~~& 
$x_{12}$ & $=$ & 
$(\ag_{5}-\bg_{6}-\cg_{7}-\dg_{8}-5\eg_{1}+\vpg_{2}+\eta_{3}+\lgg_{4})/12$
&~~,~~~\\
$x_{13}$ & $=$ & 
$(\ag_{6}+\bg_{5}-\cg_{8}+\dg_{7}-\eg_{2}-5\vpg_{1}-\eta_{4}+\lgg_{3})/12$
&~~,~~~& 
$x_{14}$ & $=$ & 
$(\ag_{7}+\bg_{8}+\cg_{5}-\dg_{6}-\eg_{3}+\vpg_{4}-5\eta_{1}-\lgg_{2})/12$
&~~,~~~\\
$x_{15}$ & $=$ & 
$(\ag_{8}+\bg_{7}+\cg_{6}+\dg_{5}-\eg_{4}-\vpg_{3}+\eta_{2}-3\lgg_{1})/8$
&~~,~~~& 
$x_{16}$ & $=$ & 
$(-\ag_{1}-5\bg_{2}+\cg_{3}+\dg_{4}+\eg_{5}+\vpg_{6}+\eta_{7}+\lgg_{8})/12$
&~~,~~~\\
$x_{17}$ & $=$ & 
$(-\ag_{1}+\bg_{2}-5\cg_{3}+\dg_{4}+\eg_{5}+\vpg_{6}+\eta_{7}+\lgg_{8})/12$
&~~,~~~& 
$x_{18}$ & $=$ & 
$(-\ag_{1}+\bg_{2}+\cg_{3}-5\dg_{4}+\eg_{5}+\vpg_{6}+\eta_{7}+\lgg_{8})/12$
&~~,~~~ \\
$x_{19}$ & $=$ & 
$(-\ag_{1}+\bg_{2}+\cg_{3}+\dg_{4}-5\eg_{5}+\vpg_{6}+\eta_{7}+\lgg_{8})/12$
&~~,~~~& 
$x_{20}$ & $=$ & 
$(-\ag_{1}+\bg_{2}+\cg_{3}+\dg_{4}+\eg_{5}-5\vpg_{6}+\eta_{7}+\lgg_{8})/12$
&~~,~~~ \\
$x_{21}$ & $=$ & 
$(-\ag_{1}+\bg_{2}+\cg_{3}+\dg_{4}+\eg_{5}+\vpg_{6}-5\eta_{7}+\lgg_{8})/12$
&~~,~~~ &
$x_{22}$ & $=$ & 
$(-\ag_{1}+\bg_{2}+\cg_{3}+\dg_{4}+\eg_{5}+\vpg_{6}+\eta_{7}-5\lgg_{8})/12$
&~~.~~~ \\
 & & & & & & & \\
\hline
\et
}
\ec
%\vs{1cm}

\new

\bc
{\tt TABLE A2-L}\\
{\fs Real coefficients for the octonionic left-barred operators}\\
\vs{.2cm}
{\fs
\bt{|rclcrclc|}
\hline
 & & & & & & & \\
~~~$x_{23}$ & $=$ & 
$(-\ag_{4}-\bg_{3}-5\cg_{2}-\dg_{1}-\eg_{8}+\vpg_{7}-\eta_{6}+\lgg_{5})/12$
&~~,~~~& 
$x_{24}$ & $=$ & 
$(\ag_{3}-\bg_{4}+\cg_{1}-5\dg_{2}+\eg_{7}+\vpg_{8}-\eta_{5}-\lgg_{6})/12$
&~~,~~~\\
$x_{25}$ & $=$ & 
$(-\ag_{6}-\bg_{5}+\cg_{8}-\dg_{7}-5\eg_{2}-\vpg_{1}+\eta_{4}-\lgg_{3})/12$
&~~,~~~& 
$x_{26}$ & $=$ & 
$(\ag_{5}-\bg_{6}-\cg_{7}-\dg_{8}+\eg_{1}-5\vpg_{2}+\eta_{3}+\lgg_{4})/12$
&~~,~~~\\
$x_{27}$ & $=$ & 
$(\ag_{8}+\bg_{7}+\cg_{6}+\dg_{5}-\eg_{4}-\vpg_{3}-3\eta_{2}+\lgg_{1})/12$
&~~,~~~& 
$x_{28}$ & $=$ & 
$(-\ag_{7}-\bg_{8}-\cg_{5}+\dg_{6}+\eg_{3}-\vpg_{4}-\eta_{1}-5\lgg_{2})/12$
&~~,~~~\\
$x_{29}$ & $=$ & 
$(\ag_{4}-5\bg_{3}-\cg_{2}+\dg_{1}+\eg_{8}-\vpg_{7}+\eta_{6}-\lgg_{5})/12$
&~~,~~~& 
$x_{30}$ & $=$ & 
$(-\ag_{2}-\bg_{1}-\cg_{4}-5\dg_{3}-\eg_{6}+\vpg_{5}+\eta_{8}-\lgg_{7})/12$
&~~,~~~\\
$x_{31}$ & $=$ & 
$(-\ag_{7}-\bg_{8}-\cg_{5}+\dg_{6}-5\eg_{3}-\vpg_{4}-\eta_{1}+\lgg_{2})/12$
&~~,~~~& 
$x_{32}$ & $=$ & 
$(\bg_{7}-\vpg_{3})/2$
&~~,~~~\\
$x_{33}$ & $=$ & 
$(\ag_{5}-\bg_{6}-\cg_{7}-\dg_{8}+\eg_{1}+\vpg_{2}-5\eta_{3}+\lgg_{4})/12$
&~~,~~~& 
$x_{34}$ & $=$ & 
$(\ag_{6}+\bg_{5}-\cg_{8}+\dg_{7}-\eg_{2}+\vpg_{1}-\eta_{4}-5\lgg_{3})/12$
&~~,~~~\\
$x_{35}$ & $=$ & 
$(-\ag_{3}-5\bg_{4}-\cg_{1}-\dg_{2}-\eg_{7}-\vpg_{8}+\eta_{5}+\lgg_{6})/12$
&~~,~~~& 
$x_{36}$ & $=$ & 
$(\ag_{2}+\bg_{1}-4\cg_{4}+\eg_{6}-\vpg_{5}-\eta_{8}+\lgg_{7})/10$
&~~,~~~\\
$x_{37}$ & $=$ & 
$(-\ag_{8}-\bg_{7}-\cg_{6}-\dg_{5}-3\eg_{4}+\vpg_{3}-\eta_{2}-\lgg_{1})/8$
&~~,~~~& 
$x_{38}$ & $=$ & 
$(\ag_{7}+\bg_{8}+\cg_{5}-\dg_{6}-\eg_{3}-5\vpg_{4}+\eta_{1}-\lgg_{2})/12$
&~~,~~~\\
$x_{39}$ & $=$ & 
$(-\ag_{6}-\bg_{5}+\cg_{8}-\dg_{7}+\eg_{2}-\vpg_{1}-5\eta_{4}-\lgg_{3})/12$
&~~,~~~& 
$x_{40}$ & $=$ & 
$(\ag_{5}-\bg_{6}-\cg_{7}-\dg_{8}+\eg_{1}+\vpg_{2}+\eta_{3}-5\lgg_{4})/12$
&~~,~~~\\
$x_{41}$ & $=$ & 
$(\ag_{6}-5\bg_{5}-\cg_{8}+\dg_{7}-\eg_{2}+\vpg_{1}-\eta_{4}+\lgg_{3})/12$
&~~,~~~& 
$x_{42}$ & $=$ & 
$(\ag_{7}+\bg_{8}-5\cg_{5}-\dg_{6}-\eg_{3}+\vpg_{4}+\eta_{1}-\lgg_{2})/12$
&~~,~~~\\
$x_{43}$ & $=$ & 
$(-\bg_{7}-\dg_{5})/2$
&~~,~~~& 
$x_{44}$ & $=$ & 
$(-\ag_{2}-\bg_{1}-\cg_{4}-\eg_{6}-4\vpg_{5}+\eta_{8}-\lgg_{7})/10$
&~~,~~~\\
$x_{45}$ & $=$ & 
$(-\ag_{3}+\bg_{4}-\cg_{1}-\dg_{2}-\eg_{7}-\vpg_{8}-5\eta_{5}+\lgg_{6})/12$
&~~,~~~& 
$x_{46}$ & $=$ & 
$(-\ag_{4}-\bg_{3}+\cg_{2}-\dg_{1}-\eg_{8}+\vpg_{7}-\eta_{6}-5\lgg_{5})/12$
&~~,~~~\\
$x_{47}$ & $=$ & 
$(-\ag_{5}-5\bg_{6}+\cg_{7}+\dg_{8}-\eg_{1}-\vpg_{2}-\eta_{3}-\lgg_{4})/12$
&~~,~~~& 
$x_{48}$ & $=$ & 
$(\ag_{8}+\bg_{7}-3\cg_{6}+\dg_{5}-\eg_{4}-\vpg_{3}+\eta_{2}+\lgg_{1})/8$
&~~,~~~\\
$x_{49}$ & $=$ & 
$(-\ag_{7}-\bg_{8}-\cg_{5}-5\dg_{6}+\eg_{3}-\vpg_{4}-\eta_{1}+\lgg_{2})/12$
&~~,~~~& 
$x_{50}$ & $=$ & 
$(\ag_{2}+\bg_{1}+\cg_{4}-4\eg_{6}-\vpg_{5}-\eta_{8}+\lgg_{7})/10$
&~~,~~~\\
$x_{51}$ & $=$ & 
$(\ag_{4}+\bg_{3}-\cg_{2}+\dg_{1}+\eg_{8}-\vpg_{7}-5\eta_{6}-\lgg_{5})/12$
&~~,~~~& 
$x_{52}$ & $=$ & 
$(-\ag_{3}+\bg_{4}-\cg_{1}-\dg_{2}-\eg_{7}-\vpg_{8}+\eta_{5}-5\lgg_{6})/12$
&~~,~~~\\
$x_{53}$ & $=$ & 
$(-\ag_{8}-5\bg_{7}-\cg_{6}-\dg_{5}+\eg_{4}+\vpg_{3}-\eta_{2}-\lgg_{1})/8$
&~~,~~~& 
$x_{54}$ & $=$ & 
$(-\ag_{5}+\bg_{6}-5\cg_{7}+\dg_{8}-\eg_{1}-\vpg_{2}-\eta_{3}-\lgg_{4})/12$
&~~,~~~\\
$x_{55}$ & $=$ & 
$(\ag_{6}+\bg_{5}-\cg_{8}-5\dg_{7}-\eg_{2}+\vpg_{1}-\eta_{4}+\lgg_{3})/12$
&~~,~~~& 
$x_{56}$ & $=$ & 
$(\ag_{3}-\bg_{4}+\cg_{1}+\dg_{2}-5\eg_{7}+\vpg_{8}-\eta_{5}-\lgg_{6})/12$
&~~,~~~\\
$x_{57}$ & $=$ & 
$(-\ag_{4}-\bg_{3}+\cg_{2}-\dg_{1}-\eg_{8}-5\vpg_{7}-\eta_{6}+\lgg_{5})/12$
&~~,~~~& 
$x_{58}$ & $=$ & 
$(\ag_{2}+\bg_{1}+\cg_{4}+\eg_{6}-\vpg_{5}-\eta_{8}-4\lgg_{7})/10$
&~~,~~~\\
$x_{59}$ & $=$ & 
$(\ag_{7}-5\bg_{8}+\cg_{5}-\dg_{6}-\eg_{3}+\vpg_{4}+\eta_{1}-\lgg_{2})/12$
&~~,~~~& 
$x_{60}$ & $=$ & 
$(-\ag_{6}-\bg_{5}-5\cg_{8}-\dg_{7}+\eg_{2}-\vpg_{1}+\eta_{4}-\lgg_{3})/12$
&~~,~~~\\
$x_{61}$ & $=$ & 
$(-\ag_{5}+\bg_{6}+\cg_{7}-5\dg_{8}-\eg_{1}-\vpg_{2}-\eta_{3}-\lgg_{4})/12$
&~~,~~~& 
$x_{62}$ & $=$ & 
$(\ag_{4}+\bg_{3}-\cg_{2}+\dg_{1}-5\eg_{8}-\vpg_{7}+\eta_{6}-\lgg_{5})/12$
&~~,~~~\\
$x_{63}$ & $=$ & 
$(\ag_{3}-\bg_{4}+\cg_{1}+\dg_{2}+\eg_{7}-5\vpg_{8}-\eta_{5}-\lgg_{6})/12$
&~~,~~~& 
$x_{64}$ & $=$ & 
$(\dg_{3}-\eta_{8})/2$
&~~.~~~\\
 & & & & & & & \\
\hline
\et
}
\ec
\vs{.5cm}

We also give the translation rules by right-barred operators. Obviously,  
we must modify eq.~(\ref{a2}) ($\rho_{23,...,64}$ will represent right-barred 
operators).

\vs{1cm}
\bc
{\tt TABLE A2-R}\\
{\fs Real coefficients for the octonionic right-barred operators}\\
\vs{.2cm}
{\fs
\bt{|rclcrclc|}
\hline
 & & & & & & & \\
~~~$x_{23}$ & $=$ & 
$(-\ag_{4}-5\bg_{3}-\cg_{2}-\dg_{1}+\eg_{8}-\vpg_{7}+\eta_{6}-\lgg_{5})/12$
&~~,~~~& 
$x_{24}$ & $=$ & 
$(\ag_{3}-5\bg_{4}+\cg_{1}-\dg_{2}-\eg_{7}-\vpg_{8}+\eta_{5}+\lgg_{6})/12$
&~~,~~~\\
$x_{25}$ & $=$ & 
$(-\ag_{6}-5\bg_{5}-\cg_{8}+\dg_{7}-\eg_{2}-\vpg_{1}-\eta_{4}+\lgg_{3})/12$
&~~,~~~& 
$x_{26}$ & $=$ & 
$(\ag_{5}-5\bg_{6}+\cg_{7}+\dg_{8}+\eg_{1}-\vpg_{2}-\eta_{3}-\lgg_{4})/12$
&~~,~~~\\
$x_{27}$ & $=$ & 
$(\ag_{8}-5\bg_{7}-\cg_{6}-\dg_{5}+\eg_{4}+\vpg_{3}-\eta_{2}+\lgg_{1})/12$
&~~,~~~& 
$x_{28}$ & $=$ & 
$(-\ag_{7}-5\bg_{8}+\cg_{5}-\dg_{6}-\eg_{3}+\vpg_{4}-\eta_{1}-\lgg_{2})/12$
&~~,~~~\\
$x_{29}$ & $=$ & 
$(\ag_{4}-\bg_{3}-5\cg_{2}+\dg_{1}-\eg_{8}+\vpg_{7}-\eta_{6}+\lgg_{5})/12$
&~~,~~~& 
$x_{30}$ & $=$ & 
$(-\ag_{2}-\bg_{1}-5\cg_{4}-\dg_{3}+\eg_{6}-\vpg_{5}-\eta_{8}+\lgg_{7})/12$
&~~,~~~\\
$x_{31}$ & $=$ & 
$(-\ag_{7}+\bg_{8}-5\cg_{5}-\dg_{6}-\eg_{3}+\vpg_{4}-\eta_{1}-\lgg_{2})/12$
&~~,~~~& 
$x_{32}$ & $=$ & 
$(-\ag_{8}-\bg_{7}-5\cg_{6}+\dg_{5}-\eg_{4}-\vpg_{3}+\eta_{2}-\lgg_{1})/12$
&~~,~~~\\
$x_{33}$ & $=$ & 
$(\ag_{5}+\bg_{6}-5\cg_{7}+\dg_{8}+\eg_{1}-\vpg_{2}-\eta_{3}-\lgg_{4})/12$
&~~,~~~& 
$x_{34}$ & $=$ & 
$(\ag_{6}-\bg_{5}-5\cg_{8}-\dg_{7}+\eg_{2}+\vpg_{1}+\eta_{4}-\lgg_{3})/12$
&~~,~~~\\
$x_{35}$ & $=$ & 
$(-\ag_{3}-\bg_{4}-\cg_{1}-5\dg_{2}+\eg_{7}+\vpg_{8}-\eta_{5}-\lgg_{6})/12$
&~~,~~~& 
$x_{36}$ & $=$ & 
$(\ag_{2}+\bg_{1}-\cg_{4}-5\dg{3}-\eg_{6}+\vpg_{5}+\eta_{8}-\lgg_{7})/12$
&~~,~~~\\
$x_{37}$ & $=$ & 
$(-\ag_{8}-\bg_{7}+\cg_{6}-5\dg_{5}-\eg_{4}-\vpg_{3}+\eta_{2}-\lgg_{1})/12$
&~~,~~~& 
$x_{38}$ & $=$ & 
$(\ag_{7}-\bg_{8}-\cg_{5}-5\dg_{6}+\eg_{3}-\vpg_{4}+\eta_{1}+\lgg_{2})/12$
&~~,~~~\\
$x_{39}$ & $=$ & 
$(-\ag_{6}+\bg_{5}-\cg_{8}-5\dg_{7}-\eg_{2}-\vpg_{1}-\eta_{4}+\lgg_{3})/12$
&~~,~~~& 
$x_{40}$ & $=$ & 
$(\ag_{5}+\bg_{6}+\cg_{7}-5\dg_{8}+\eg_{1}-\vpg_{2}-\eta_{3}-\lgg_{4})/12$
&~~,~~~\\
$x_{41}$ & $=$ & 
$(\ag_{6}-\bg_{5}+\cg_{8}-\dg_{7}-5\eg_{2}+\vpg_{1}+\eta_{4}-\lgg_{3})/12$
&~~,~~~& 
$x_{42}$ & $=$ & 
$(\ag_{7}-\bg_{8}-\cg_{5}+\dg_{6}-5\eg_{3}-\vpg_{4}+\eta_{1}+\lgg_{2})/12$
&~~,~~~\\
$x_{43}$ & $=$ & 
$(\ag_{8}+\bg_{7}-\cg_{6}-\dg_{5}-5\eg_{4}+\vpg_{3}-\eta_{2}+\lgg_{1})/12$
&~~,~~~& 
$x_{44}$ & $=$ & 
$(-\ag_{2}-\bg_{1}+\cg_{4}-\dg{3}-5\eg_{6}-\vpg_{5}-\eta_{8}+\lgg_{7})/12$
&~~,~~~\\
$x_{45}$ & $=$ & 
$(-\ag_{3}-\bg_{4}-\cg_{1}+\dg_{2}-5\eg_{7}+\vpg_{8}-\eta_{5}-\lgg_{6})/12$
&~~,~~~& 
$x_{46}$ & $=$ & 
$(-\ag_{4}+\bg_{3}-\cg_{2}-\dg_{1}-5\eg_{8}-\vpg_{7}+\eta_{6}-\lgg_{5})/12$
&~~,~~~\\
$x_{47}$ & $=$ & 
$(-\ag_{5}-\bg_{6}-\cg_{7}-\dg_{8}-\eg_{1}-5\vpg_{2}+\eta_{3}+\lgg_{4})/12$
&~~,~~~& 
$x_{48}$ & $=$ & 
$(\ag_{8}+\bg_{7}-\cg_{6}-\dg_{5}+\eg_{4}-5\vpg_{3}-\eta_{2}+\lgg_{1})/12$
&~~,~~~\\
$x_{49}$ & $=$ & 
$(-\ag_{7}+\bg_{8}+\cg_{5}-\dg_{6}-\eg_{3}-5\vpg_{4}-\eta_{1}-\lgg_{2})/12$
&~~,~~~& 
$x_{50}$ & $=$ & 
$(\ag_{2}+\bg_{1}-\cg_{4}+\dg{3}-\eg_{6}-5\vpg_{5}+\eta_{8}-\lgg_{7})/12$
&~~,~~~\\
$x_{51}$ & $=$ & 
$(\ag_{4}-\bg_{3}+\cg_{2}+\dg_{1}-\eg_{8}-5\vpg_{7}-\eta_{6}+\lgg_{5})/12$
&~~,~~~& 
$x_{52}$ & $=$ & 
$(-\ag_{3}-\bg_{4}-\cg_{1}+\dg_{2}+\eg_{7}-5\vpg_{8}-\eta_{5}-\lgg_{6})/12$
&~~,~~~\\
$x_{53}$ & $=$ & 
$(-\ag_{8}-\bg_{7}+\cg_{6}+\dg_{5}-\eg_{4}-\vpg_{3}-5\eta_{2}-\lgg_{1})/12$
&~~,~~~& 
$x_{54}$ & $=$ & 
$(-\ag_{5}-\bg_{6}-\cg_{7}-\dg_{8}-\eg_{1}+\vpg_{2}-5\eta_{3}+\lgg_{4})/12$
&~~,~~~\\
$x_{55}$ & $=$ & 
$(\ag_{6}-\bg_{5}+\cg_{8}-\dg_{7}+\eg_{2}+\vpg_{1}-5\eta_{4}-\lgg_{3})/12$
&~~,~~~& 
$x_{56}$ & $=$ & 
$(\ag_{3}+\bg_{4}+\cg_{1}-\dg_{2}-\eg_{7}-\vpg_{8}-5\eta_{5}+\lgg_{6})/12$
&~~,~~~\\
$x_{57}$ & $=$ & 
$(-\ag_{4}+\bg_{3}-\cg_{2}-\dg_{1}+\eg_{8}-\vpg_{7}-5\eta_{6}-\lgg_{5})/12$
&~~,~~~& 
$x_{58}$ & $=$ & 
$(\ag_{2}+\bg_{1}-\cg_{4}+\dg_{3}-\eg_{6}+\vpg_{5}-5\eta_{8}-\lgg_{7})/12$
&~~,~~~\\
$x_{59}$ & $=$ & 
$(\ag_{7}-\bg_{8}-\cg_{5}+\dg_{6}+\eg_{3}-\vpg_{4}+\eta_{1}-5\lgg_{2})/12$
&~~,~~~& 
$x_{60}$ & $=$ & 
$(-\ag_{6}+\bg_{5}-\cg_{8}+\dg_{7}-\eg_{2}-\vpg_{1}-\eta_{4}-5\lgg_{3})/12$
&~~,~~~\\
$x_{61}$ & $=$ & 
$(-\ag_{5}-\bg_{6}-\cg_{7}-\dg_{8}-\eg_{1}+\vpg_{2}+\eta_{3}-5\lgg_{4})/12$
&~~,~~~& 
$x_{62}$ & $=$ & 
$(\ag_{4}-\bg_{3}+\cg_{2}+\dg_{1}-\eg_{8}+\vpg_{7}-\eta_{6}-5\lgg_{5})/12$
&~~,~~~\\
$x_{63}$ & $=$ & 
$(\ag_{3}+\bg_{4}+\cg_{1}-\dg_{2}-\eg_{7}-\vpg_{8}+\eta_{5}-5\lgg_{6})/12$
&~~,~~~& 
$x_{64}$ & $=$ & 
$(-\ag{2}-\bg{1}+\cg{4}-\dg_{3}+\eg{6}-\vpg{5}-\eta_{8}-5\lgg_{7})/12$
&~~.~~~\\
 & & & & & & & \\
\hline
\et
}
\ec

\new

\section*{Appendix B1}
We give the action of barred operators on octonionic functions
\[ \psi = \psi_{1} +e_{2}\psi_{2}+e_{4}\psi_{3}+e_{6}\psi_{4} \qq
[~\psi_{1, ...,4} \in \co (1, \; e_{1})~] \q .
\]
In the following tables we use the notation
\[
e_{2} ~\raw ~ \{ -\psi_{2}, \; \psi_{1}, \;
-\psi_{4}^{*}, \; \psi_{3}^{*} \} \q ,
\]
to indicate
\[ 
e_{2}\psi~=~ -\psi_{2}+e_{2}\psi_{1}-e_{4}\psi_{4}^{*}+e_{6}\psi_{3}^{*} \q .
\]
\vs{.5cm}
\bc
{\tt TABLE B1a}\\
{\fs Action on $\psi$, of the octonionic barred operators, $e_{m}$ and 
$1\mid e_{m}$}\\
\vs{.2cm}
\bt{|rcrrrrcrcrrrrc|}
\hline 
 & & & & & & & & & & & & & \\ 
~~~~~~~~\bfm{e_{1}} & $~\raw~\{$ & $e_{1}\psi_{1}$, & ~$-e_{1}\psi_{2}$, & 
~$-e_{1}\psi_{3}$, & 
~$-e_{1}\psi_{4} ~\}$
&~~~,~~~~& 
\bfm{1\mid e_{1}} & $~\raw~\{$ & $e_{1}\psi_{1}$, & ~$e_{1}\psi_{2}$, & 
~$e_{1}\psi_{3}$, & 
~$e_{1}\psi_{4} ~\}$
&~~~,~~~~\\
\bfm{e_{2}} & $~\raw~\{$ & $ -\psi_{2}$, & ~$\psi_{1}$, 
& ~$-\psi_{4}^{*}$, & ~$\psi_{3}^{*} ~\}$
&~~~,~~~~& 
\bfm{1\mid e_{2}} & $~\raw~\{$ & $ -\psi_{2}^{*}$, & ~$\psi_{1}^{*}$, 
& ~$\psi_{4}^{*}$, & 
~$-\psi_{3}^{*} ~\}$
&~~~,~~~~\\
\bfm{e_{3}} & $~\raw~\{$ & $ -e_{1}\psi_{2}$, & ~$-e_{1}\psi_{1}$, 
& ~$-e_{1}\psi_{4}^{*}$, & 
~$e_{1}\psi_{3}^{*} ~\}$
&~~~,~~~~& 
\bfm{1\mid e_{3}} & $~\raw~\{$ & $ e_{1}\psi_{2}^{*}$, & ~$-e_{1}\psi_{1}^{*}$, 
& ~$e_{1}\psi_{4}^{*}$, & 
~$-e_{1}\psi_{3}^{*} ~\}$
&~~~,~~~~\\
\bfm{e_{4}} & $~\raw~\{$ & $ -\psi_{3}$, & ~$\psi_{4}^{*}$, 
& ~$\psi_{1}$, & ~$-\psi_{2}^{*} ~\}$
&~~~,~~~~& 
\bfm{1\mid e_{4}} & $~\raw~\{$ & $ -\psi_{3}^{*}$, & ~$-\psi_{4}^{*}$, 
& ~$\psi_{1}^{*}$, & 
~$\psi_{2}^{*} ~\}$
&~~~,~~~~\\
\bfm{e_{5}} & $~\raw~\{$ & $ -e_{1}\psi_{3}$, & ~$e_{1}\psi_{4}^{*}$, 
& ~$-e_{1}\psi_{1}$, & 
~$-e_{1}\psi_{2}^{*} ~\}$
&~~~,~~~~& 
\bfm{1\mid e_{5}} & $~\raw~\{$ & $ e_{1}\psi_{3}^{*}$, & ~$-e_{1}\psi_{4}^{*}$, 
&~$-e_{1}\psi_{1}^{*}$, & 
~$e_{1}\psi_{2}^{*} ~\}$
&~~~,~~~~\\
\bfm{e_{6}} & $~\raw~\{$ & $ -\psi_{4}$, & ~$-\psi_{3}^{*}$, &
 ~$\psi_{2}^{*}$, & 
~$\psi_{1} ~\}$
&~~~,~~~~&
\bfm{1\mid e_{6}} & $~\raw~\{$ & $ -\psi_{4}^{*}$, & ~$\psi_{3}^{*}$, 
& ~$-\psi_{2}^{*}$, & 
~$\psi_{1}^{*} ~\}$
&~~~,~~~~\\
\bfm{e_{7}} & $~\raw~\{$ & $ e_{1}\psi_{4}$, & ~$e_{1}\psi_{3}^{*}$, 
& ~$-e_{1}\psi_{2}^{*}$, & ~$e_{1}\psi_{1} ~\}$
&~~~,~~~~& 
\bfm{1\mid e_{7}} & $~\raw~\{$ & $ -e_{1}\psi_{4}^{*}$, 
& ~$-e_{1}\psi_{3}^{*}$, 
& ~$e_{1}\psi_{2}^{*}$, & 
~$e_{1}\psi_{1}^{*} ~\}$
&~~~,~~~~\\
 & & & & & & & & & & & & & \\
\hline
\et
\ec
\vs{1cm}
\bc
{\tt TABLE B1b}\\
{\fs Action on $\psi$, of the octonionic barred operators, 
$e_{m} \mid e_{m}$}\\
\vs{.2cm}
\bt{|rcrrrrcrcrrrrc|}
\hline 
 & & & & & & & & & & & & & \\ 
~~~~\bfm{e_{1}\mid e_{1}} & $~\raw~\{$ & $ -\psi_{1}$, & ~$\psi_{2}$,
 & ~$\psi_{3}$, & ~$\psi_{4} ~\}$
&~~~,~~~~& 
\bfm{e_{2}\mid e_{2}} & $~\raw~\{$ & $ -\psi_{1}^{*}$, & ~$-\psi_{2}^{*}$, & 
~$\psi_{3}$, & ~$\psi_{4} ~\}$
&~~~,~~~~\\
\bfm{e_{3}\mid e_{3}} & $~\raw~\{$ & $ -\psi_{1}^{*}$, & ~$\psi_{2}^{*}$, & 
~$\psi_{3}$, & ~$\psi_{4} ~\}$
&~~~,~~~~& 
\bfm{e_{4}\mid e_{4}} & $~\raw~\{$ & $ -\psi_{1}^{*}$, & ~$\psi_{2}$, 
& ~$-\psi_{3}^{*}$, & 
~$\psi_{4} ~\}$
&~~~,~~~~\\
\bfm{e_{5}\mid e_{5}} & $~\raw~\{$ & $ -\psi_{1}^{*}$, & ~$\psi_{2}$, & 
~$\psi_{3}^{*}$, & ~$\psi_{4} ~\}$
&~~~,~~~~& 
\bfm{e_{6}\mid e_{6}} & $~\raw~\{$ & $ -\psi_{1}^{*}$, & ~$\psi_{2}$, 
& ~$\psi_{3}$, & 
~$-\psi_{4}^{*} ~\}$
&~~~,~~~~\\
\bfm{e_{7}\mid e_{7}} & $~\raw~\{$ & $ -\psi_{1}^{*}$, & ~$\psi_{2}$, 
& ~$\psi_{3}$, & 
~$\psi_{4}^{*} ~\}$
&~~~.~~~~& & & & & & & \\
 & & & & & & & & & & & & & \\
\hline
\et
\ec
\new
%\vs{1.5cm}
\bc
{\tt TABLE B1-L}\\
{\fs Octonionic left-barred operators action on $\psi$}\\
\vs{.2cm}
\bt{|rcrrrrcrcrrrrc|}
\hline 
 & & & & & & & & & & & & & \\
~~~~\bfm{e_{1}~)~e_{2}} & 
$~\raw~\{$ & $ -e_{1}\psi_{2}^{*}$, & ~$-e_{1}\psi_{1}^{*}$, 
& ~$e_{1}\psi_{4}^{*}$, & 
~$-e_{1}\psi_{3}^{*} ~\}$
&~~~,~~~~& 
\bfm{e_{1}~)~e_{3}} & $~\raw~\{$ & $ -\psi_{2}^{*}$, & ~$-\psi_{1}^{*}$, 
& ~$-\psi_{4}^{*}$, & ~$\psi_{3}^{*} ~\}$
&~~~,~~~~\\
\bfm{e_{1}~)~e_{4}} & $~\raw~\{$ & $ -e_{1}\psi_{3}^{*}$, & 
~$-e_{1}\psi_{4}^{*}$, 
& ~$-e_{1}\psi_{1}^{*}$, & 
~$e_{1}\psi_{2}^{*} ~\}$
&~~~,~~~~& 
\bfm{e_{1}~)~e_{5}} & $~\raw~\{$ & $ -\psi_{3}^{*}$, & ~$\psi_{4}^{*}$, 
& ~$-\psi_{1}^{*}$, & ~$-\psi_{2}^{*} ~\}$
&~~~,~~~~\\
\bfm{e_{1}~)~e_{6}} & $~\raw~\{$ & $ -e_{1}\psi_{4}^{*}$, 
& ~$e_{1}\psi_{3}^{*}$, 
& ~$-e_{1}\psi_{2}^{*}$, & 
~$-e_{1}\psi_{1}^{*} ~\}$
&~~~,~~~~& 
\bfm{e_{1}~)~e_{7}} & $~\raw~\{$ & $ \psi_{4}^{*}$, & ~$\psi_{3}^{*}$, 
& ~$-\psi_{2}^{*}$, & 
~$\psi_{1}^{*} ~\}$
&~~~,~~~~\\
\bfm{e_{2}~)~e_{1}} & $~\raw~\{$ & $ -e_{1}\psi_{2}$, & ~$e_{1}\psi_{1}$, 
& ~$-e_{1}\psi_{4}^{*}$, & 
~$e_{1}\psi_{3}^{*} ~\}$
&~~~,~~~~& 
\bfm{e_{2}~)~e_{3}} & $~\raw~\{$ & $ e_{1}\psi_{1}^{*}$, 
& ~$e_{1}\psi_{2}^{*}$, 
& ~$e_{1}\psi_{3}$, & 
~$e_{1}\psi_{4} ~\}$
&~~~,~~~~\\
\bfm{e_{2}~)~e_{4}} & $~\raw~\{$ & $ \psi_{4}$, & ~$-\psi_{3}$, 
& ~$-\psi_{2}^{*}$, & 
~$\psi_{1}^{*} ~\}$
&~~~,~~~~& 
\bfm{e_{2}~)~e_{5}} & $~\raw~\{$ & $ -e_{1}\psi_{4}$, 
& ~$-e_{1}\psi_{3}$, 
& ~$e_{1}\psi_{2}^{*}$, & 
~$e_{1}\psi_{1}^{*} ~\}$
&~~~,~~~~\\
\bfm{e_{2}~)~e_{6}} & $~\raw~\{$ & $ -\psi_{3}$, & ~$-\psi_{4}$, 
& ~$-\psi_{1}^{*}$, & 
~$-\psi_{2}^{*} ~\}$
&~~~,~~~~& 
\bfm{e_{2}~)~e_{7}} & $~\raw~\{$ & $ -e_{1}\psi_{3}$, 
& ~$e_{1}\psi_{4}$, 
& ~$e_{1}\psi_{1}^{*}$, & 
~$-e_{1}\psi_{2}^{*} ~\}$
&~~~,~~~~\\
\bfm{e_{3}~)~e_{1}} & $~\raw~\{$ & $ \psi_{2}$, & ~$\psi_{1}$, 
& ~$\psi_{4}^{*}$, & 
~$-\psi_{3}^{*} ~\}$
&~~~,~~~~& 
\bfm{e_{3}~)~e_{2}} & $~\raw~\{$ & $ -e_{1}\psi_{1}^{*}$, 
& ~$e_{1}\psi_{2}^{*}$, 
& ~$-e_{1}\psi_{3}$, & 
~$-e_{1}\psi_{4} ~\}$
&~~~,~~~~\\
\bfm{e_{3}~)~e_{4}} & $~\raw~\{$ & $ -e_{1}\psi_{4}$, & ~$e_{1}\psi_{3}$, 
& ~$e_{1}\psi_{2}^{*}$, & 
~$e_{1}\psi_{1}^{*} ~\}$
&~~~,~~~~& 
\bfm{e_{3}~)~e_{5}} & $~\raw~\{$ & $ -\psi_{4}$, & ~$-\psi_{3}$, 
& ~$\psi_{2}^{*}$, & 
~$-\psi_{1}^{*} ~\}$
&~~~,~~~~\\
\bfm{e_{3}~)~e_{6}} & $~\raw~\{$ & $ e_{1}\psi_{3}$, & ~$e_{1}\psi_{4}$, 
& ~$-e_{1}\psi_{1}^{*}$, & 
~$e_{1}\psi_{2}^{*} ~\}$
&~~~,~~~~& 
\bfm{e_{3}~)~e_{7}} & $~\raw~\{$ & $ -\psi_{3}$, & ~$\psi_{4}$, 
& ~$-\psi_{1}^{*}$, & 
~$-\psi_{2}^{*} ~\}$
&~~~,~~~~\\
\bfm{e_{4}~)~e_{1}} & $~\raw~\{$ & $ -e_{1}\psi_{3}$, & ~$e_{1}\psi_{4}^{*}$, 
& ~$e_{1}\psi_{1}$, & 
~$-e_{1}\psi_{2}^{*} ~\}$
&~~~,~~~~& 
\bfm{e_{4}~)~e_{2}} & $~\raw~\{$ & $ -\psi_{4}$, & ~$-\psi_{3}^{*}$, 
& ~$-\psi_{2}$, & ~$-\psi_{1}^{*} ~\}$
&~~~,~~~~\\
\bfm{e_{4}~)~e_{3}} & $~\raw~\{$ & $ e_{1}\psi_{4}$, & 
~$e_{1}\psi_{3}^{*}$, 
& ~$-e_{1}\psi_{2}$, & 
~$-e_{1}\psi_{1}^{*} ~\}$
&~~~,~~~~& 
\bfm{e_{4}~)~e_{5}} & $~\raw~\{$ & $ e_{1}\psi_{1}^{*}$, & ~$e_{1}\psi_{2}$, 
& ~$e_{1}\psi_{3}^{*}$, & 
~$e_{1}\psi_{4} ~\}$
&~~~,~~~~\\
\bfm{e_{4}~)~e_{6}} & $~\raw~\{$ & $ \psi_{2}$, & ~$\psi_{1}^{*}$, 
& ~$-\psi_{4}$, & 
~$-\psi_{3}^{*} ~\}$
&~~~,~~~~& 
\bfm{e_{4}~)~e_{7}} & $~\raw~\{$ & $ e_{1}\psi_{2}$, & 
~$-e_{1}\psi_{1}^{*}$, 
& ~$e_{1}\psi_{4}$, & 
~$-e_{1}\psi_{3}^{*} ~\}$
&~~~,~~~~\\
\bfm{e_{5}~)~e_{1}} & $~\raw~\{$ & $ \psi_{3}$, & ~$-\psi_{4}^{*}$, 
& ~$\psi_{1}$, & 
~$\psi_{2}^{*} ~\}$
&~~~,~~~~& 
\bfm{e_{5}~)~e_{2}} & $~\raw~\{$ & $ e_{1}\psi_{4}$, & 
~$e_{1}\psi_{3}^{*}$, 
& ~$e_{1}\psi_{2}$, & 
~$-e_{1}\psi_{1}^{*} ~\}$
&~~~,~~~~\\
\bfm{e_{5}~)~e_{3}} & $~\raw~\{$ & $\psi_{4}$, & ~$\psi_{3}^{*}$, 
& ~$-\psi_{2}$, & 
~$\psi_{1}^{*} ~\}$
&~~~,~~~~& 
\bfm{e_{5}~)~e_{4}} & $~\raw~\{$ & $ -e_{1}\psi_{1}^{*}$, & 
~$-e_{1}\psi_{2}$, 
& ~$e_{1}\psi_{3}^{*}$, & 
~$-e_{1}\psi_{4} ~\}$
&~~~,~~~~\\
\bfm{e_{5}~)~e_{6}} & $~\raw~\{$ & $ -e_{1}\psi_{2}$, & ~$e_{1}\psi_{1}^{*}$, 
& ~$e_{1}\psi_{4}$, & 
~$e_{1}\psi_{3}^{*} ~\}$
&~~~,~~~~& 
\bfm{e_{5}~)~e_{7}} & $~\raw~\{$ & $ \psi_{2}$, & ~$\psi_{1}^{*}$, 
& ~$\psi_{4}$, & 
~$-\psi_{3}^{*} ~\}$
&~~~,~~~~\\
\bfm{e_{6}~)~e_{1}} & $~\raw~\{$ & $ -e_{1}\psi_{4}$, & ~$-e_{1}\psi_{3}^{*}$, 
& ~$e_{1}\psi_{2}^{*}$, & 
~$e_{1}\psi_{1} ~\}$
&~~~,~~~~& 
\bfm{e_{6}~)~e_{2}} & $~\raw~\{$ & $ \psi_{3}$, & ~$-\psi_{4}^{*}$, 
& ~$\psi_{1}^{*}$, & 
~$-\psi_{2} ~\}$
&~~~,~~~~\\
\bfm{e_{6}~)~e_{3}} & $~\raw~\{$ & $ -e_{1}\psi_{3}$, & ~$e_{1}\psi_{4}^{*}$, 
& ~$e_{1}\psi_{1}^{*}$, & 
~$-e_{1}\psi_{2} ~\}$
&~~~,~~~~& 
\bfm{e_{6}~)~e_{4}} & $~\raw~\{$ & $ -\psi_{2}$, & ~$-\psi_{1}^{*}$, 
& ~$-\psi_{4}^{*}$, & 
~$-\psi_{3} ~\}$
&~~~,~~~~\\
\bfm{e_{6}~)~e_{5}} & $~\raw~\{$ & $ e_{1}\psi_{2}$, 
& ~$-e_{1}\psi_{1}^{*}$, 
& ~$e_{1}\psi_{4}^{*}$, & 
~$-e_{1}\psi_{3} ~\}$
&~~~,~~~~& 
\bfm{e_{6}~)~e_{7}} & $~\raw~\{$ & $ -e_{1}\psi_{1}^{*}$, & ~$-e_{1}\psi_{2}$, 
& ~$-e_{1}\psi_{3}$, & 
~$-e_{1}\psi_{4}^{*} ~\}$
&~~~,~~~~\\
\bfm{e_{7}~)~e_{1}} & $~\raw~\{$ & $ -\psi_{4}$, & ~$-\psi_{3}^{*}$, 
& ~$\psi_{2}^{*}$, & 
~$-\psi_{1} ~\}$
&~~~,~~~~& 
\bfm{e_{7}~)~e_{2}} & $~\raw~\{$ & $ e_{1}\psi_{3}$, & 
~$-e_{1}\psi_{4}^{*}$, 
& ~$-e_{1}\psi_{1}^{*}$, & 
~$-e_{1}\psi_{2} ~\}$
&~~~,~~~~\\
\bfm{e_{7}~)~e_{3}} & $~\raw~\{$ & $ \psi_{3}$, & ~$-\psi_{4}^{*}$, 
& ~$\psi_{1}^{*}$, & 
~$\psi_{2} ~\}$
&~~~,~~~~& 
\bfm{e_{7}~)~e_{4}} & $~\raw~\{$ & $ -e_{1}\psi_{2}$, 
& ~$e_{1}\psi_{1}^{*}$, 
& ~$-e_{1}\psi_{4}^{*}$, & 
~$-e_{1}\psi_{3} ~\}$
&~~~,~~~~\\
\bfm{e_{7}~)~e_{5}} & $~\raw~\{$ & $ -\psi_{2}$, & ~$-\psi_{1}^{*}$, 
& ~$-\psi_{4}^{*}$, & 
~$\psi_{3} ~\}$
&~~~,~~~~& 
\bfm{e_{7}~)~e_{6}} & $~\raw~\{$ & $ e_{1}\psi_{1}^{*}$, & ~$e_{1}\psi_{2}$, 
& ~$e_{1}\psi_{3}$, & 
~$-e_{1}\psi_{4}^{*} ~\}$
&~~~.~~~~\\
 & & & & & & & & & & & & & \\
\hline
\et
\ec
\vs{1.5cm}
\bc
{\tt TABLE B1-R}\\
{\fs Octonionic right-barred operators action on $\psi$}\\
\vs{.2cm}
\bt{|rcrrrrcrcrrrrc|}
\hline
 & & & & & & & & & & & & & \\
~~~~\bfm{e_{1}~(~e_{2}} & 
$~\raw~\{$ & $-e_{1}\psi_{2}^{*}$, & ~$-e_{1}\psi_{1}^{*}$, 
& ~$-e_{1}\psi_{4}^{*}$, & 
~$e_{1}\psi_{3}^{*} ~\}$
&~~~,~~~~& 
\bfm{e_{1}~(~e_{3}} & $~\raw~\{$ & $ -\psi_{2}^{*}$, & ~$-\psi_{1}^{*}$, 
& ~$-\psi_{3}^{*}$, & 
~$\psi_{4}^{*} ~\}$
&~~~,~~~~\\
\bfm{e_{1}~(~e_{4}} & $~\raw~\{$ & $ -e_{1}\psi_{3}^{*}$, 
& ~$e_{1}\psi_{4}^{*}$, 
& ~$-e_{1}\psi_{1}^{*}$, & 
~$-e_{1}\psi_{2}^{*} ~\}$
&~~~,~~~~& 
\bfm{e_{1}~(~e_{5}} & $~\raw~\{$ & $ -\psi_{3}^{*}$, & ~$-\psi_{4}^{*}$, 
& ~$-\psi_{1}^{*}$, & ~$\psi_{2}^{*} ~\}$
&~~~,~~~~\\
\bfm{e_{1}~(~e_{6}} & $~\raw~\{$ & $ -e_{1}\psi_{4}^{*}$, &
 ~$-e_{1}\psi_{3}^{*}$, 
& ~$e_{1}\psi_{2}^{*}$, & 
~$-e_{1}\psi_{1}^{*} ~\}$
&~~~,~~~~& 
\bfm{e_{1}~(~e_{7}} & $~\raw~\{$ & $ \psi_{4}^{*}$, & ~$-\psi_{3}^{*}$, 
& ~$\psi_{2}^{*}$, & 
~$\psi_{1}^{*} ~\}$
&~~~,~~~~\\
\bfm{e_{2}~(~e_{1}} & $~\raw~\{$ & $ -e_{1}\psi_{2}$, & ~$e_{1}\psi_{1}$, 
& ~$e_{1}\psi_{4}^{*}$, & 
~$-e_{1}\psi_{3}^{*} ~\}$
&~~~,~~~~& 
\bfm{e_{2}~(~e_{3}} & $~\raw~\{$ & $ e_{1}\psi_{1}^{*}$, & 
~$e_{1}\psi_{2}^{*}$, 
& ~$-e_{1}\psi_{3}$, & 
~$-e_{1}\psi_{4} ~\}$
&~~~,~~~~\\
\bfm{e_{2}~(~e_{4}} & $~\raw~\{$ & $ \psi_{4}^{*}$, & ~$-\psi_{3}^{*}$, 
& ~$-\psi_{2}$, & ~$\psi_{1} ~\}$
&~~~,~~~~& 
\bfm{e_{2}~(~e_{5}} & $~\raw~\{$ & $ e_{1}\psi_{4}^{*}$, 
& ~$e_{1}\psi_{3}^{*}$, 
& ~$e_{1}\psi_{2}$, & 
~$\psi_{1} ~\}$
&~~~,~~~~\\
\bfm{e_{2}~(~e_{6}} & $~\raw~\{$ & $ -\psi_{3}^{*}$, & ~$-\psi_{4}^{*}$, 
& ~$-\psi_{1}$, & 
~$-\psi_{2} ~\}$
&~~~,~~~~& 
\bfm{e_{2}~(~e_{7}} & $~\raw~\{$ & $ e_{1}\psi_{3}^{*}$, 
& ~$-e_{1}\psi_{4}^{*}$, 
& ~$e_{1}\psi_{1}$, & 
~$-e_{1}\psi_{2} ~\}$
&~~~,~~~~\\
\bfm{e_{3}~(~e_{1}} & $~\raw~\{$ & $ \psi_{2}$, & ~$\psi_{1}$, 
& ~$-\psi_{4}^{*}$, & ~$\psi_{3}^{*} ~\}$
&~~~,~~~~& 
\bfm{e_{3}~(~e_{2}} & $~\raw~\{$ & $ -e_{1}\psi_{1}^{*}$, 
& ~$e_{1}\psi_{2}^{*}$, 
& ~$e_{1}\psi_{3}$, & 
~$e_{1}\psi_{4} ~\}$
&~~~,~~~~\\
\bfm{e_{3}~(~e_{4}} & $~\raw~\{$ & $ e_{1}\psi_{4}^{*}$, 
& ~$e_{1}\psi_{3}^{*}$, 
& ~$-e_{1}\psi_{2}$, & 
~$e_{1}\psi_{1} ~\}$
&~~~,~~~~& 
\bfm{e_{3}~(~e_{5}} & $~\raw~\{$ & $ -\psi_{4}^{*}$, & ~$\psi_{3}^{*}$, 
& ~$-\psi_{2}$, & 
~$-\psi_{1} ~\}$
&~~~,~~~~\\
\bfm{e_{3}~(~e_{6}} & $~\raw~\{$ & $ -e_{1}\psi_{3}^{*}$, & 
~$e_{1}\psi_{4}^{*}$, 
& ~$-e_{1}\psi_{2}$, & 
~$-e_{1}\psi_{1} ~\}$
&~~~,~~~~& 
\bfm{e_{3}~(~e_{7}} & $~\raw~\{$ & $ -\psi_{3}^{*}$, & ~$-\psi_{4}^{*}$, 
& ~$-\psi_{1}$, & 
~$\psi_{2} ~\}$
&~~~,~~~~\\
\bfm{e_{4}~(~e_{1}} & $~\raw~\{$ & $ -e_{1}\psi_{3}$, & 
~$-e_{1}\psi_{4}^{*}$, 
& ~$e_{1}\psi_{1}$, & 
~$e_{1}\psi_{2}^{*} ~\}$
&~~~,~~~~& 
\bfm{e_{4}~(~e_{2}} & $~\raw~\{$ & $ -\psi_{4}^{*}$, & ~$-\psi_{3}$, 
& ~$-\psi_{2}^{*}$, & 
~$-\psi_{1} ~\}$
&~~~,~~~~\\
\bfm{e_{4}~(~e_{3}} & $~\raw~\{$ & $ -e_{1}\psi_{4}^{*}$, & ~$e_{1}\psi_{3}$, 
& ~$e_{1}\psi_{2}^{*}$, & 
~$-e_{1}\psi_{1} ~\}$
&~~~,~~~~& 
\bfm{e_{4}~(~e_{5}} & $~\raw~\{$ & $ e_{1}\psi_{1}^{*}$, & ~$-e_{1}\psi_{2}$, 
& ~$e_{1}\psi_{3}^{*}$, & 
~$-e_{1}\psi_{4} ~\}$
&~~~,~~~~\\
\bfm{e_{4}~(~e_{6}} & $~\raw~\{$ & $ \psi_{2}^{*}$, & ~$\psi_{1}$, 
& ~$-\psi_{3}$, & 
~$-\psi_{4}^{*} ~\}$
&~~~,~~~~& 
\bfm{e_{4}~(~e_{7}} & $~\raw~\{$ & $ -e_{1}\psi_{2}^{*}$, 
& ~$-e_{1}\psi_{1}$, 
& ~$-e_{1}\psi_{4}^{*}$, & 
~$-e_{1}\psi_{3} ~\}$
&~~~,~~~~\\
\bfm{e_{5}~(~e_{1}} & $~\raw~\{$ & $ \psi_{3}$, & ~$\psi_{4}^{*}$, &
 ~$\psi_{1}$, & 
~$-\psi_{2}^{*} ~\}$
&~~~,~~~~& 
\bfm{e_{5}~(~e_{2}} & $~\raw~\{$ & $ -e_{1}\psi_{4}^{*}$, & ~$-e_{1}\psi_{3}$, 
& ~$e_{1}\psi_{2}^{*}$, & 
~$-e_{1}\psi_{1} ~\}$
&~~~,~~~~\\
\bfm{e_{5}~(~e_{3}} & $~\raw~\{$ & $ \psi_{4}^{*}$, & ~$-\psi_{3}$, 
& ~$\psi_{2}^{*}$, & ~$\psi_{1} ~\}$
&~~~,~~~~& 
\bfm{e_{5}~(~e_{4}} & $~\raw~\{$ & $ -e_{1}\psi_{1}^{*}$, & ~$e_{1}\psi_{2}$, 
& ~$e_{1}\psi_{3}^{*}$, & 
~$e_{1}\psi_{4} ~\}$
&~~~,~~~~\\
\bfm{e_{5}~(~e_{6}} & $~\raw~\{$ & $ e_{1}\psi_{2}^{*}$, & ~$e_{1}\psi_{1}$, 
& ~$e_{1}\psi_{4}^{*}$, & 
~$-e_{1}\psi_{3} ~\}$
&~~~,~~~~& 
\bfm{e_{5}~(~e_{7}} & $~\raw~\{$ & $ \psi_{2}^{*}$, & ~$\psi_{1}$, &
 ~$-\psi_{4}^{*}$, & ~$\psi_{3} ~\}$
&~~~,~~~~\\
\bfm{e_{6}~(~e_{1}} & $~\raw~\{$ & $ -e_{1}\psi_{4}$, & 
~$e_{1}\psi_{3}^{*}$, 
& ~$-e_{1}\psi_{2}^{*}$, & 
~$e_{1}\psi_{1} ~\}$
&~~~,~~~~& 
\bfm{e_{6}~(~e_{2}} & $~\raw~\{$ & $ \psi_{3}^{*}$, & ~$-\psi_{4}$, 
& ~$\psi_{1}$, & ~$-\psi_{2}^{*} ~\}$
&~~~,~~~~\\
\bfm{e_{6}~(~e_{3}} & $~\raw~\{$ & $ e_{1}\psi_{3}^{*}$, & ~$e_{1}\psi_{4}$, 
& ~$e_{1}\psi_{1}$, & 
~$e_{1}\psi_{2}^{*} ~\}$
&~~~,~~~~& 
\bfm{e_{6}~(~e_{4}} & $~\raw~\{$ & $ -\psi_{2}^{*}$, & ~$-\psi_{1}$, 
& ~$-\psi_{4}$, & 
~$-\psi_{3}^{*} ~\}$
&~~~,~~~~\\
\bfm{e_{6}~(~e_{5}} & $~\raw~\{$ & $ -e_{1}\psi_{2}^{*}$, & ~$-e_{1}\psi_{1}$, 
& ~$e_{1}\psi_{4}$, & 
~$e_{1}\psi_{3}^{*} ~\}$
&~~~,~~~~& 
\bfm{e_{6}~(~e_{7}} & $~\raw~\{$ & $ -e_{1}\psi_{1}^{*}$, & ~$e_{1}\psi_{2}$, 
& ~$e_{1}\psi_{3}$, & 
~$-e_{1}\psi_{4}^{*} ~\}$
&~~~,~~~~\\
\bfm{e_{7}~(~e_{1}} & $~\raw~\{$ & $ -\psi_{4}$, 
& ~$\psi_{3}^{*}$, & ~$-\psi_{2}^{*}$, & 
~$-\psi_{1} ~\}$
&~~~,~~~~& 
\bfm{e_{7}~(~e_{2}} & $~\raw~\{$ & $ -e_{1}\psi_{3}^{*}$, & ~$e_{1}\psi_{4}$, 
& ~$-e_{1}\psi_{1}$, & 
~$-e_{1}\psi_{2}^{*} ~\}$
&~~~,~~~~\\
\bfm{e_{7}~(~e_{3}} & $~\raw~\{$ & $ \psi_{3}^{*}$, & ~$\psi_{4}$, 
& ~$\psi_{1}$, & 
~$-\psi_{2}^{*} ~\}$
&~~~,~~~~& 
\bfm{e_{7}~(~e_{4}} & $~\raw~\{$ & $ e_{1}\psi_{2}^{*}$, & 
~$e_{1}\psi_{1}$, 
& ~$e_{1}\psi_{4}$, & 
~$-e_{1}\psi_{3}^{*} ~\}$
&~~~,~~~~\\
\bfm{e_{7}~(~e_{5}} & $~\raw~\{$ & $ -\psi_{2}^{*}$, & ~$-\psi_{1}$, 
& ~$\psi_{4}$, & 
~$-\psi_{3}^{*} ~\}$
&~~~,~~~~& 
\bfm{e_{7}~(~e_{6}} & $~\raw~\{$ & $ e_{1}\psi_{1}^{*}$, & ~$-e_{1}\psi_{2}$, 
& ~$-e_{1}\psi_{3}$, & 
~$-e_{1}\psi_{4}^{*} ~\}$
&~~~.~~~~\\
 & & & & & & & & & & & & & \\
\hline
\et
\ec

\new

\section*{Appendix B2}

In the following charts we establish the connection between $4\times 4$ 
complex matrices and octonionic left/right-barred operators. We indicate with 
${\cal R}_{mn}$ (${\cal C}_{mn}$)  the $4\times 4$ real (complex) matrices 
with 1 ($i$) in $mn$-element and zeros elsewhere.\\ 
\bc
{\tt $4 \times 4$ complex matrices and left-barred operators:}
\ec
\bean
{\cal R}_{11} & ~\lrw~ & \frac{1}{2}~[~1-e_{1}\mid e_{1}~] \\ 
{\cal R}_{12} & ~\lrw~ & \frac{1}{6}~[~2 e_{1}~)~e_{3} + e_{3}~)~e_{1} - 
2 \mid e_{2} - e_{2} + e_{4}~)~e_{6} - e_{6}~)~e_{4} + 
e_{5}~)~e_{7} - e_{7}~)~e_{5} ~] \\ 
{\cal R}_{13} & ~\lrw~ & \frac{1}{6}~[~2 e_{1}~)~e_{5} + e_{5}~)~e_{1} - 
2 \mid e_{4} - e_{4} + e_{6}~)~e_{2} - e_{2}~)~e_{6} + 
e_{7}~)~e_{3} - e_{3}~)~e_{7} ~] \\ 
{\cal R}_{14} & ~\lrw~ & \frac{1}{6}~[~2 e_{1}~)~e_{7} + e_{7}~)~e_{1} - 
2 \mid e_{6} - e_{6} + e_{2}~)~e_{4} - e_{4}~)~e_{2} + 
e_{5}~)~e_{3} - e_{3}~)~e_{5} ~] \\ 
{\cal R}_{21} & ~\lrw~ & \frac{1}{2}~[~e_{2} + e_{3}~)~e_{1} ~] \\ 
{\cal R}_{22} & ~\lrw~ & \frac{1}{6}~[~1+e_{1}\mid e_{1}+e_{4}\mid e_{4}+
e_{5}\mid e_{5}+e_{6}\mid e_{6}+e_{7}\mid e_{7}~] -
\frac{1}{3}~[~e_{2}\mid e_{2}+e_{3}\mid e_{3}~]\\
{\cal R}_{23} & ~\lrw~ & \frac{1}{2}~[~-e_{2}~)~e_{4} - e_{3}~)~e_{5} ~] \\ 
{\cal R}_{24} & ~\lrw~ & \frac{1}{2}~[~e_{3}~)~e_{7} - e_{2}~)~e_{6} ~] \\ 
{\cal R}_{31} & ~\lrw~ & \frac{1}{2}~[~e_{4} + e_{5}~)~e_{1} ~] \\ 
{\cal R}_{32} & ~\lrw~ & \frac{1}{2}~[~-e_{5}~)~e_{3} - e_{4}~)~e_{2} ~] \\ 
{\cal R}_{33} & ~\lrw~ & \frac{1}{6}~[~1+e_{1}\mid e_{1}+e_{2}\mid e_{2}+
e_{3}\mid e_{3}+e_{6}\mid e_{6}+e_{7}\mid e_{7}~] -
\frac{1}{3}~[~e_{4}\mid e_{4}+e_{5}\mid e_{5}~]\\
{\cal R}_{34} & ~\lrw~ & \frac{1}{2}~[~e_{5}~)~e_{7} - e_{4}~)~e_{6} ~] \\ 
{\cal R}_{41} & ~\lrw~ & \frac{1}{2}~[~e_{6} - e_{7}~)~e_{1} ~] \\ 
{\cal R}_{42} & ~\lrw~ & \frac{1}{2}~[~e_{7}~)~e_{3} - e_{6}~)~e_{2} ~] \\ 
{\cal R}_{43} & ~\lrw~ & \frac{1}{2}~[~e_{7}~)~e_{5} - e_{6}~)~e_{4} ~] \\ 
{\cal R}_{44} & ~\lrw~ & \frac{1}{6}~[~1+e_{1}\mid e_{1}+e_{2}\mid e_{2}+
e_{3}\mid e_{3}+e_{4}\mid e_{4}+e_{5}\mid e_{5}~] -
\frac{1}{3}~[~e_{6}\mid e_{6}+e_{7}\mid e_{7}~]\\
{\cal C}_{11} & ~\lrw~ & \frac{1}{2}~[~1\mid e_{1}+e_{1}~] \\ 
{\cal C}_{12} & ~\lrw~ & \frac{1}{6}~[~-2 e_{1}~)~e_{2} - e_{3} - 2 \mid e_{3}
- e_{2}~)~e_{1}   + e_{4}~)~e_{7} + e_{6}~)~e_{5} - e_{5}~)~e_{6} - 
e_{7}~)~e_{4}  ~] \\ 
{\cal C}_{13} & ~\lrw~ & \frac{1}{6}~[~-2 e_{1}~)~e_{4} - e_{5} - 2 \mid e_{5} 
- e_{4}~)~e_{1} - e_{6}~)~e_{3} - e_{2}~)~e_{7}  + e_{7}~)~e_{2} 
+ e_{3}~)~e_{6} ~] \\ 
{\cal C}_{14} & ~\lrw~ & \frac{1}{6}~[~-2 e_{1}~)~e_{6} + e_{7} + 2 \mid e_{7} 
- e_{6}~)~e_{1} - e_{2}~)~e_{5} + 
e_{4}~)~e_{3} + e_{5}~)~e_{2}  - e_{3}~)~e_{4} ~] \\ 
{\cal C}_{21} & ~\lrw~ & \frac{1}{2}~[~-e_{3} + e_{2}~)~e_{1} ~] \\ 
{\cal C}_{22} & ~\lrw ~ & \frac{1}{6}~[~1\mid e_{1} - e_{1} + e_{4}~)~e_{5}
-e_{5}~)~e_{4}-e_{6}~)~ e_{7}+e_{7}~)~ e_{6}~] -
\frac{1}{3}~[~e_{2}~)~ e_{3}-e_{3}~)~e_{2}~]\\
{\cal C}_{23} & ~\lrw~ & \frac{1}{2}~[~ - e_{2}~)~e_{5} + e_{3}~)~e_{4} ~] \\ 
{\cal C}_{24} & ~\lrw~ & \frac{1}{2}~[~ e_{3}~)~e_{6} + e_{2}~)~e_{7}~] \\ 
{\cal C}_{31} & ~\lrw~ & \frac{1}{2}~[~- e_{5} + e_{4}~)~e_{1} ~] \\ 
{\cal C}_{32} & ~\lrw~ & \frac{1}{2}~[~e_{5}~)~e_{2} - e_{4}~)~e_{3} ~] \\ 
{\cal C}_{33} & ~\lrw~ & \frac{1}{6}~[~1 \mid e_{1} - e_{1} +e_{2}~)~ e_{3}
-e_{3}~)~ e_{2}-e_{6}~)~ e_{7}+e_{7}~)~e_{6}~] -
\frac{1}{3}~[~e_{4}~)~ e_{5}-e_{5}~)~e_{4}~]\\
{\cal C}_{34} & ~\lrw~ & \frac{1}{2}~[~e_{5}~)~e_{6} + e_{4}~)~e_{7} ~] \\ 
{\cal C}_{41} & ~\lrw~ & \frac{1}{2}~[~ e_{7} + e_{6}~)~e_{1} ~] \\ 
{\cal C}_{42} & ~\lrw~ & \frac{1}{2}~[~- e_{7}~)~e_{2} - e_{6}~)~e_{3}  ~] \\ 
{\cal C}_{43} & ~\lrw~ & \frac{1}{2}~[~- e_{7}~)~e_{4} -e_{6}~)~e_{5}  ~] \\ 
{\cal C}_{44} & ~\lrw~ & \frac{1}{6}~[~1\mid e_{1} - e_{1} +e_{2}~)~e_{3}
-e_{3}~)~ e_{2}+e_{4}~)~ e_{5}-e_{5}~)~ e_{4}~] -
\frac{1}{3}~[~e_{7}~)~ e_{6}-e_{6}~)~ e_{7}~]
\eean
\\
\bc
{\tt $4 \times 4$ complex matrices and right-barred operators:}
\ec
\bean
{\cal R}_{11} & ~\lrw~ & \frac{1}{2}~[~1-e_{1}\mid e_{1}~] \\ 
{\cal R}_{12} & ~\lrw~ & \frac{1}{2}~[~-e_{2} + e_{3}~(~e_{1} ~] \\ 
{\cal R}_{13} & ~\lrw~ & \frac{1}{2}~[~-e_{4} + e_{5}~(~e_{1} ~] \\ 
{\cal R}_{14} & ~\lrw~ & \frac{1}{2}~[~-e_{6} - e_{7}~(~e_{1} ~] \\ 
{\cal R}_{21} & ~\lrw~ & \frac{1}{6}~[~2 e_{1}~(~e_{3} + e_{3}~(~e_{1} +
2 \mid e_{2} + e_{2} + e_{4}~(~e_{6} - e_{6}~(~e_{4} + 
e_{5}~(~e_{7} - e_{7}~(~e_{5} ~] \\ 
{\cal R}_{22} & ~\lrw~ & \frac{1}{6}~[~1+e_{1}\mid e_{1}+e_{4}\mid e_{4}+
e_{5}\mid e_{5}+e_{6}\mid e_{6}+e_{7}\mid e_{7}~] -
\frac{1}{3}~[~e_{2}\mid e_{2}+e_{3}\mid e_{3}~]\\
{\cal R}_{23} & ~\lrw~ & \frac{1}{2}~[~-e_{5}~(~e_{3} - e_{4}~(~e_{2} ~] \\ 
{\cal R}_{24} & ~\lrw~ & \frac{1}{2}~[~e_{7}~(~e_{3} - e_{6}~(~e_{2} ~] \\ 
{\cal R}_{31} & ~\lrw~ & \frac{1}{6}~[~2 e_{1}~(~e_{5} + e_{5}~(~e_{1} + 
2 \mid e_{4} + e_{4} + e_{6}~(~e_{2} - e_{2}~(~e_{6} + 
e_{7}~(~e_{3} - e_{3}~(~e_{7} ~] \\ 
{\cal R}_{32} & ~\lrw~ & \frac{1}{2}~[~-e_{2}~(~e_{4} - e_{3}~(~e_{5} ~] \\ 
{\cal R}_{33} & ~\lrw~ & \frac{1}{6}~[~1+e_{1}\mid e_{1}+e_{2}\mid e_{2}+
e_{3}\mid e_{3}+e_{6}\mid e_{6}+e_{7}\mid e_{7}~] -
\frac{1}{3}~[~e_{4}\mid e_{4}+e_{5}\mid e_{5}~]\\
{\cal R}_{34} & ~\lrw~ & \frac{1}{2}~[~e_{7}~(~e_{5} - e_{6}~(~e_{4} ~] \\ 
{\cal R}_{41} & ~\lrw~ & \frac{1}{6}~[~2 e_{1}~(~e_{7} + e_{7}~(~e_{1} + 
2 \mid e_{6} + e_{6} + e_{2}~(~e_{4} - e_{4}~(~e_{2} + 
e_{5}~(~e_{3} - e_{3}~(~e_{5} ~] \\ 
{\cal R}_{42} & ~\lrw~ & \frac{1}{2}~[~e_{3}~(~e_{7} - e_{2}~(~e_{6} ~] \\ 
{\cal R}_{43} & ~\lrw~ & \frac{1}{2}~[~e_{5}~(~e_{7} - e_{4}~(~e_{6} ~] \\ 
{\cal R}_{44} & ~\lrw~ & \frac{1}{6}~[~1+e_{1}\mid e_{1}+e_{2}\mid e_{2}+
e_{3}\mid e_{3}+e_{4}\mid e_{4}+e_{5}\mid e_{5}~] -
\frac{1}{3}~[~e_{6}\mid e_{6}+e_{7}\mid e_{7}~]\\
{\cal C}_{11} & ~\lrw~ & \frac{1}{2}~[~1\mid e_{1}+e_{1}~] \\ 
{\cal C}_{12} & ~\lrw~ & \frac{1}{2}~[~-e_{2}~(~e_{1} - e_{3} ~] \\ 
{\cal C}_{13} & ~\lrw~ & \frac{1}{2}~[~-e_{4}~(~e_{1} - e_{5} ~] \\ 
{\cal C}_{14} & ~\lrw~ & \frac{1}{2}~[~-e_{6}~(~e_{1} + e_{7} ~] \\ 
{\cal C}_{21} & ~\lrw~ & \frac{1}{6}~[~2 e_{1}~(~e_{2} - e_{3} +
-2 \mid e_{3} + e_{2}~(~e_{1} + e_{4}~(~e_{7} + e_{6}~(~e_{5} - 
e_{5}~(~e_{6} - e_{7}~(~e_{4} ~] \\ 
{\cal C}_{22} & ~\lrw~ & \frac{1}{6}~[~1\mid e_{1}-e_{1}-e_{4}~(~e_{5}+
e_{5}~(~ e_{4}+e_{6}~(~ e_{7}-e_{7}~(~ e_{6}~] -
\frac{1}{3}~[~-e_{2}~(~e_{3}+e_{3}~(~e_{2}~]\\
{\cal C}_{23} & ~\lrw~ & \frac{1}{2}~[~-e_{5}~(~e_{2} + e_{4}~(~e_{3} ~] \\ 
{\cal C}_{24} & ~\lrw~ & \frac{1}{2}~[~e_{7}~(~e_{2} + e_{6}~(~e_{3} ~] \\ 
{\cal C}_{31} & ~\lrw~ & \frac{1}{6}~[~2 e_{1}~(~e_{4} - e_{5} - 
2 \mid e_{5} + e_{4}~(~e_{1} - e_{6}~(~e_{3} - e_{2}~(~e_{7} + 
e_{7}~(~e_{2} + e_{3}~(~e_{6} ~] \\ 
{\cal C}_{32} & ~\lrw~ & \frac{1}{2}~[~e_{2}~(~e_{5} - e_{3}~(~e_{4} ~] \\ 
{\cal C}_{33} & ~\lrw~ & \frac{1}{6}~[~1\mid e_{1}-e_{1}-e_{2}~(~ e_{3}+
e_{3}~(~e_{2}+e_{6}~(~ e_{7}-e_{7}~(~e_{6}~] -
\frac{1}{3}~[~-e_{4}~(~ e_{5}+e_{5}\mid e_{4}~]\\
{\cal C}_{34} & ~\lrw~ & \frac{1}{2}~[~e_{7}~(~e_{4} + e_{6}~(~e_{5} ~] \\ 
{\cal C}_{41} & ~\lrw~ & \frac{1}{6}~[~-2 e_{1}~(~e_{6} - e_{7} + 
2 \mid e_{7} + e_{6}~(~e_{1} - e_{2}~(~e_{5} + e_{4}~(~e_{3} + 
e_{5}~(~e_{2} - e_{3}~(~e_{4} ~] \\ 
{\cal C}_{42} & ~\lrw~ & \frac{1}{2}~[~-e_{3}~(~e_{6} - e_{2}~(~e_{7} ~] \\ 
{\cal C}_{43} & ~\lrw~ & \frac{1}{2}~[~-e_{5}~(~e_{6} - e_{4}~(~e_{7} ~] \\ 
{\cal C}_{44} & ~\lrw~ & \frac{1}{6}~[~1\mid e_{1}-e_{1}-e_{2}~(~e_{3}+
e_{3}~(~e_{2}-e_{4}~(~ e_{5}+e_{5}\mid e_{4}~] -
\frac{1}{3}~[~e_{6}~(~ e_{7}-e_{7}~(~e_{6}~]
\eean

\vs{1cm}

The previous tables could be very useful in order to extract octonionic 
operators multiplication rules or left/right barred operators connection.  
For example, we quickly find
\[ 
~[~e_{2}~)~e_{7} + e_{3}~)~e_{6} ~] ~~ \lrw ~~ 
2 \; {\cal C}_{24} ~~\lrw ~~ [~e_{7}~(~e_{2} + e_{6}~(~e_{3} ~] 
\q , \qq \mbox{and so on} \q .
\]

\new

\section*{References}
\bref
\bi{jor}
P.~Jordan, Z.~Phys. {\bf 80}, 285 (1933).
\bi{wig}
P.~Jordan and E.~P.~Wigner, Ann.~Math. {\bf 35}, 29 (1934).
\bi{alb1}
A.~A.~Albert, Ann.~Math. {\bf 35}, 65 (1934).
\bi{om1}
C.~Chevalley and R.~D.~Schafer, Proc.~Nat.~Acad.~Sci. {\bf 36}, 137 (1950).
\bi{om2}
J.~Tits, Proc.~Colloq.~Utrecht, 135 (1962).
\bi{om3}
H.~Freudenthal, Advances in Math.~I, 145 (1965).
\bi{om4}
R.~D.~Schafer,  {\it An introduction to Non-Associative Algebras} 
(Academic Press, New York, 1966).
\bi{gur1}
F.~G\"ursey, {\it Symmetries in Physics (1600-1980): Proc.~of the 1st 
International Meeting on the History of Scientific Ideas}, Seminari 
d'~Hist\`oria de les Ci\`ences, Barcelona, Spain, 1987, p.~557.
\bi{pais}
A.~Pais, \pxh{7}{291}{61}.
\bi{gur2}
M.~G\"unaydin and F.~G\"ursey, \jxe{14}{1651}{73}; \pxf{9}{3387}{74}.
\bi{mor}
K.~Morita, \pxxa{65}{787}{81}.
\bi{dix}
G.~Dixon, \nxd{B105}{349}{90}.
\bi{gur3}
F.~G\"ursey, {\it Yale Preprint C00-3075-178} (1978).
\bi{edm}
J.~D.~Edmonds, \pxa{5}{56}{92}.
\bi{jos1}
A.~Waldron and G.~C.~Joshi, {\it Melbourne Preprint UM-P-92/60} (1992).
\bi{jos2}
G.~C.~Lassig and G.~C.~Joshi, {\it Melbourne Preprint UM-P-95/09} (1995).\\
A.~Ritz and G.~C.~Joshi, {\it Melbourne Preprint UM-P-95/69} (1995).
\bi{dav}
A.~J.~Davies and G.~C.~Joshi, \jxe{27}{3036}{86}.
\bi{sup1}
T.~Kugo and P.~Townsend, \nxb{B221}{357}{87}.
\bi{sup2}
B.~Julia, {\it Lptens Preprint 82/14} (1982).
\bi{adl}
S.~L.~Adler, {\it Quaternionic Quantum Mechanics and Quantum Fields} 
(Oxford, New York, 1995).
\bi{adl1}
S.~L.~Adler,  \nxb{B415}{195}{94}.
\bi{qua1}
S.~De Leo and P.~Rotelli, \pxf{45}{575}{92}; \nxd{B110}{33}{95};\\
S.~De Leo, \pxxa{94}{11}{95}; {\it Quaternions for GUTs}, 
Int.~J.~Theor.~Phys. (submitted).
\bi{qua2}
S.~De Leo and P.~Rotelli, \pxxa{92}{917}{94}.
\bi{qua3} 
S.~De Leo and P.~Rotelli, \ixa{10}{4359}{95}.\\
S.~De Leo and P.~Rotelli, {\it Quaternionic Dirac Lagrangian}, 
Mod.~Phys.~Lett.~A (to be published).\\
S.~De Leo and P.~Rotelli, {\it Quaternionic Electroweak Theory}, 
J.~Phys.~G (submitted).
\bi{dir1}
S.~L.~Adler, \pxi{221B}{39}{89}.
\bi{dir2}
P.~Rotelli, \mxb{4}{933}{89}.
\bi{dir3}
A.~J.~Davies, \pxf{41}{2628}{90}.
\bi{dir4}
S.~De Leo, {\it One-component Dirac Equation}, Int.~J.~Mod.~Phys.~A 
(to be published).
\bi{pen}
In literature we find a Dirac equation formulation by {\tt complexified} 
octonions with an embarrassing doubling of solutions: {\sl ``... the wave 
functions $\tilde{\psi}$ is not a column matrix, but must be taken as an 
octonion. $\tilde{\psi}$ therefore consists of eight wave functions, rather 
that the four wave functions of the Dirac equation''}\\
R.~Penney, \nxd{B3}{95}{71}.
\bi{hor}
L.~P.~Horwitz and L.~C.~Biedenharn, \axp{157}{432}{84}. 
\bi{rem}
J.~Rembieli\'nski, \jxg{11}{2323}{78}.
\bi{alb2}
A.~A.~Albert, Ann.~Math. {\bf 48}, 495 (1947).
\bi{ham}
W.~R.~Hamilton, {\it Elements of Quaternions} (Chelsea Publishing Co., New 
York, 1969).
\bi{gra}
J.~T.~Graves, {\it Mathematical Papers} (1843).
\bi{cay}
A.~Cayley, Phil.~Mag.~(London) {\bf 26}, 210 (1845).
A.~Cayley, {\it Papers Collected Mathematical Papers} (Cambridge 1889). 
\bi{fro}
G.~Frobenius, J.~Reine Angew.~Math. {\bf 84}, 59 (1878).
\bi{hur}
A.~Hurwitz, Nachr.~Gesell.~Wiss.~G\"ottingeg, Math.~Phys.~Kl., 309.\\
A.~Hurwitz, {\it Mathematische Werke Band II, Zahlentheorie, Algebra und 
Geometrie}, pag.~565 (Birkha\"user, Basel, 1933).
\bi{bot}
R.~Bott and J.~Milnor, Bull.~Amer.~Math.~Soc. {\bf 64}, 87 (1958). 
\bi{ker}
M.~Kervaire, Proc.~Nat.~Acad.~Sci. {\bf 44}, 280 (1958).
\bi{oku}
S.~Okubo, {\it Introduction to Octonion and Other Non-Associative Algebras 
in Physics} (Cambridge University Press, Cambridge, to be published).
\bi{mor2}
K.~Morita, \pxxa{67}{1860}{81}; \xxx{68}{2159}{82}; \xxx{70}{1648}{83}; 
\xxx{72}{1056}{84}; \xxx{73}{999}{84}; \xxx{75}{220}{85}; \xxx{90}{219}{93}. 
\bi{rel}
S.~De Leo, {\it Quaternions and Special Relativity}, J.~Math.~Phys. 
(to be published).
\bi{math}
S.~Wolfram, {\it Mathematica} (Addison-Wesley Publishing Co., Redwood City, 
1991).
\bi{itz}
The eqs.~(\ref{odgm1}-d)~[(\ref{odgm2}-d), (\ref{odgm3}-d)] 
represent the octonionic 
counterpart of the complex matrices given on pag.~49~[694] of the book:\\
C.~Itzykson and J.~B.~Zuber, {\it Quantum Field Theory} 
(McGraw-Hill, New York, 1985).
\bi{penr}
R.~Penrose and W.~Rindler, {\it Spinors and Space-Time} 
(Cambridge UP, Cambridge, 1984).
\bi{adl2}
S.~Adler, \pxi{332B}{358}{94}.

\eref

%%%%%%%%%%%%%%%%%%
\end{document}